\documentclass{egpubl}
\usepackage{sca2022}

\SpecialIssuePaper         

\usepackage[T1]{fontenc}
\usepackage{dfadobe}  
\usepackage{array}
\usepackage{booktabs}
\usepackage{makecell}

\usepackage{cite}  
\BibtexOrBiblatex
\electronicVersion
\PrintedOrElectronic
\ifpdf \usepackage[pdftex]{graphicx} \pdfcompresslevel=9
\else \usepackage[dvips]{graphicx} \fi

\usepackage{egweblnk} 

\usepackage{xcolor}
\usepackage{amsmath}
\usepackage{multirow}
\usepackage[normalem]{ulem}

\newcommand*{\final}{}

\ifdefined\final
\newcommand{\added}[1]{{{#1}}}
\newcommand{\deleted}[1]{}
\else
\newcommand{\added}[1]{\textcolor{blue}{{#1}}}
\newcommand{\deleted}[1]{\textcolor{red}{\sout{#1}}}
\fi

\title[High-Order Elasticity Interpolants for Microstructure Simulation]{High-Order Elasticity Interpolants for Microstructure Simulation}

\ifdefined\final
\author[A. Chan-Lock, J. Pérez \& M. A. Otaduy]
{
\parbox{\textwidth}{\centering Antoine Chan-Lock\orcid{0000-0002-4632-9579}, Jesús Pérez\orcid{0000-0001-5085-6681} and Miguel A. Otaduy\orcid{0000-0002-3880-7622}}
        \\
{
\parbox{\textwidth}{\centering Universidad Rey Juan Carlos, Madrid, Spain}
}
}
\else
\author{SUBMISSION 8549}
\fi

\begin{document}

\let\rm=\rmfamily
\let\sf=\sffamily
\let\tt=\ttfamily
\let\it=\itshape
\let\sl=\slshape
\let\sc=\scshape
\let\bf=\bfseries

\renewcommand{\vec}[1]{\mathbf{#1}}
\newcommand{\mat}[1]{\mathbf{#1}}
\newcommand{\set}[1]{\mathcal{#1}}
\newcommand{\func}[1]{\mathrm{#1}}

\def \N {\mbox{\rm \hbox{I\kern-.15em\hbox{N}}}}
\def \R {\mbox{\rm \hbox{I\kern-.15em\hbox{R}}}}
\def \laplace {\Delta}
\def \grad {\nabla}
\newcommand{\of}[1]{\!\left( #1 \right)}
\newcommand{\abs}[1]{\left| #1 \right|}
\newcommand{\norm}[1]{\left\Vert {#1} \right\Vert}
\renewcommand{\matrix}[2]{\left(\begin{array}{#1}#2\end{array}\right)}
\newcommand{\twovec}[2]{\left(\begin{array}{c}#1\\#2\end{array}\right)}
\newcommand{\threevec}[3]{\left(\begin{array}{c}#1\\#2\\#3\end{array}\right)}
\newcommand{\fourvec}[4]{\left(\begin{array}{c}#1\\#2\\#3\\#4\end{array}\right)}
\newcommand{\DD}[2] {\frac{\partial{#1}}{\partial{#2}}}
\newcommand{\DDTwo}[2] {\frac{\partial^2{#1}}{\partial{#2}^2}}
\newcommand{\DDThree}[2] {\frac{\partial^3{#1}}{\partial{#2}^3}}
\newcommand{\DDfull}[2] {\frac{{\text d}{#1}}{{\text d}{#2}}}
\newcommand{\DDp}[2] {\frac{\partial\left({#1}\right)}{\partial{#2}}}
\newcommand{\DDfullTwo}[2] {\frac{{\text d}^2{#1}}{{\text d}{#2}^2}}
\newcommand{\Sr}[1] {{#1}^*}
\newcommand{\Srp}[1] {\left({#1}\right)^*}
\newcommand{\sk}[1] {{\text{skew}\left({#1}\right)}}

\newcommand{\refapp}[1]{Appendix~\ref{sec:#1}}
\newcommand{\refsec}[1]{Section~\ref{sec:#1}}
\newcommand{\refeq}[1]{(\ref{eq:#1})}
\newcommand{\refeqs}[2]{(\ref{eq:#1})-(\ref{eq:#2})}
\newcommand{\reffig}[1]{Fig.~\ref{fig:#1}}
\newcommand{\reftab}[1]{Table~\ref{tab:#1}}
\newcommand{\cf}[1]{cf.\ Fig.~\ref{fig:#1}}
\newcommand{\eq}[1]{(\ref{eq:#1})}

\newcommand{\Ithree} {{\mat{I}_{3\times3}}}

\newcommand{\cm} {$\backslash\backslash$}

\renewcommand{\labelenumii}{\arabic{enumii}$)$ }
\renewcommand{\labelenumiii}{\arabic{enumiii}$)$ }

\definecolor{amethyst}{rgb}{0.6, 0.4, 0.8}
\definecolor{darkpastelgreen}{rgb}{0.01, 0.75, 0.24}
\newcommand{\Antoine}[1]{\textcolor{darkpastelgreen}{[\textbf{Antoine}: {#1}]}} 
\newcommand{\Miguel}[1]{\textcolor{amethyst}{[\textbf{Miguel}: {#1}]}}

\teaser{
  \includegraphics[width=\linewidth]{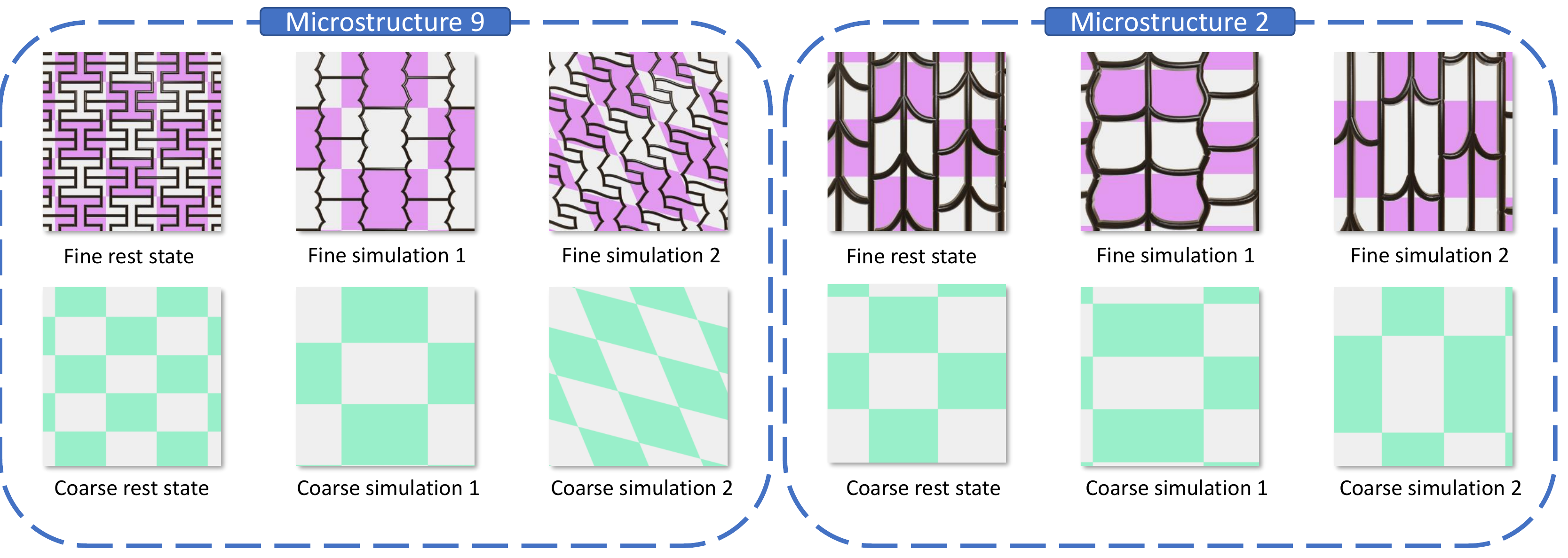}
 \centering
  \caption{
  We use our elasticity model, based on high-order interpolants, to homogenize 2D microstructures.
  As shown in these images of two different microstructures, the coarse homogenized materials match accurately the mesoscale deformation response of the microstructures.
  In these examples, we simulate the microstructures with periodic boundary conditions; this results in uniform mesoscale strain, which we use to deform the purple-white pattern.
  The left microstructure exhibits auxetic behavior, which is accurately captured by our model.}
\label{fig:teaser}
}

\maketitle
\begin{abstract}
We propose a novel formulation of elastic materials based on high-order interpolants, which fits accurately complex elastic behaviors, but remains conservative.
The proposed high-order interpolants can be regarded as a high-dimensional extension of radial basis functions, and they allow the interpolation of derivatives of elastic energy, in particular stress and stiffness.
Given the proposed parameterization of elasticity models, we devise an algorithm to find optimal model parameters based on training data.
We have tested our methodology for the homogenization of 2D microstructures, and we show that it succeeds to match complex behaviors with high accuracy.


\end{abstract}  

\section{Introduction}
\label{sec:introduction}

Accurately representing the elastic response of complex materials is an ongoing challenge across computer graphics and computational mechanics.
This problem has application in fitting material models to physical tests of real-world objects~\cite{Bickel2009,Wang2011,sperl2022eylsmpf}, developing mesoscale models for microscale materials~\cite{Schumacher:2015:MCE:2809654.2766926}, or designing simulation models with nonlinear response~\cite{Xu2015}.

A common approach to designing complex elastic material behaviors is to define elastic energy or parameters of stress-strain functions using weighted scalar basis functions~\cite{Bickel2009,Wang2011,Miguel2016,sperl2020hylc,Wang2020}.
However, as we demonstrate in this paper, this approach suffers various problems.
Some variants fail to represent the elastic behavior accurately, while other variants lack fundamental properties of elasticity, such as energy conservation.

In this paper, we develop a novel formulation of elastic materials based on high-order interpolants, which fits accurately complex elastic behaviors, but remains conservative.
The contributions of our work are:
\begin{enumerate}
    \item 
    The design of tensor basis functions to interpolate derivatives of elastic energy (\refsec{interp}). These basis functions can be regarded as a high-dimensional extension of radial basis functions (RBFs).
    \item
    Based on the tensor interpolants, we design a parameterization of elasticity models (\refsec{energy}). This paramterization provides suitable degrees of freedom to fit both the stress and stiffness behavior of complex materials.
    \item
    An algorithm to optimize the parametric elasticity model based on training data (\refsec{estimation}), which finds the control points and coefficients of the elasticity interpolants.
    \item
    The application of the methodology to homogenization of 2D microstructures (\refsec{micro}). This includes the generation of representative training data and the application of the estimation algorithm mentioned above.
\end{enumerate}

In the paper, we evaluate the accuracy of our method, we compare it to other variants, and we analyze the effect of various design choices.
As a conclusion, the proposed methodology for the design of elasticity models succeeds at capturing complex behaviors, such as those shown in \reffig{teaser}.
We have tested the methodology on 11 2D microstructures with different deformation behaviors, and we discuss the full results.

\section{Related Work}
\label{sec:prev}

\subsection{Elasticity Interpolation}

The baseline approach to model elastic behaviors is to design expressive constitutive models.
Research in this direction is ample, covering both the ability to reproduce interesting behaviors (nonlinearity, anisotropy, volume conservation), as well as robustness~\cite{Li2015,Smith2018,Kim2020b,Kim2020}.
However, designing constitutive models is built on the inherent assumption of homogeneous materials, and is not meant to accurately represent the complex nonlinearities of heterogeneous materials.

The common approach in computer graphics to represent complex nonlinearities and anisotropy is to interpolate elasticity models.
There is a large variety of methods to do so, with different features.
Some methods model nonlinear stress-strain relationships through interpolation.
Examples include RBF interpolation of material parameters \cite{Bickel2009}, interpolation of stiffness values at control points in strain domain \cite{Wang2011}, or stress interpolation based on RBFs \cite{Wang2020}.
Unfortunately, modeling the stress-strain function through interpolation lacks energy conservation, as the stress-strain function is not integrable.
This can produce artifacts through energy gain or loss, and prevents the use of attractive optimization-based numerical integrators~\cite{Gast2015}.
One exception~\cite{Miguel2012} models the stress-strain curve for individual strain values, hence it remains conservative, but it largely limits the expressiveness of the material.

Other methods model the elastic energy function through interpolation, and therefore remain conservative by construction.
Examples include formulating energy addends that depend on different subdomains of strain \cite{Miguel2016}, and modeling such energy addends using spline interpolation \cite{sperl2020hylc}.
Xu et al.~\cite{Xu2015} used spline interpolation to model energy addends within the Valanis-Landel isotropy assumption.
They handled anisotropy separately, but with limited expressiveness.

When modeling microscale heterogeneous materials, numerical coarsening~\cite{Nesme:2009:PTA,Kharevych:2009:NCO,Torres:2016:HIC:2980179.2982414,Chen2018} is an alternative to elasticity model design.
In numerical coarsening, the material models are evaluated at high-resolution spatial discretization, respecting the heterogeneous material distribution.
However, the simulation is computed at a coarse mesoscale and interpolated to the microscale through complex nonlinear shape functions.

\subsection{Microstructure Simulation}

2D and 3D microstructures are a powerful way of controlling mesoscale deformation behavior under limited material choices, and have therefore become a major tool in computational fabrication of deformable objects~\cite{Bickel2010,Schumacher:2015:MCE:2809654.2766926,Panetta2015,Konakovic2018}.
However, the simulation of large objects at microscale resolution is computationally costly, and it challenges the use of microstructures within design optimization algorithms.

Homogenization is a powerful tool for computational design with microstructures, as it fits mesoscale material models that accurately represent the aggregate microscale behavior~\cite{LIU201671}.
We test our material modeling approach in the context of material homogenization for microstructures, and in this regard we follow a popular homogenization methodology.
Same as previous works~\cite{Schumacher2018,sperl2020hylc}, we simulate microstructures under periodic boundary conditions.
This makes the mesoscale strain uniform, and enables easy transfer of training data from microstructure simulation to mesoscale.
\begin{figure*}
\centering
\includegraphics[width =\linewidth]{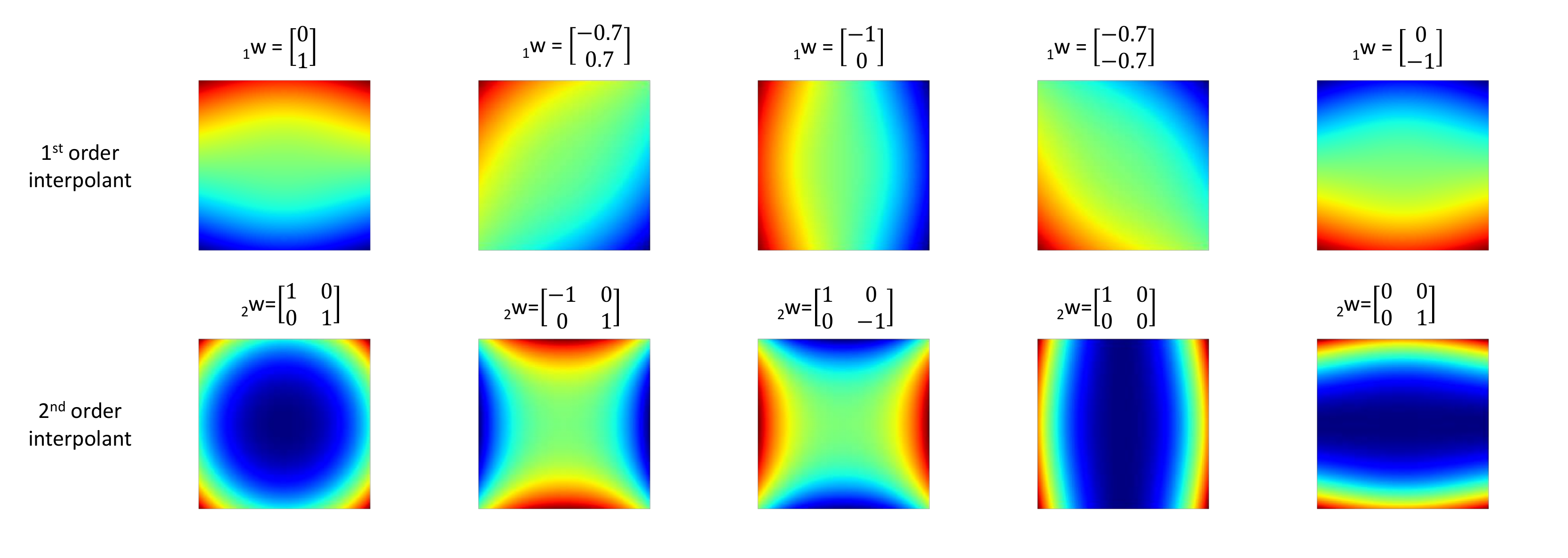}
\caption{
\label{fig:interpolants}
Our proposed first-order interpolant \refeq{GradientInterp} and second-order interpolant \refeq{HessianInterp} provide local control, respectively, of the gradient (stress) and curvature (stiffness) of energy functions.
The images show energies in the neighborhood of an RBF center, for representative choices of the RBF coefficients.
We used a multiquadric RBF, and blue denotes low energy while red denotes high energy.
Notice how the direction of the gradient is controlled on the top row, and this dictates the local stress.
Notice also how the second-order interpolant allows modeling isotropic stiffness (left), directional stiffness (right), or also saddle-point configurations (center).}
\end{figure*}

\section{Conservative Derivative Interpolation}
\label{sec:interp}

We want to design a parametric function (the elastic energy) such that it interpolates given values of its derivatives, i.e., its gradient and Hessian (stress and stiffness).
To do this, we leverage RBF interpolation, but we face the question of designing a good parameterization such that the resulting function is conservative and interpolates derivative values.

To answer this question, in this section we analyze matrix-valued RBFs for gradient interpolation. We conclude that this formulation can be generalized and extended to the interpolation of arbitrary higher-order derivatives. By leveraging these conclusions, we will later show how to design a good parameterization for RBF energies.

\subsection{Matrix-Valued RBFs for Gradient Interpolation}

In our exposition of the fundamentals of high-order RBF interpolants, we denote the domain of RBF interpolation as $x$.
With RBF center $x_i$, radial vector $\Delta x_i = x - x_i$, and RBF radius $r_i = \| \Delta x_i \|$, we express the corresponding RBF as $\phi_i \equiv \phi(r_i)$.
Appendix A lists some derivatives of RBFs that we use throughout the paper.

Matrix-valued RBFs can be constructed from scalar-valued RBFs $\phi$ through a double differentiation process, $\left( \alpha \, \nabla^2 \, I + \beta \, \nabla \nabla^T \right) \, \phi$, with $\alpha$ and $\beta$ scalar coefficients.
Vector-valued RBF coefficients $w_i$ yield a vector field:
\begin{equation}
v(x) = \sum_i \left( \alpha \, \nabla^2 \, I + \beta \, \nabla \nabla^T \right) \, \phi_i \, w_i.
\end{equation}
When interpolating vector values, matrix-valued RBFs yield positive-definite systems~\cite{Narcowich1994}.

Thanks to a Helmholtz-Hodge decomposition~\cite{Bhatia2013}, the matrix-valued RBF interpolation can be decomposed into curl-free and divergence-free vector fields~\cite{Fuselier2008}: 
\begin{align}
\label{eq:freeRBFs}
& v_\text{curl-free}(x) = \sum_i \nabla \nabla^T \phi_i \, w_i, \\
& v_\text{div-free}(x) = \sum_i \left( \nabla^2 \, I - \nabla \nabla^T \right) \, \phi_i \, w_i.
\end{align}

Moreover, it is easy to show that the curl-free vector field can be derived from a potential function $f(x)$, hence concluding that the vector field is also conservative:
\begin{equation}
\label{eq:curlfree}
v_\text{curl-free}(x) = \nabla f(x), \text{   with } f(x) = \sum_i w_i^T \, \nabla \phi_i.
\end{equation}

In Appendix B, we demonstrate that RBF interpolants based on RBF gradients can be recast based on RBFs directly. Then, the interpolant $w_i^T \, \nabla \phi_i$ in $f(x)$ in \refeq{curlfree} can be recast as $\phi_i \, w_i^T \, \Delta x_i$, with some other choice of RBF.
As a result, the curl-free vector field in \refeq{curlfree} can be rewritten as:
\begin{equation}
\label{eq:curlfree2}
v_\text{curl-free}(x) = \nabla f(x), \text{   with } f(x) = \sum_i \phi_i \, w_i^T \, \Delta x_i.
\end{equation}

\subsection{Generalization to High-Order Derivatives}

In the previous section, we observe that the key property to interpolate gradients with conservative functions is that the RBF interpolants are expressed as inner product of the radial vector $\Delta x_i$ and a vector of RBF coefficients $w_i$ with the same dimensionality as the target gradients.
In fact, this observation can be generalized to arbitrary high-order n-th derivatives. The sufficient and necessary condition for interpolation of n-th derivatives with a conservative function is that the RBF interpolants are expressed as the tensor contraction of n tensor products of the radial vector $\Delta x_i$ with an n-th dimensional tensor of RBF coefficients ${}_{n}w_i$. Formally:
\begin{equation}
f(x) = \sum_i \phi_i \, {}_{n}w_i \colon \underbrace{\left( \Delta x_i \otimes \Delta x_i \dots \otimes \Delta x_i \right)}_\text{n times}
\end{equation}
is a function whose n-th derivative can interpolate n-th dimensional tensor data, i.e., the target n-th dimensional derivatives.

Conservative interpolation of gradients (first derivatives) and Hessians (second derivatives), for example, reduce to defining interpolants of the form:
\begin{align}
\label{eq:GradientInterp}
& \text{Gradient:} \quad \phi_i \, {}_{1}w_i \colon \Delta x_i = \phi_i \, {}_{1}w_i^T \Delta x_i, \\
\label{eq:HessianInterp}
& \text{Hessian:} \quad \phi_i \, {}_{2}w_i \colon \left( \Delta x_i \otimes \Delta x_i \right) = \phi_i \, \Delta x_i^T {}_{2}w_i \, \Delta x_i,
\end{align}
with ${}_{1}w_i$ a vector of RBF coefficients and ${}_{2}w_i$ a matrix of RBF coefficients, respectively.

\reffig{interpolants} shows examples of first-order and second-order interpolants for some representative choices of the RBF coefficients ${}_{1}w_i$ and ${}_{2}w_i$.
We can see that the first-order interpolants provide local control of the gradient (both value and direction) of the energy, and the second-order interpolants provide local control of the curvature of the energy.
In the next section, we leverage these interpolants in the definition of RBF elastic energy functions.

\section{RBF Elastic Energy}
\label{sec:energy}

We want to define nonlinear and anisotropic elastic materials that are parameterized by the current deformation, and we do this following a scattered data interpolation strategy using RBFs.
We start the section with some definitions and a discussion of desired properties.
Then, we introduce our energy parameterization, leveraging the high-order RBF interpolants derived in \refsec{interp}.

\subsection{Definitions and Desiderata}

When designing elastic materials, we want to preserve the stress-strain response. This includes both the stress value at a certain strain, and its derivative or tangent stiffness.

We choose Green strain $\epsilon = \frac{1}{2} \left( F^T \, F - I \right)$ as representation of strain or deformation, with $F$ the deformation gradient. For convenience, we write the strain in Voigt notation $E$, which becomes the interpolation domain $E = x$ for our RBF interpolation method. In our 2D examples, we have $E = \left( \epsilon_{xx}, \epsilon_{yy}, 2 \, \epsilon_{xy} \right)$.

Following the Voigt notation of Green strain $E$, and with elastic energy density $\Psi$, we define stress as the energy gradient wrt strain, $s = \nabla \Psi = \DD{\Psi}{E}^T$, which is a vector form of the 2nd Piola-Kirchhoff stress.
We also define the tangent stiffness as the Hessian of the energy wrt strain, $K = \nabla \nabla^T \Psi = \DDTwo{\Psi}{E}$.

We seek a material model $\Psi = f \left( E, \{ w_i \} \right)$ that relates strain to energy according to some material parameters $\{ w_i \}$.
In designing a good parameterization for elasticity models, we pay attention to the properties of the magnitudes we wish to match, namely the stress and the tangent stiffness.
A na\"ive solution for the design of a stress(strain) function would be to formulate scalar basis functions in the strain domain (e.g., RBFs), together with vector-type basis coefficients. Unfortunately, the resulting function is not guaranteed to produce a conservative field. Most importantly, conservativeness cannot be enforced through an appropriate choice of basis coefficients; the lack of conservativeness is an inherent limitation of the formulation.

The key to enforce conservativeness of the stress field is to regard stress as the gradient of an energy field. Then, fitting a stress field can be posed as a gradient interpolation problem, with the stress the gradient of the underlying energy field.
Similarly, fitting a stiffness field can be posed as a Hessian interpolation problem, with the stiffness the Hessian of the underlying energy field.
To this end, we look at the high-order RBF interpolants of \refsec{interp}.

\subsection{Energy and its Derivatives}
\label{sec:energy:energy}

We design an RBF energy formulation that is equipped with conservative gradient interpolants \refeq{GradientInterp}, to fit a target stress field, and with conservative Hessian interpolants \refeq{HessianInterp}, to fit a target tangent stiffness field.
We denote each gradient interpolant as $\Psi_{\text{GI},i}$, with RBF center $E_i$ and vector RBF coefficients $w_i$ (the ${}_{1}w_i$ in \refeq{GradientInterp}). Similarly, we denote each Hessian interpolant as $\Psi_{\text{HI},i}$, with RBF center $E_i$ and matrix RBF coefficients $W_i$ (the ${}_{2}w_i$ in \refeq{HessianInterp}).
We also add to the energy formulation two offset terms $\Psi_{\text{GO}}$ and $\Psi_{\text{HO}}$ that produce, respectively, a stress offset $s_\text{O}$ and a stiffness offset $K_\text{O}$.
We add the stress offset to easily enforce zero stress at zero strain, and the stiffness offset to easily fit the average stiffness.
In this way, the RBF interpolants act as corrections with respect to offset terms.

The full energy formulation is summarized as:
\begin{align}
\label{eq:energy}
& \Psi = \Psi_{\text{GO}} + \Psi_{\text{HO}} + \sum_i \Psi_{\text{GI},i} + \sum_i \Psi_{\text{HI},i}, \\
\nonumber
& \Psi_{\text{GO}} = s_\text{O}^T \, E,\\
\nonumber
& \Psi_{\text{HO}} = \frac{1}{2} \, E^T \, K_\text{O} \, E,\\
\nonumber
& \Psi_{\text{GI},i} = \phi_i \, w_i^T \, \Delta E_i,\\
\nonumber
& \Psi_{\text{HI},i} = \phi_i \, \Delta E_i^T \, W_i \, \Delta E_i.
\end{align}
Note that the formulation above (and also our implementation) uses the same RBF function $\phi$ and RBF centers $\{ E_i \}$ for gradient and Hessian interpolants, but these could be different in practice.

From the energy definition \refeq{energy}, we obtain the stress and the tangent stiffness.
\begin{align}
\label{eq:stress}
& s = \nabla \Psi_{\text{GO}} + \nabla \Psi_{\text{HO}} + \sum_i \nabla \Psi_{\text{GI},i} + \sum_i \nabla \Psi_{\text{HI},i}, \\
\nonumber
& \nabla \Psi_{\text{GO}} = s_\text{O},\\
\nonumber
& \nabla \Psi_{\text{HO}} = K_\text{O} \, E,\\
\nonumber
& \nabla \Psi_{\text{GI},i} = \phi_i \, w_i + w_i^T \, \Delta E_i \, \DD{\phi_i}{E}^T,\\
\nonumber
& \nabla \Psi_{\text{HI},i} = 2 \, \phi_i \, W_i \, \Delta E_i + \Delta E_i^T \, W_i \, \Delta E_i \, \DD{\psi_i}{E}^T.
\end{align}
\begin{align}
\label{eq:stiffness}
& K = \nabla \nabla^T \Psi_{\text{HO}} + \sum_i \nabla \nabla^T \Psi_{\text{GI},i} + \sum_i \nabla \nabla^T \Psi_{\text{HI},i}, \\
\nonumber
& \nabla \nabla^T \Psi_{\text{HO}} = K_\text{O},\\
\nonumber
& \nabla \nabla^T \Psi_{\text{GI},i} = w_i \, \DD{\phi_i}{E} + \DD{\phi_i}{E}^T \, w_i^T + w_i^T \, \Delta E_i \, \DDTwo{\phi_i}{E},\\
\nonumber
& \nabla \nabla^T \Psi_{\text{HI},i} = 2 \, W_i \, \Delta E_i \, \DD{\phi_i}{E} + 2 \, \DD{\phi_i}{E}^T \, \Delta E_i^T \, W_i \\
\nonumber
& ~~~~~~~~~~~~~~~~~~ + 2 \, \phi_i \, W_i + \Delta E_i^T \, W_i \, \Delta E_i \, \DDTwo{\phi_i}{E}.
\end{align}
The first and second partial derivatives of the RBFs are listed in Appendix A.

Our energy model \refeq{energy} is parameterized by the stress and stiffness offsets $s_\text{O}, K_\text{O}$, and the gradient and Hessian interpolant centers and coefficients $\{E_i, w_i, W_i\}$.
Note that, thanks to our conservative RBF interpolants, both the stress \refeq{stress} and the tangent stiffness \refeq{stiffness} are expressed as the sum of weighted basis functions (i.e., they are linear with respect to the basis coefficients), each RBF introduces degrees of freedom with the same dimensionality as the stress and/or the stiffness, and the formulation remains conservative by construction.

\section{Material Fitting Algorithm}
\label{sec:estimation}

Once we have defined our RBF energy model in the previous section, we describe how we estimate the parameters of this model.
Our algorithm includes two aspects: one is the optimization of energy coefficients, the other one is the optimization of metaparameters (i.e., RBF centers and radius/smoothness parameters).

\subsection{Optimization of RBF Coefficients}
\label{sec:estimation:coeffs}

We assume we have target stress and stiffness data $\{ s_j, K_j \}$ available for a set of known strains $\{ E_j \}$.
In \refsec{micro}, we describe how we obtain representative target data for 2D microstructures.
And at this point we also assume that the energy RBF centers $\{ E_i \}$ are given.
In the next subsection, we discuss how these centers are optimized.

The energy function includes the following parameters to be optimized (see \refsec{energy:energy}): stress and stiffness offsets $s_\text{O}, K_\text{O}$, and coefficients of stress and stiffness interpolants $\{w_i\}, \{W_i\}$.
The stress offset is implicitly defined by constraining the stress to be zero at zero strain.
From~\refeq{stress}, we get:
\begin{align}
\nonumber
    & s(0) = s_\text{O} + \sum_i \nabla \Psi_{\text{GI},i}(0) + \sum_i \nabla \Psi_{\text{HI},i}(0) = 0 \Rightarrow \\
    & s_\text{O} = -\sum_i \nabla \Psi_{\text{GI},i}(0) -\sum_i \nabla \Psi_{\text{HI},i}(0).
\end{align}

We denote the remaining parameter set as $p = \left( K_\text{O}, \{w_i\}, \{W_i\} \right)$.
We compute these parameters by minimizing the difference between target and estimated stress and stiffness values.
This is expressed formally as:
\begin{equation}
\label{eq:optim}
p = \arg\min_p \sum_j \frac{1}{s_\text{RMS}^2} \, \| s( E_j, p ) - s_j \|^2 + \frac{1}{K_\text{RMS}^2} \, \| K ( E_j, p ) - K_j \|^2.
\end{equation}
Note that we normalize the stress and stiffness errors by the root-mean square of target stress and target stiffness values, respectively.

This optimization is a simple linear least squares problem, which yields a positive definite linear system for the solution of the parameters $p$.

\subsection{RBF Metaparameters}
\label{sec:estimation:meta}

In addition to RBF coefficients, the elastic energy function is also parameterized by the number of RBFs, their centers, and other RBF-specific smoothness or support parameters (e.g., the variance of Gaussian RBFs).
We have followed a greedy algorithm to optimize these metaparameters.

\begin{figure}[t!]
\centering
\includegraphics[width =.9\linewidth]{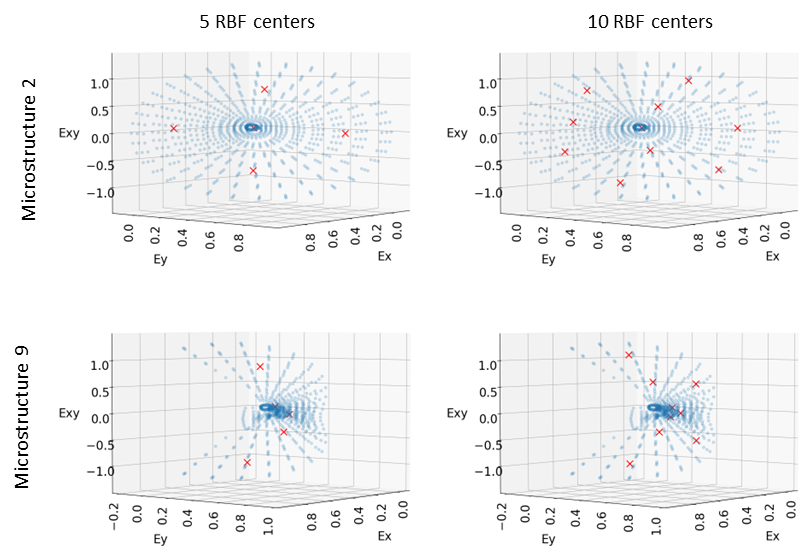}
\caption{
\label{fig:clustering}
Several results of $k$-means clustering for the computation of RBF centers. The plots show the distribution of training strains for two different microstructures, and RBF centers with 5 vs. 10 clusters.
Microstructure 9 exhibits buckling effects that make the training data discontinuous, and it requires more RBFs to cover well the domain.
}
\end{figure}

We start with no RBFs, and we progressively add RBFs until the energy fitting error as defined in \refeq{optim} is smaller than a target threshold.
Given $k$ RBFs, we first optimize the locations of the RBF centers $\{E_i, 1 \leq i \leq k \}$.
We do this by clustering the target strain values $\{ E_j \}$ into $k$ clusters using $k$-means clustering.
Then we solve the optimization \refeq{optim} while sweeping smoothness or support parameters, and we choose the optimal result.
\reffig{clustering} shows example results of $k$-means clustering for two different microstructures.
In some cases (e.g., top of \reffig{clustering}), the target strains $\{ E_j \}$ are evenly distributed and a small number of RBFs may cover well the domain.
In other cases (e.g., bottom of \reffig{clustering}), the target strains show discontinuties, e.g., due to buckling of the microstructures, and a larger number of RBFs may be necessary.

Our approach for selecting the RBF centers is not optimal, as the clustering algorithm does not account for local error. There are other possible approaches for optimizing the metaparameters of the RBFs, such as recursive orthogonal least squares~\cite{Gomm2000,Chen1991}, but we leave this to future work.
In \refsec{experiments} we discuss how the fitting error is affected by the choice of metaparameters and RBF functions.

\begin{figure*}[t!]
\centering
\includegraphics[width =1.0\linewidth]{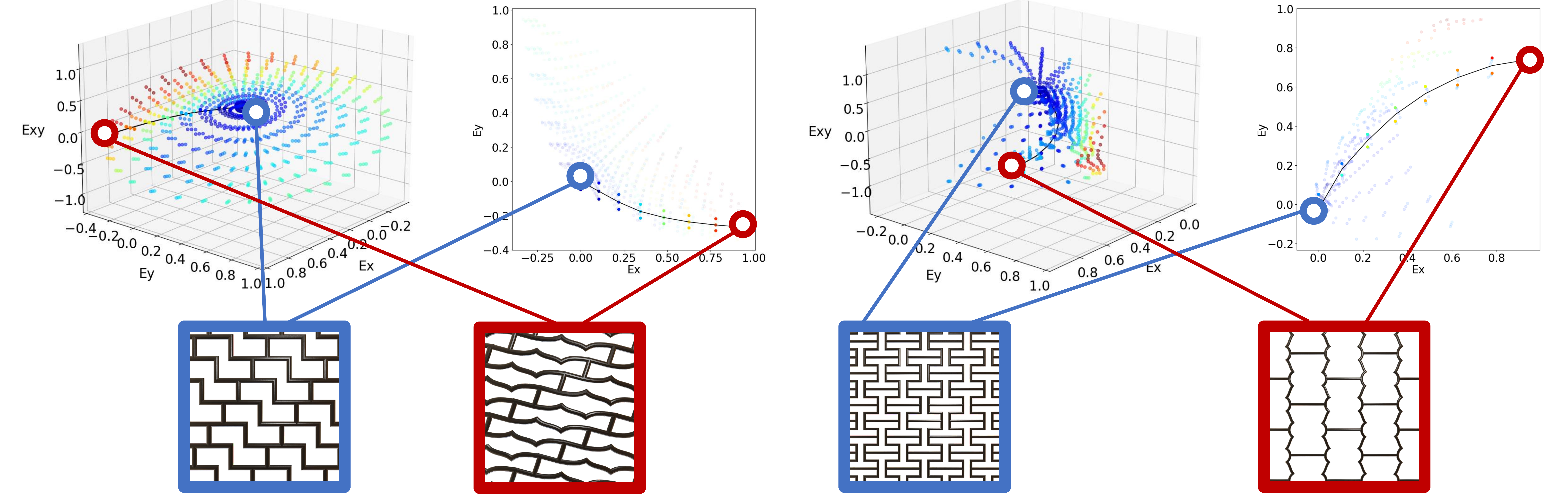}
\caption{
\label{fig:data}
The plots compare the training data (colored according to the norm of stress) in strain domain, for two different microstructures.
We also highlight a directional stretch from rest (blue) to a deformed configuration (red).
Projecting the data to 2D we can clearly see the extremely different behavior of these two microstructures; the one on the right shows negative Poisson's ratio for this stretch, and the training data populates a different region of the strain domain.}
\end{figure*}

\section{Homogenization of 2D Microstructures}
\label{sec:micro}

We apply our parametric energy model (\refsec{energy}) and estimation algorithm (\refsec{estimation}) to design homogeneous mesoscale elasticity models for 2D microstructures. In doing so, we pay special attention to the generation of representative strain, stress and stiffness data for the estimation algorithm.

To generate training data, we simulate 2D microstructures under planar deformations with periodic boundary conditions (PBCs), similar to the work of Schumacher et al.~\cite{Schumacher2018}.
Specifically, to simulate the high-resolution microstructures with PBCs, we follow the method by Sperl et al.~\cite{sperl2020hylc}. 

We model a repeatable tile of microstructure at high resolution, using a finite-element mesh. The positions of the mesh nodes are grouped as $\vec{x}$, and they are governed by the combination of a coarse homogeneous deformation $E$ and local mesh displacements $\vec{u}$.
Following Sperl et al., we apply a known coarse deformation $E$, and we solve for mesh displacements that minimize the tile's elastic energy density $\Psi$ under PBCs.
Formally, this is expressed as:
\begin{equation}
\vec{u} = \arg\min \Psi(\vec{x}(E, \vec{u})), ~~ \text{s.t. } \vec{c}(\vec{u}) = 0,
\end{equation}
where $\vec{c}(\vec{u})$ includes PBCs as well as constraints to avoid net rigid motion of the tile.

We produce training data in a controlled way, generating microstructure deformations that span planar \added{uniaxial} stretch deformations in all directions.
\added{These cover situations where there is a dominant direction of deformation, and were also the main focus of attention of several previous works~\cite{Wang2011,Schumacher2018,sperl2022eylsmpf}.}
The \added{rotation-invariant  part of the} deformation gradient can be defined as $F = \text{Rot}(\theta) \, \text{diag}(\lambda_1, \lambda_2) \, \text{Rot}(\theta)^T$, where $\lambda_1$ and $\lambda_2$ are principal stretches, and $\theta$ is the direction of stretch.
We regularly sample  the first principal stretch $\lambda_1$ in the range 0.9 \textemdash 2.0, and the stretch direction $\theta$ in the range 0 \textemdash $\pi$.
For each combination $(\lambda_1, \theta)$, we simulate the microstructure and we search for the orthogonal stretch $\lambda_2$ that produces zero orthogonal stress, i.e., $\DD{\Psi}{\lambda_2} = 0$.
Then, we add two more deformations by changing the orthogonal stretch by $\pm 0.05$.

We collect the full set of deformations and compile the coarse strain, stress, and tangent stiffness for each deformation, $\{ E_j, s_j, K_j \}$.
The training data is roughly centered around uniaxial stretches in all directions.
\reffig{data} shows training data for two microstructures with very different behavior; the one on the left shows negative Poisson's ratio, and hence the training data populates a very different region  in the strain domain.
Please watch the accompanying video for animations of the training data generation and 3D visualizations of the data in strain domain.

\begin{table}[b!]
    \centering
    \begin{tabular}{c|c}
         RBF & Stress and stiffness error (\%)\\
        \hline
        Multiquadric & $6.53$\\
        Gaussian &  $6.89$ \\
        Inverse quadratic & $6.66$ \\
        Inverse multiquadric & $6.62$ \\
    \end{tabular}
    \caption{We fit all microstructures using four different RBFs with the same number of RBF centers (10), and the error differences are minimal.}
    \label{tab:RBFcomp}
\end{table}

\section{Experiments and Discussion}
\label{sec:experiments}

\begin{figure*}[t!]
\centering
\includegraphics[width =\linewidth]{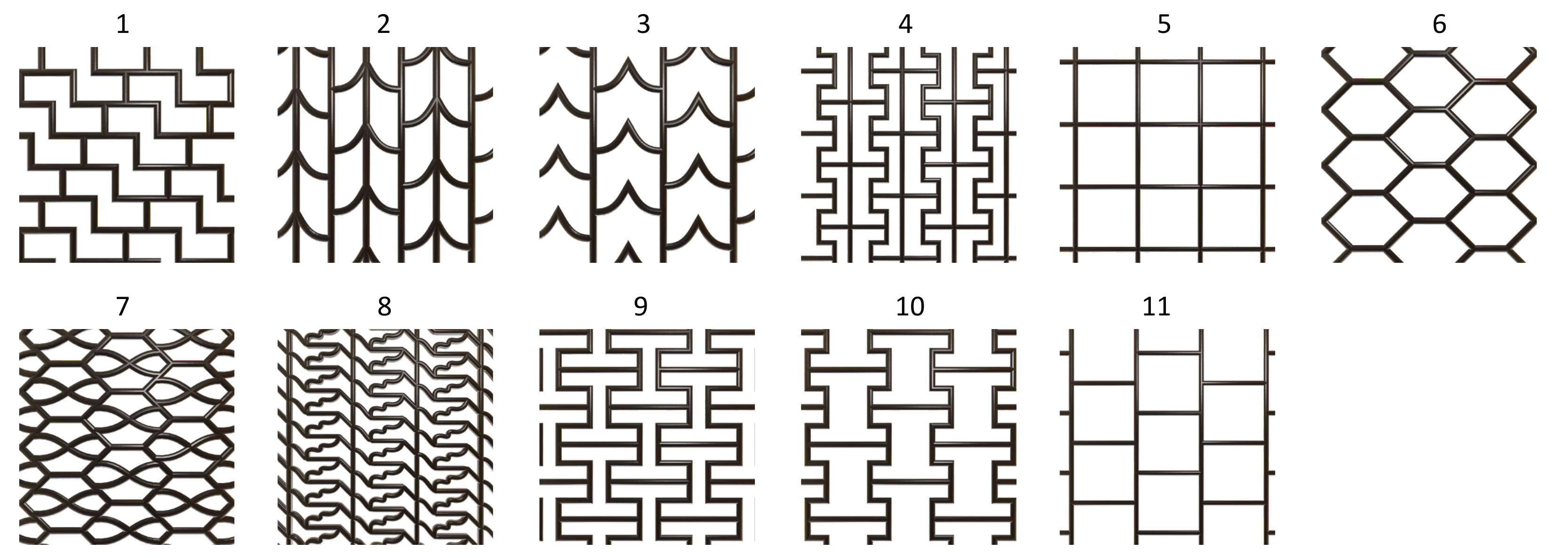}
\caption{
\label{fig:materials}
All 11 periodic microstructures tested in our experiments. They exhibit diverse nonlinearities and anisotropy.
}
\end{figure*}

We have tested our high-order elasticity interpolation methodology on the homogenization of 11 different periodic microstructures.
All microstructures are shown in \reffig{materials}.
They exhibit diverse nonlinearities and anisotropic behavior, including auxetic response.

We start the section discussing choice and estimation of metaparameters.
Then we analyze the fitting error across all microstructures, and we discuss the result of validation tests.
We conclude with a discussion of comparisons to other methods.

\begin{figure}[b!]
\centering
\includegraphics[width =\linewidth]{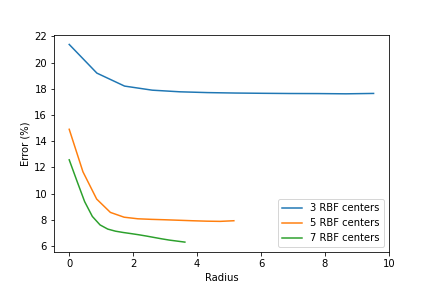}
\caption{
\label{fig:errVSradius}
This figure shows the fitting error for microstructure 1 as we sweep the smoothness radius of a multiquadric RBF. Note that the optimal radius gets smaller as we grow the number of RBFs. The plots are interrupted when the estimation problem becomes ill-conditioned.
}
\end{figure}

\subsection{Metaparameters}

Our first test evaluates what type of RBF provides highest accuracy under the same number of parameters.
We have tested four RBFs that are smooth, to ensure Hessians are well defined: multiquadric, Gaussian, inverse quadratic, and inverse multiquadric.
We have estimated all 11 microstructures following the method described in \refsec{estimation}, with 10 RBF centers.
As shown in \reftab{RBFcomp}, the differences across RBF types are minimal.
This result concurs with previous experiments~\cite{Chen92}.
Based on minimal advantage, we choose the multiquadric RBF $\phi(r) = \sqrt{r^2 + r_0^2}$.

As discussed in \refsec{estimation:meta}, as part of our estimation algorithm, we sweep radius/smoothness parameters of the RBF.
With the multiquadric RBF, this is the radius $r_0$.
\reffig{errVSradius} shows the total error for microstructure 1 as a function of $r_0$, for different numbers of RBF centers.
The error is not shown after a certain radius $r_0$, because the fitting problem becomes ill-conditioned.
Note that ill-conditioning occurs at smaller $r_0$ as we add more RBFs and they get closer.
For this reason, it is not possible to choose a single optimal value of $r_0$ for all numbers of RBFs.

\subsection{Fitting Error and Validation}

\begin{table}[b!]
    \centering
    \begin{tabular}{c|c|c|c|c}
         \multirow{2}{*}{Material} & \multirow{2}{*}{\#RBFs} & Stress & Stiffness & Orthogonal \\
         & & error (\%) & error (\%) & error (\%) \\
         \hline
          1 & 7 & 2.23 & 7.71 & 6.39\\
          2 & 4 & 2.34 & 6.28 & 8.59\\
          3 & 5 & 1.87 & 5.27 & 6.08\\
          4 & 19 & 3.99 & 6.70 & 4.30\\
          5 & 11 & 2.21 & 7.15 & 5.49\\
          6 & 9 & 2.35 & 7.15 & 6.59\\
          7 & 4 & 2.04 & 7.24 & 12.02\\
          8 & 19 & 6.02 & 21.3 & 6.96\\
          9 & 17 & 2.59 & 7.33 & 3.01\\
          10 & 19 & 5.14 & 5.86 & 4.52 \\
          11 & 19 & 5.23 & 30.8 & 6.60\\
    \end{tabular}
    \caption{Fitting error (stress and stiffness) for all 11 microstructure materials.
    The last column reports validation error in orthogonal stretch under directional stretch experiments, comparing the results with microstructures and our fitted homogenized elasticity models.}
    \label{tab:results}
\end{table}

We have fitted the training stress and stiffness of all test microstructures.
We increase the number of RBFs until we reach an average error of $5\%$ between stress and stiffness, but we stop the process if we reach 19 RBFs.
See \refeq{optim} for the error definition and normalization based on RMS values.
\reftab{results} summarizes the fitting quality across all materials.
The stress error is below or just above $5\%$ for all materials, and the stiffness error is below $10\%$ for all materials except two (which suffer error above $20\%$).
For some materials, adding more RBFs produced only a marginal gain.
Those cases probably require higher local control, with non-uniform selection of RBF centers and smoothness radius.

Detailed fitting results for all materials are shown in \reftab{bigtable} and \reftab{bigtable2}.
These tables show the norm of all values of stress and stiffness in the training data, the fitted values, and the error percentage (normalized with respect to RMS values).
Interestingly, the error in stress remains low and is spread across the domain for many materials, although it shows high local values for some materials.
On the other hand, the error in stiffness shows some high spikes for most of the materials.
This again suggests that higher local control is needed for higher accuracy.

\begin{figure*}[t!]
 \centering
  \includegraphics[width=\linewidth]{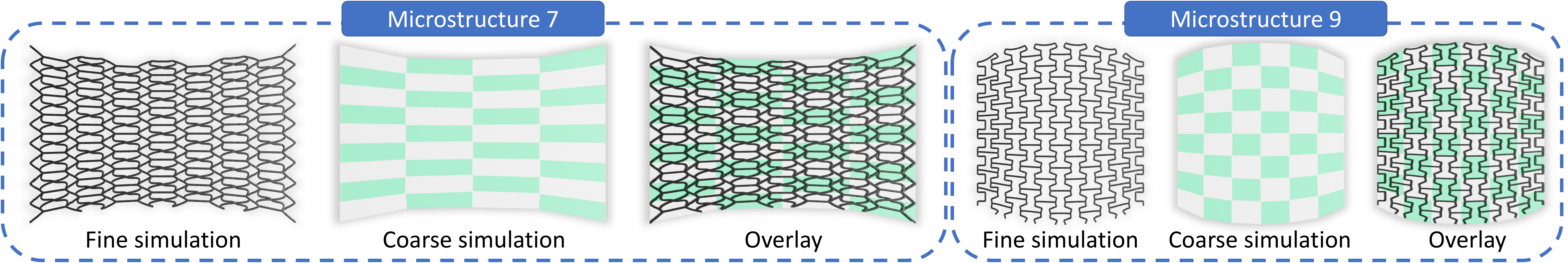}
  \caption{
  \added{Here, we evaluate our model on non-uniform strain deformations. We compare the simulation of high-resolution FEM microstructure models with coarse FEM models using our fitted energies.
  The images show two of the microstructures in the data set, under Dirichlet conditions on part of the boundary, and zero-traction Neumann conditions on the rest.
  The tested microstructure patches consist of 63 and 48 tiles, and were simulated using FEM models with 4252 and 5786 elements, respectively.
  The coarse models use 36 and 48 quad meshes. As shown in the overlays, the match between our fitted model and the full microstructure simulations is practically perfect.}}
\label{fig:nonuniform}
\end{figure*}

We have also validated the homogenized elasticity model on test simulations.
We stretch the homogenized material based on the values of principal stretch $\lambda_1$ and stretch direction $\theta$ in the training data, but we optimize for the orthogonal stretch $\lambda_2$ that minimizes energy.
This is the same procedure we apply to the microstructures to generate the training data, as described in \refsec{micro}, but we do it this time on the homogenized material.
We evaluate the error in orthogonal stretch $\lambda_2$ as validation.
The results for all 11 materials are listed in \reftab{results}.
Note that the error remained under $10\%$ in all materials except for one.
Furthermore, all simulations were robust.
By accurately fitting both the stress and the stiffness, we achieve in practice material models that are stable. 

\added{\subsection{Extrapolation}}

\added{The training data for the model consists of uniform uniaxial stretch data within a prescribed range. Therefore, we regard and test extrapolation in multiple ways. One is to extrapolate the energy behavior outside the range of strains in the training data. We have no a priori expectation for the model to succeed in this though, as the behavior of the microstructure materials may be unpredictable outside the training range. The other one is to extrapolate to non-uniform deformations. This, in contrast, is an expected and critical behavior, as it makes the model practical for real applications.}

\added{To evaluate extrapolation outside the range of strains, we have performed two tests. First, we trained using data from the lower half of the stretch range, and tested extrapolation to the upper half. The error on the training data was 5.27\% $\pm$ 3.48\% across all 11 materials, and on the test data it was 114.44\% $\pm$ 60.70\%. Second, we trained using data from half of the stretch directions, and tested extrapolation to the other half. The error on the training data was 12.39\% $\pm$ 9.10\% across all 11 materials, and on the test data it was 147.08\% $\pm$ 131.45\%.
As expected, the models fail to extrapolate.
But this is not a limitation of the methodology; it is an inherent challenge of the problem, because the behavior outside the training range may be highly nonlinear and unpredictable.
For this reason, we exhaustively sample the expected deformation range as part of training.}

\begin{table}[b!]
    \centering
    \begin{tabular}{c|c|c}
         & Stress error (\%) & Stiffness error (\%)\\
         \hline
        Stress fit & 7.67 & 30.23  \\
        Stiffness fit & 19.72  &  18.98 \\
        Stress + stiffness fit & 12.28 & 19.40
    \end{tabular}
    \caption{Under the same number of parameters, we have evaluated the accuracy of fitting stress data only, stiffness data only, or our combined stress and stiffness fitting.
    Our approach keeps the best balance in stress and stiffness error.}
    \label{tab:stressstiffness}
\end{table}

\added{To evaluate extrapolation to non-uniform strains, we have simulated large patches of microstructures, both with high-resolution FEM simulations, and with coarse simulations using our fitted energy models.
\reffig{nonuniform} shows two comparisons, for two different microstructures. We demonstrate that, thanks to the fitted energy models, we can replicate the behavior of complex microstructure patches (4252 and 5786 finite elements each) with coarse simulation meshes (36 and 48 elements each).}

\subsection{Comparisons}

Our first comparison analyzes if some terms of our elasticity model are more relevant.
To this end, we compared (a) fitting stress data only using stress interpolants only, (b) fitting stiffness data only using stiffness interpolants only, and (c) our full method fitting both stress and stiffness data using both stress and stiffness interpolants, on microstructure 1.
For a fair comparison, we used the same number of parameters (36) in all cases: (a) 11 RBFs plus stress offset, (b) 5 RBFs plus stiffness offset, and (c) 3 RBFs and both stress and stiffness offsets. 
As shown in \reftab{stressstiffness}, our method achieves the best balance in fitting both stress and stiffness. Fitting stiffness only leads to \added{higher}\deleted{huge} stress error.
Fitting stress only produces high stiffness error, but most importantly there is no control over the quality of the stiffness, which can lead to unstable material models.

\begin{table}[b!]
    \centering
    \begin{tabular}{c|c|c}
         & Stress error (\%) & Stiff. error (\%) \\
         \hline
        Energy fit, energy interp.  & 17.54 & 65.67 \\ 
        Our fit, energy interp. & 6.43 & 16.56 \\ 
        Our method & 3.06 & 9.21
    \end{tabular}
    \caption{Under the same number of parameters, we have evaluated the accuracy of fitting energy data with energy interpolation, fitting our stress and stiffness metric with energy interpolation, and our method. Our approach achieves the highest accuracy.}
    \label{tab:energyfitComparison}
\end{table}

We have also compared our method to models that interpolate the elastic energy directly, hence they do not provide direct control for stress and stiffness as in our method.
We designed an interpolated energy model of the form $\Psi = \sum_i \phi_i \, w_i$, with scalar RBF coefficients $w_i$~\cite{Miguel2016}.
We compared (a) fitting energy data with energy interpolation, (b) fitting our stress and stiffness metric with energy interpolation, and (c) our method.
For a fair comparison, we used the same number of parameters (54) in all cases: (a) and (b) 54 RBFs, and (c) 5 RBFs and both stress and stiffness offsets. 
As shown in \reftab{energyfitComparison}, our approach achieves the highest accuracy in all cases.

\begin{figure}[t!]
    \begin{tabular}{m{4.0cm}  m{4.0cm}  }
    \thead{Stress error (\%)} &  \thead{Curl / RMS Stiffness (\%)}  \\
    \includegraphics[width=40mm]{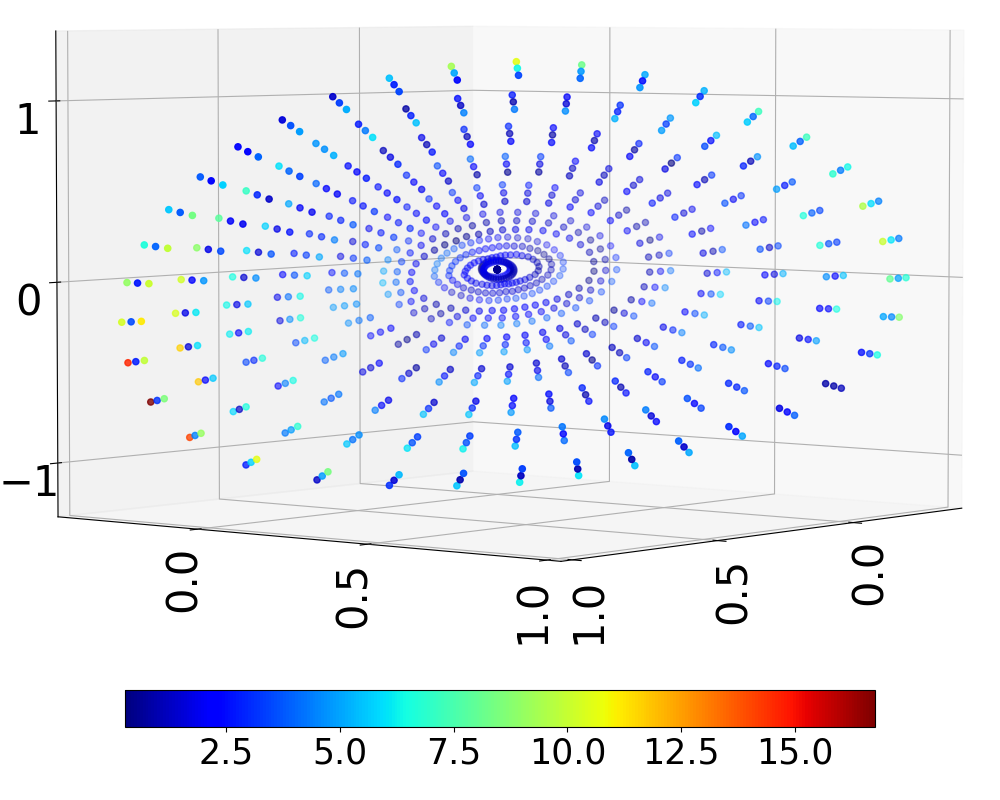} &
    \includegraphics[width=40mm]{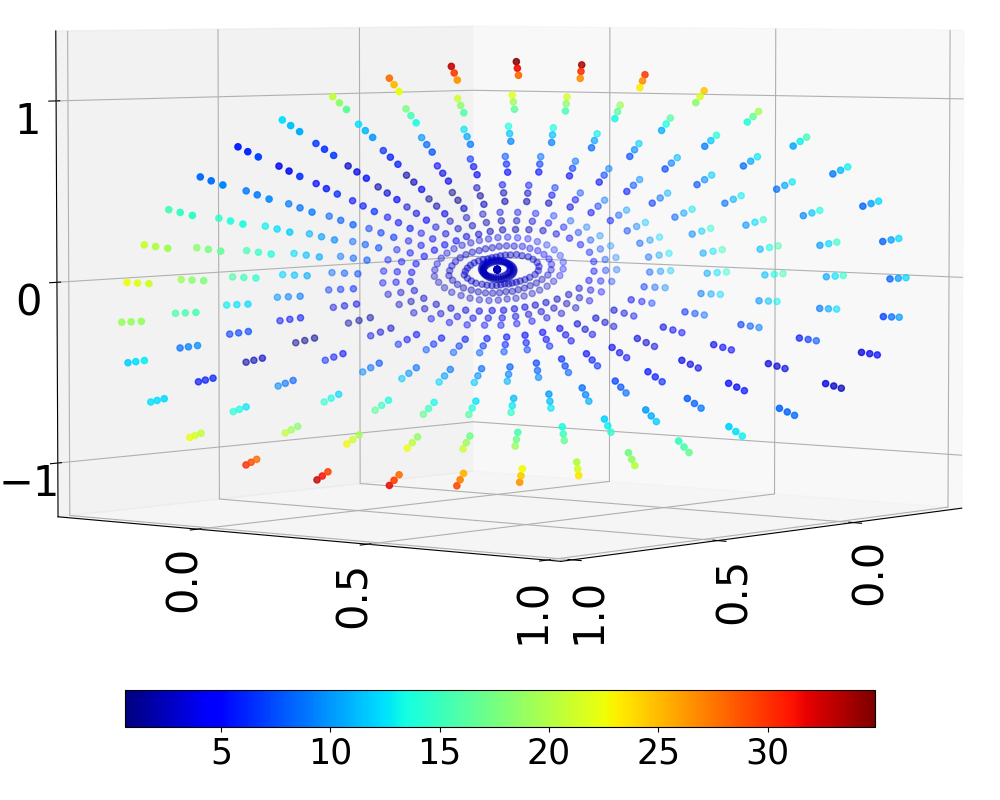} \\
\end{tabular}
\caption{
\label{fig:tablecurl2}
We tried fitting the stress of microstructure 1 using stress interpolation of the form $s = \sum_i \phi_i \, w_i$~\cite{Wang2020}.
With 20 RBFs the stress error is below $5\%$ (left).
However, the material model is far from conservative.
The curl of the stress (normalized by the RMS of stiffness) is above $30\%$ at times.
}
\end{figure}

Finally, we also tried fitting the stress of microstructure 1 using a non-conservative stress interpolation method.
In particular, we formulated the stress $s = \sum_i \phi_i \, w_i$, with vector RBF coefficients $w_i$, as done by Wang et al.~\cite{Wang2020}.
The optimization required 20 RBFs to reach a stress error below $5\%$.
\reffig{tablecurl2}-left shows the distribution of stress error.
Most importantly, we quantified the curl of stress, $\nabla \times s$, and we normalized it by the RMS of stiffness.
Note that the curl measures the non-symmetry of the Hessian.
As shown in \reffig{tablecurl2}-right, the curl of stress reached over $30\%$ of the RMS of stiffness at times.

\begin{table*}[t!]
    \begin{tabular}{m{0.5cm}  m{2.4cm}  m{2.4cm}  m{2.4cm}  m{2.4cm} m{2.4cm} m{2.4cm} }
    \thead{ID} &
    \thead{Target stress} &
    \thead{Fitted stress} &
    \thead{Stress error (\%)} &
    \thead{Target stiffness} &
    \thead{Fitted stiffness} &
    \thead{Stiff. error (\%)}\\
    \hline
    M1 &
    \includegraphics[width=28mm]{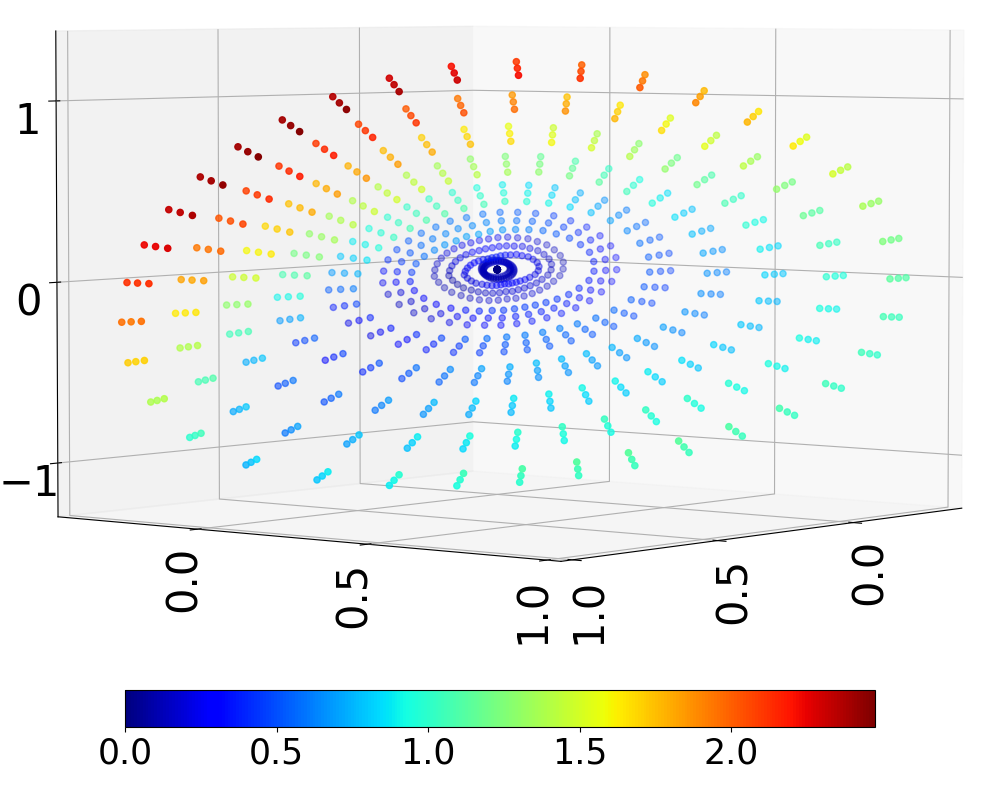}& \includegraphics[width=28mm]{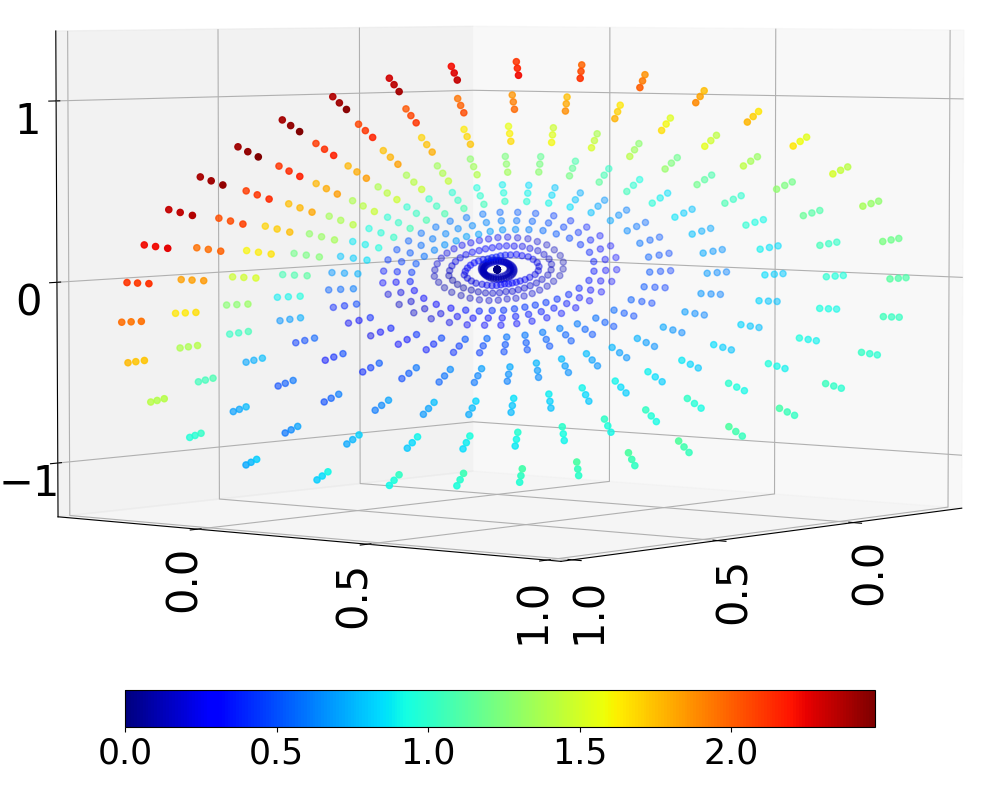}& \includegraphics[width=28mm]{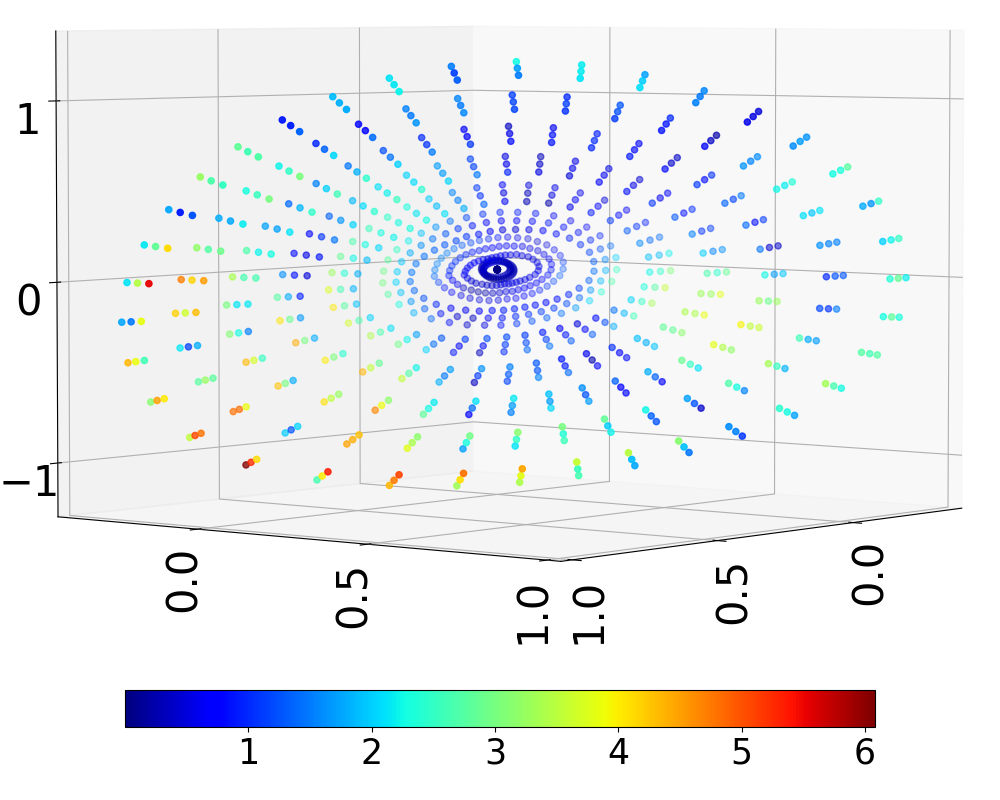}& \includegraphics[width=28mm]{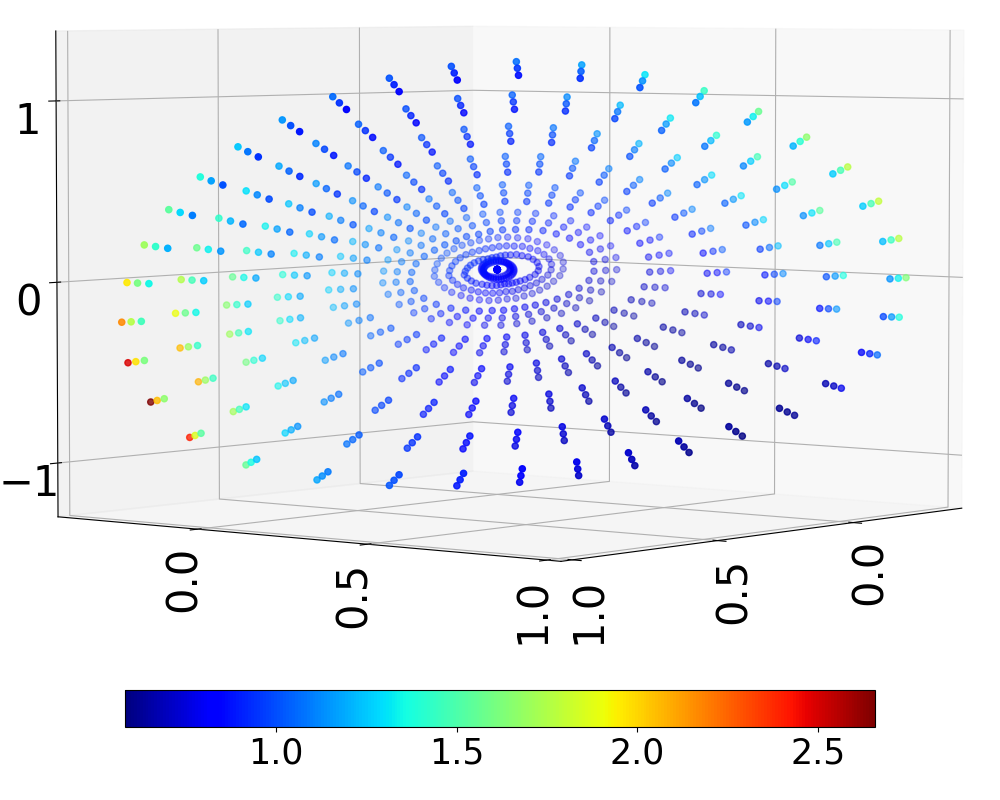}& \includegraphics[width=28mm]{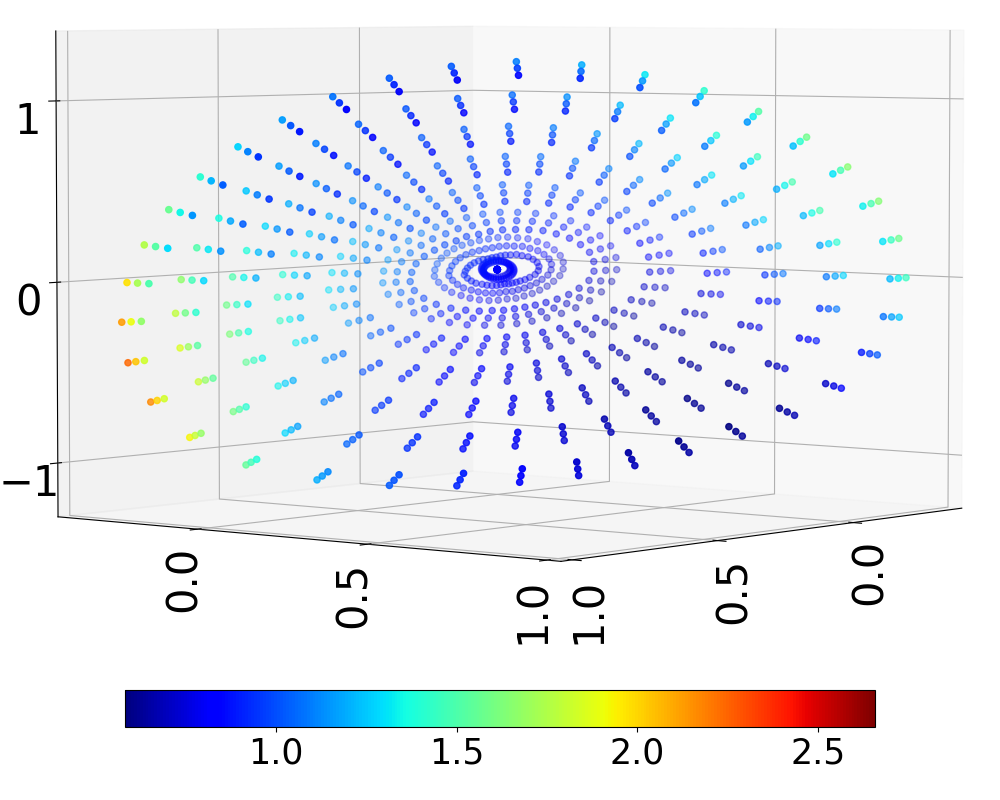}&
    \includegraphics[width=28mm]{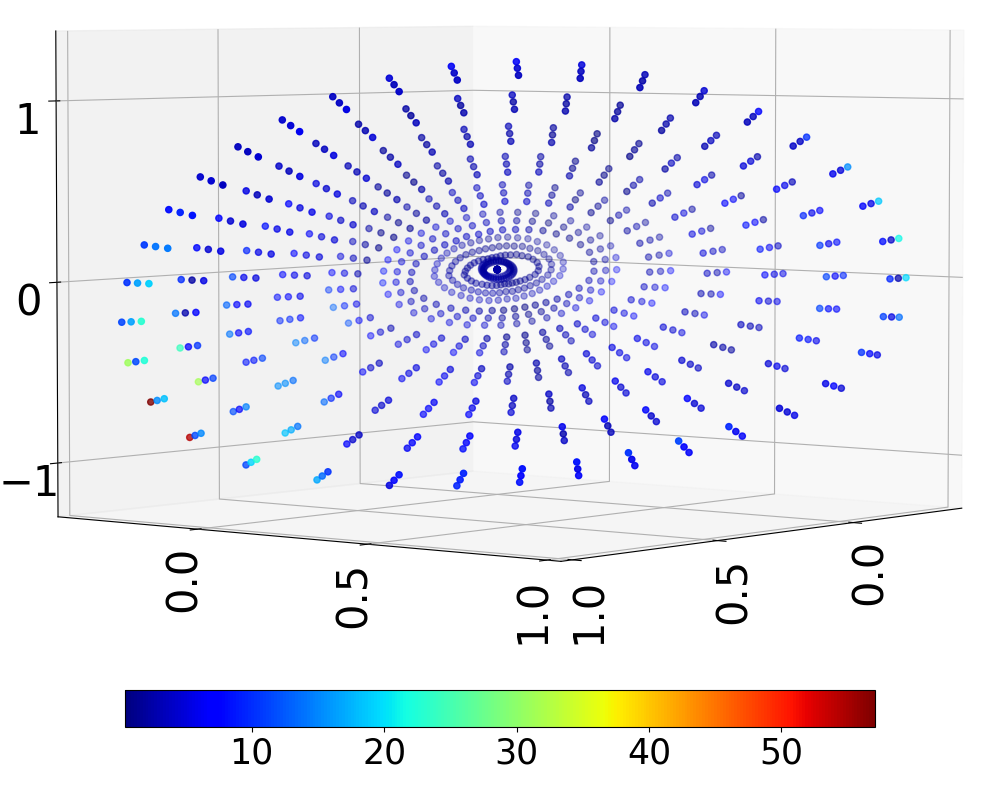}\\
    \hline
    M2 &
    \includegraphics[width=28mm]{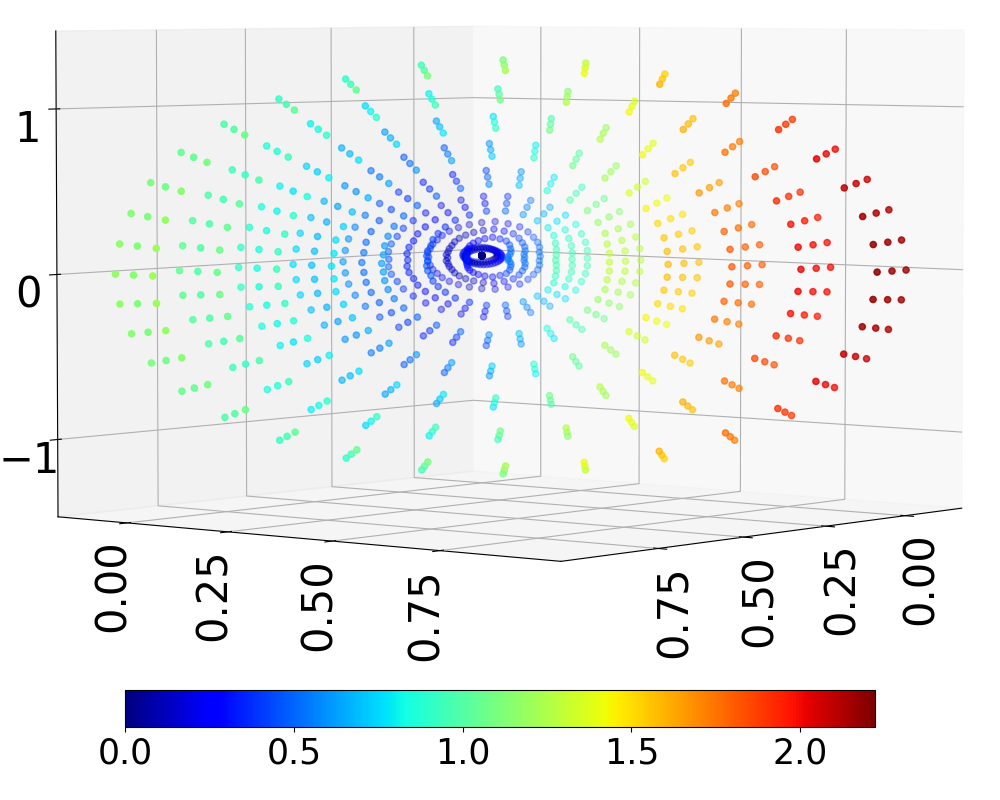}& \includegraphics[width=28mm]{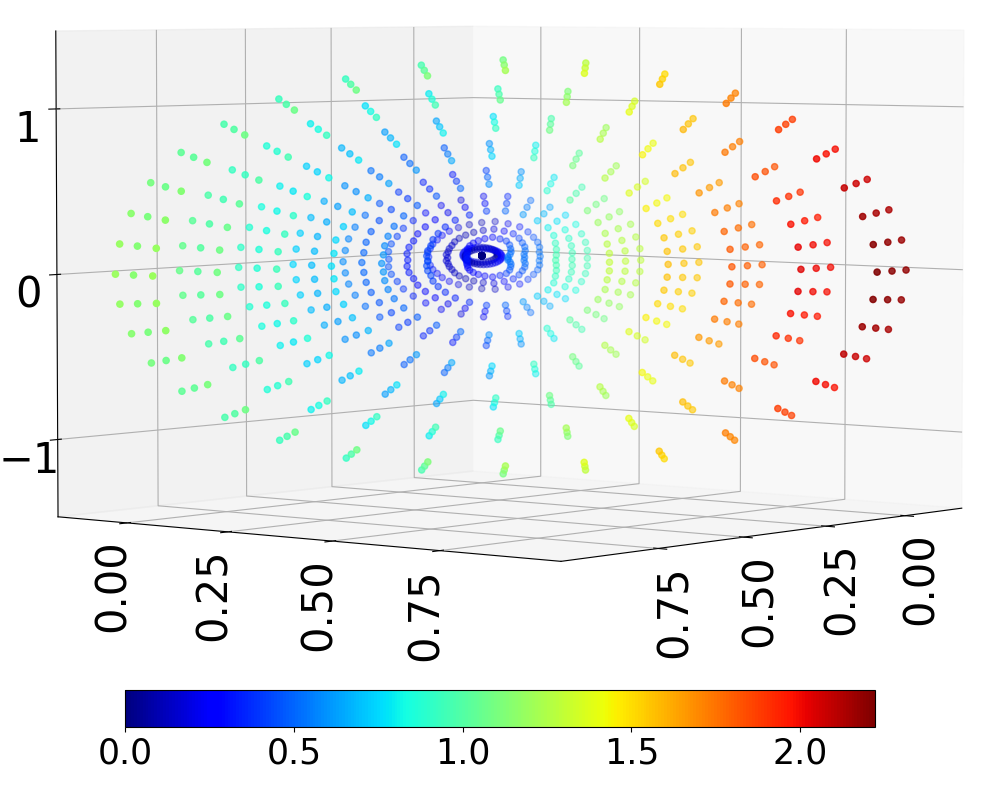}& \includegraphics[width=28mm]{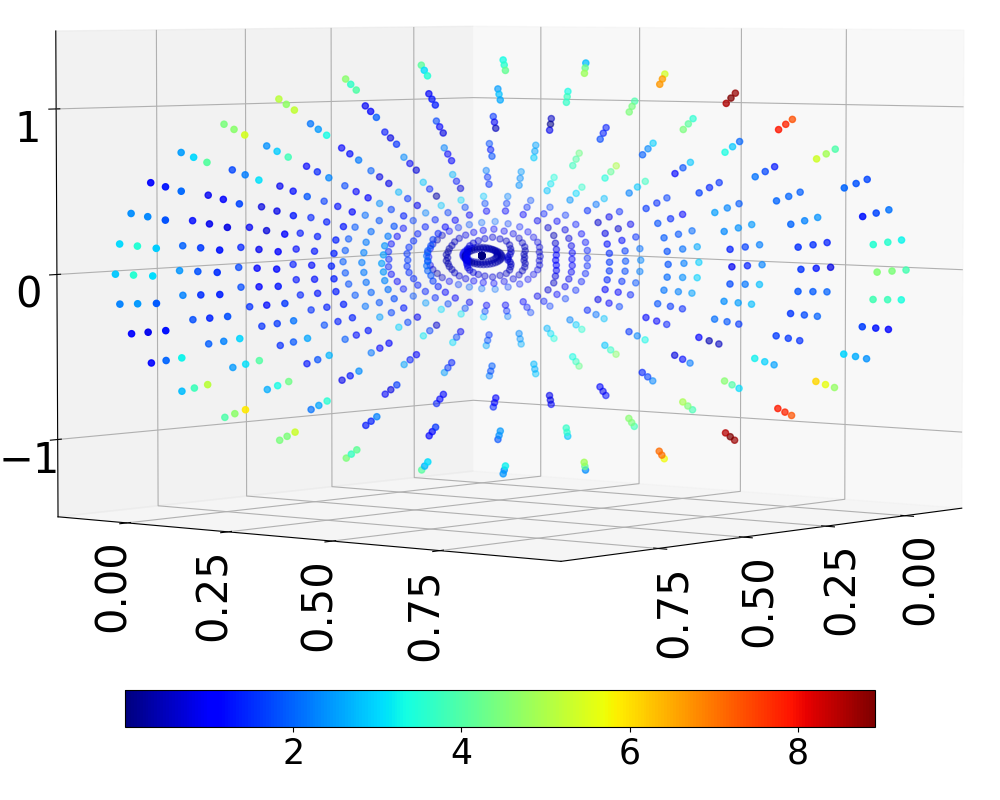}& \includegraphics[width=28mm]{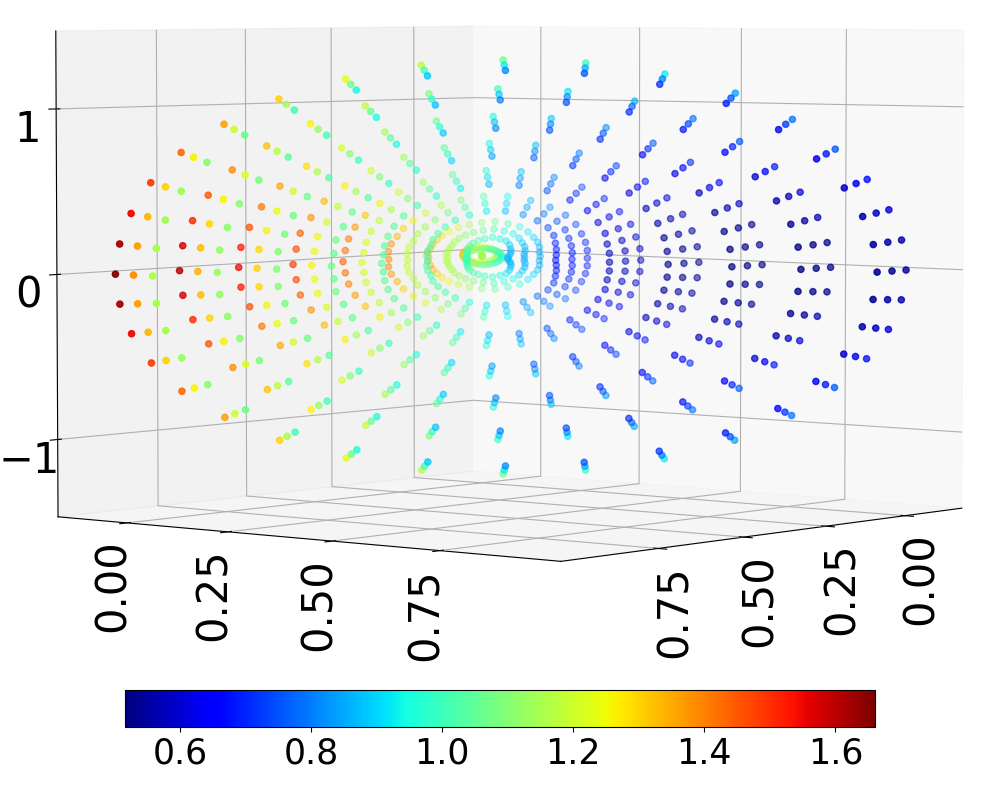}& \includegraphics[width=28mm]{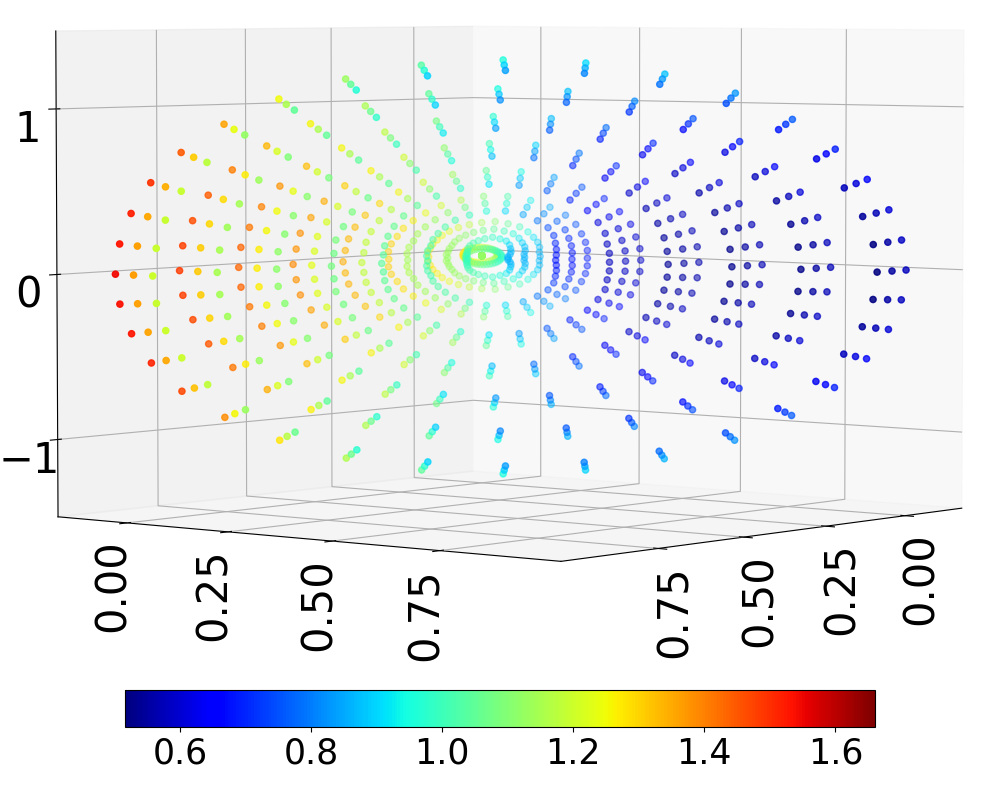}&
    \includegraphics[width=28mm]{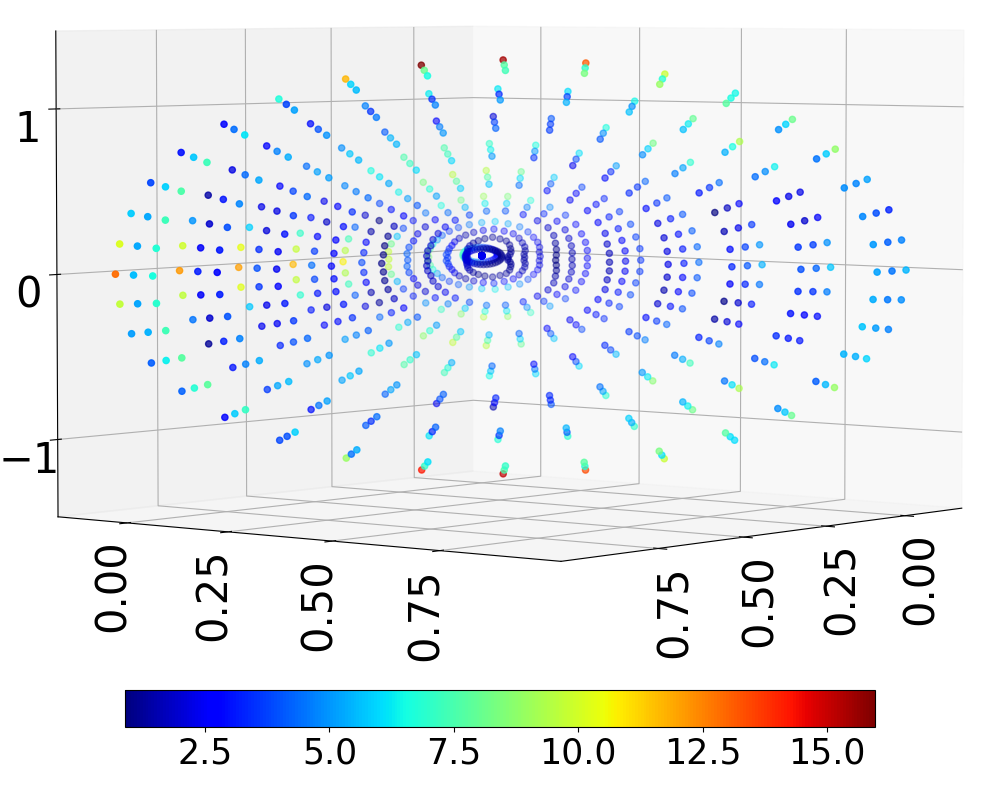}\\
    \hline
    M3 &
    \includegraphics[width=28mm]{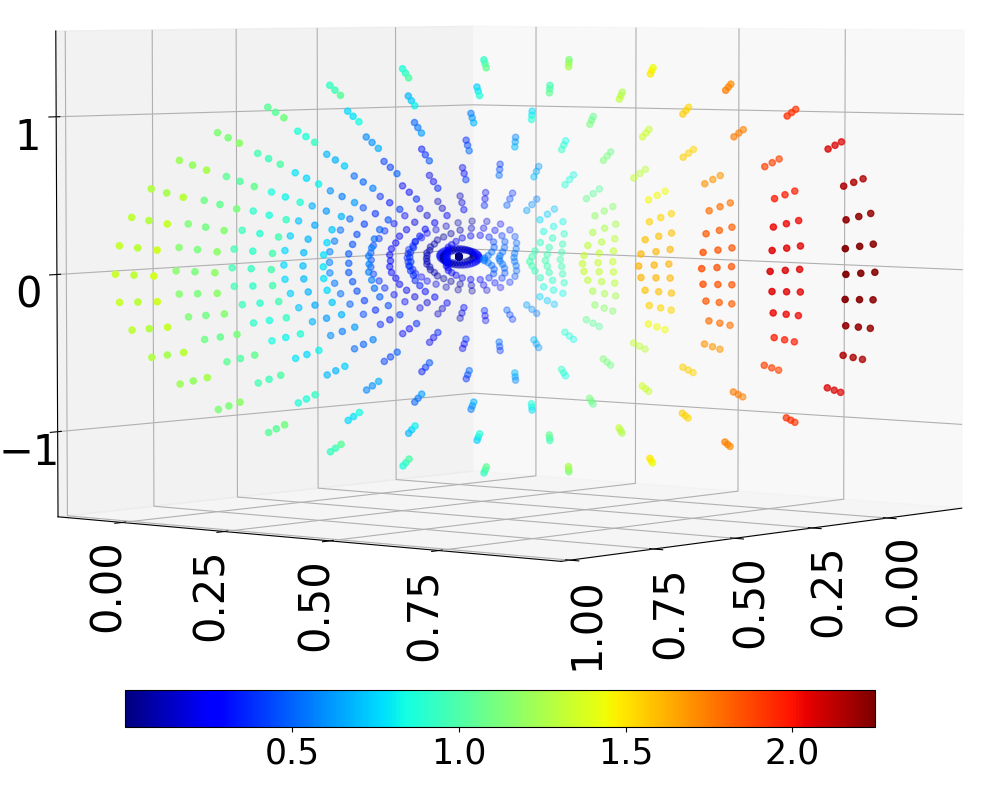}& \includegraphics[width=28mm]{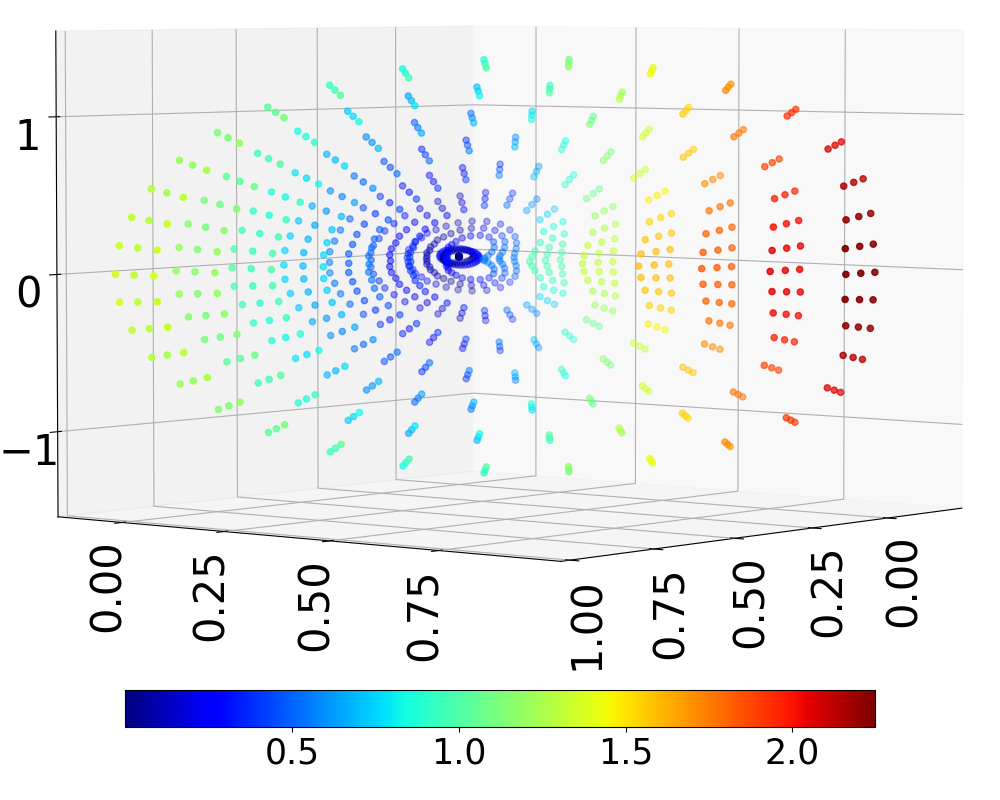}& \includegraphics[width=28mm]{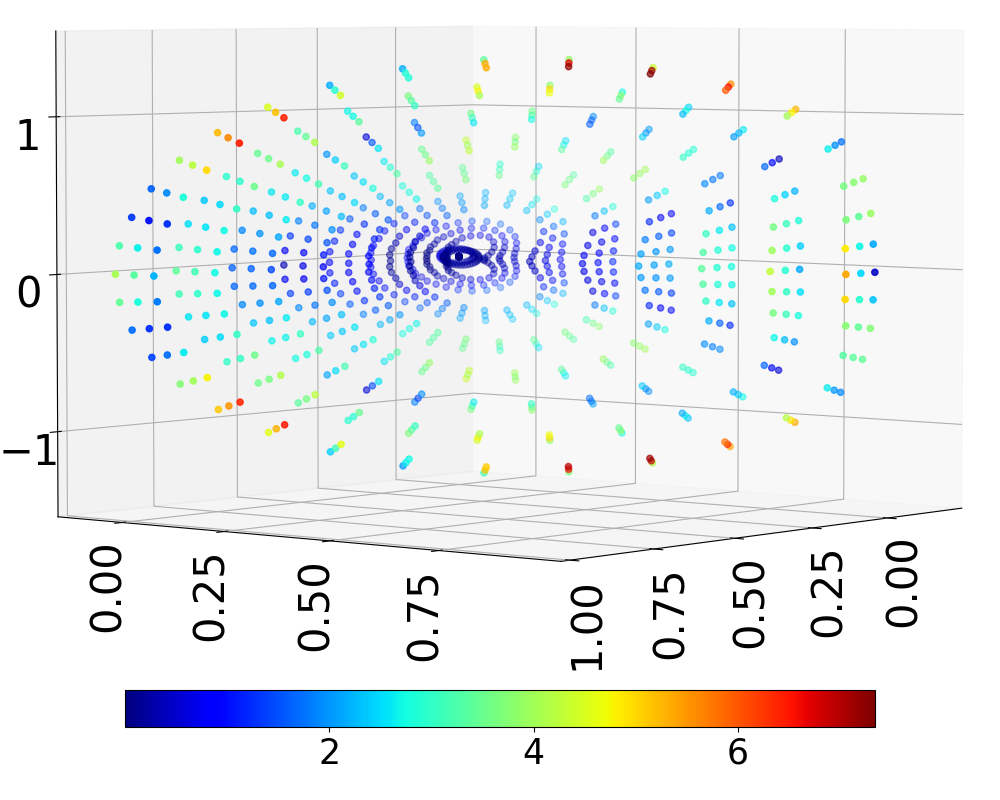}& \includegraphics[width=28mm]{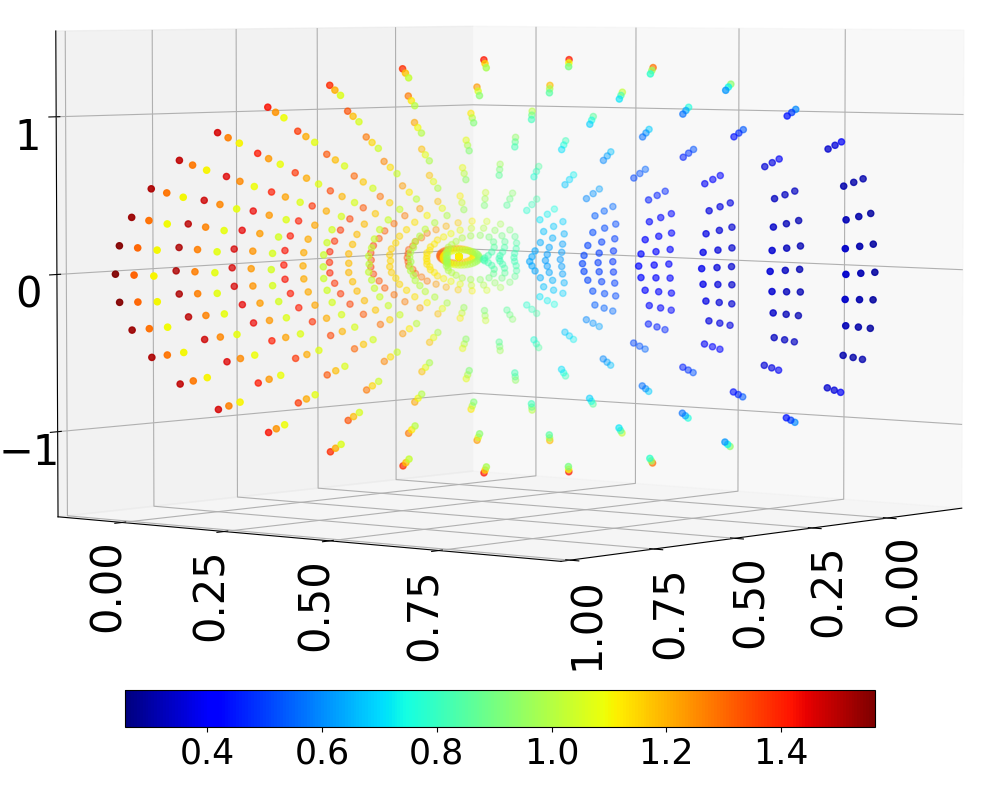}& \includegraphics[width=28mm]{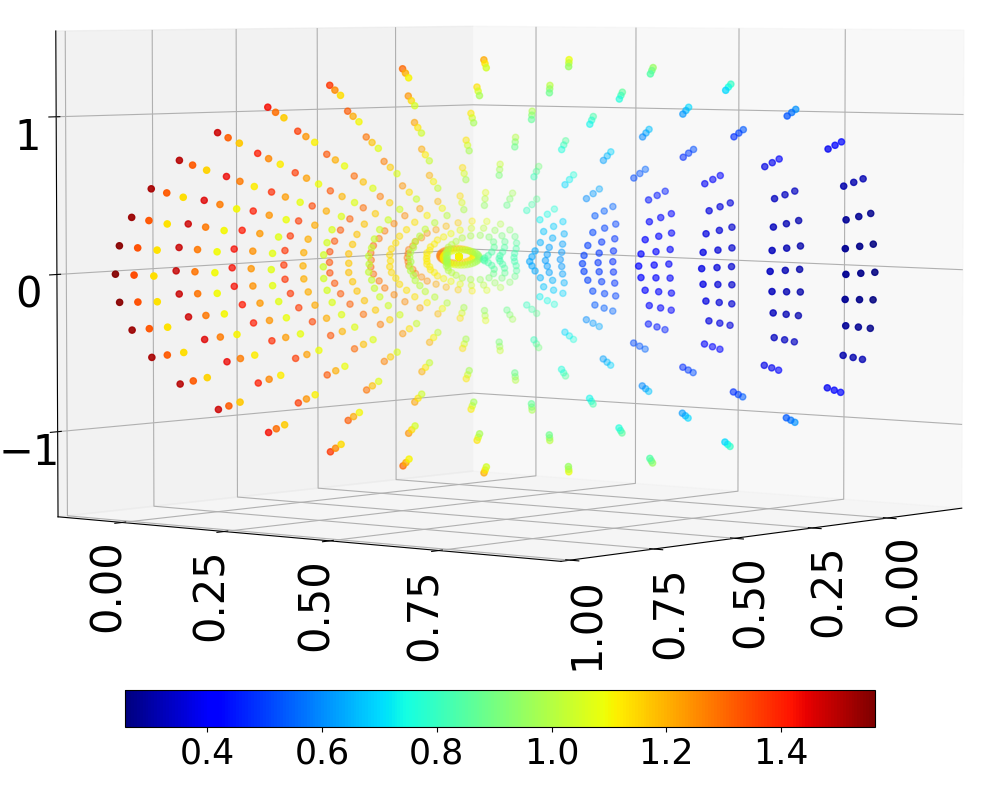}&
    \includegraphics[width=28mm]{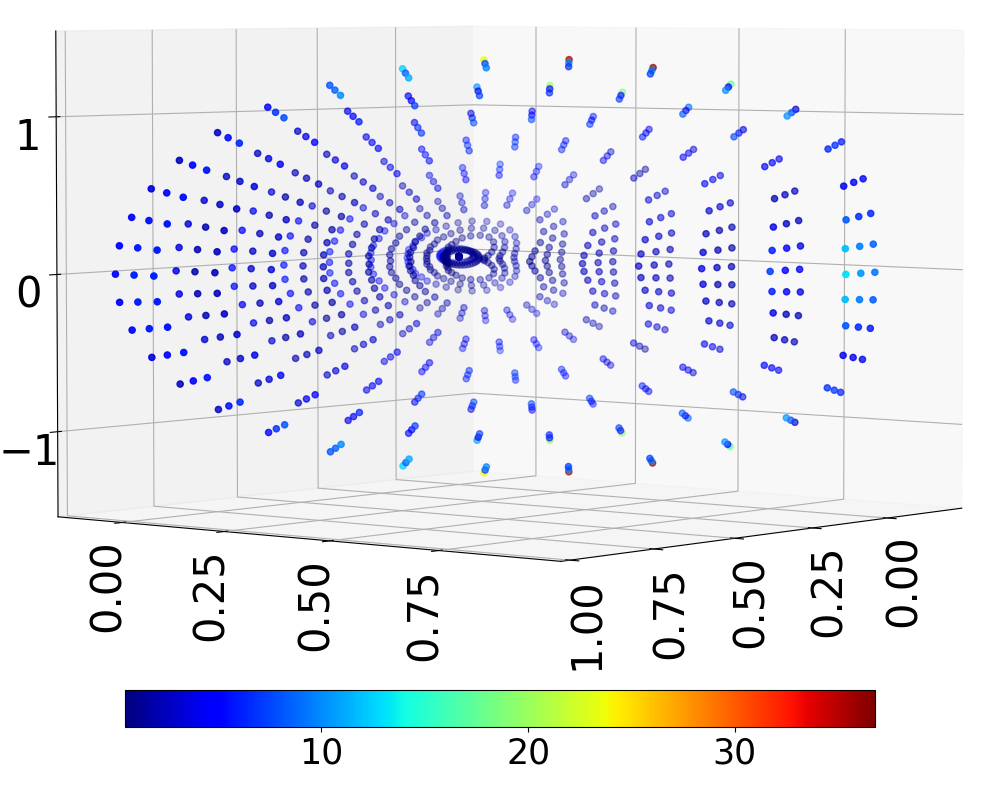}\\
    \hline
    M4 &
    \includegraphics[width=28mm]{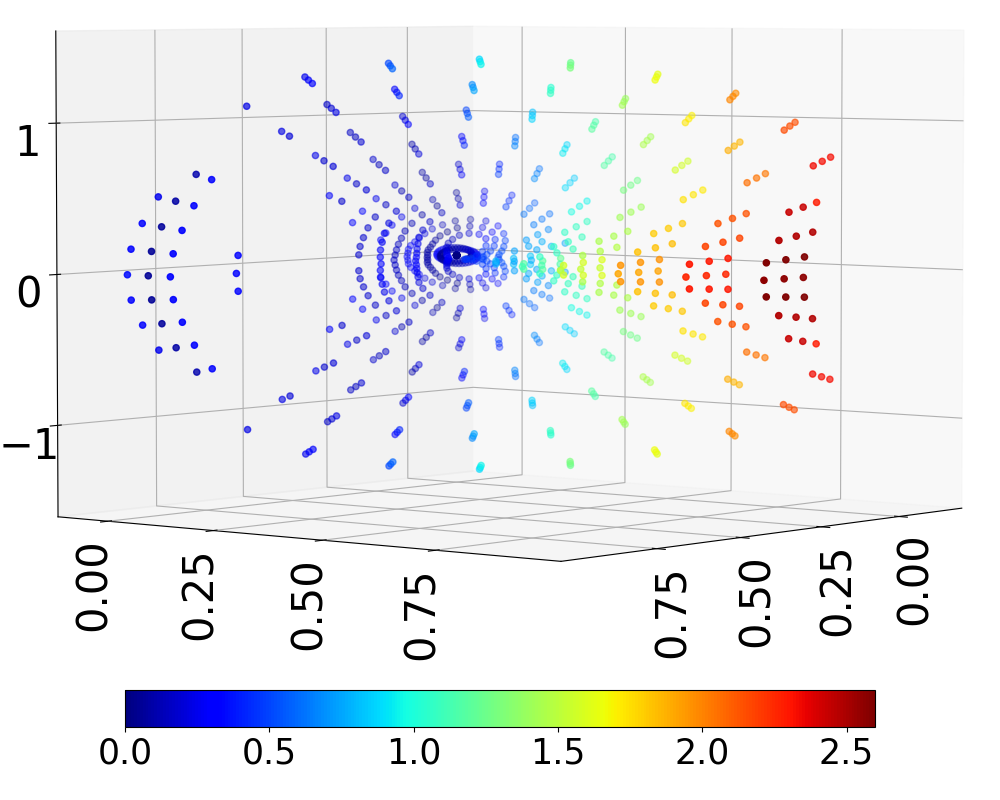}& \includegraphics[width=28mm]{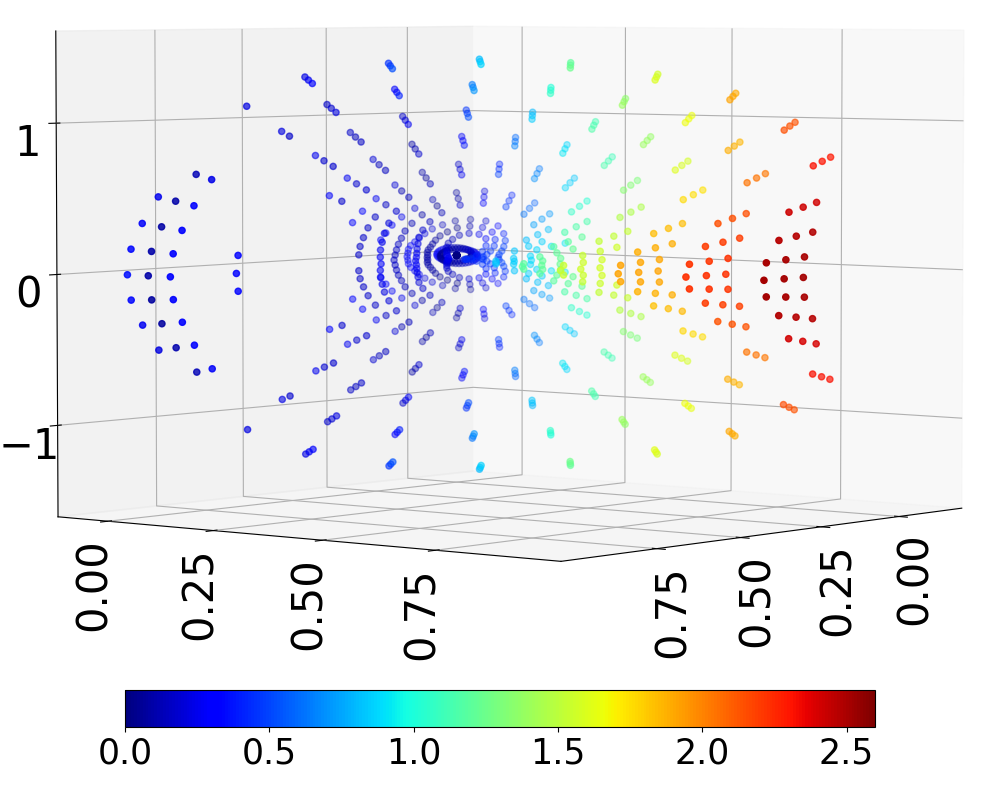}& \includegraphics[width=28mm]{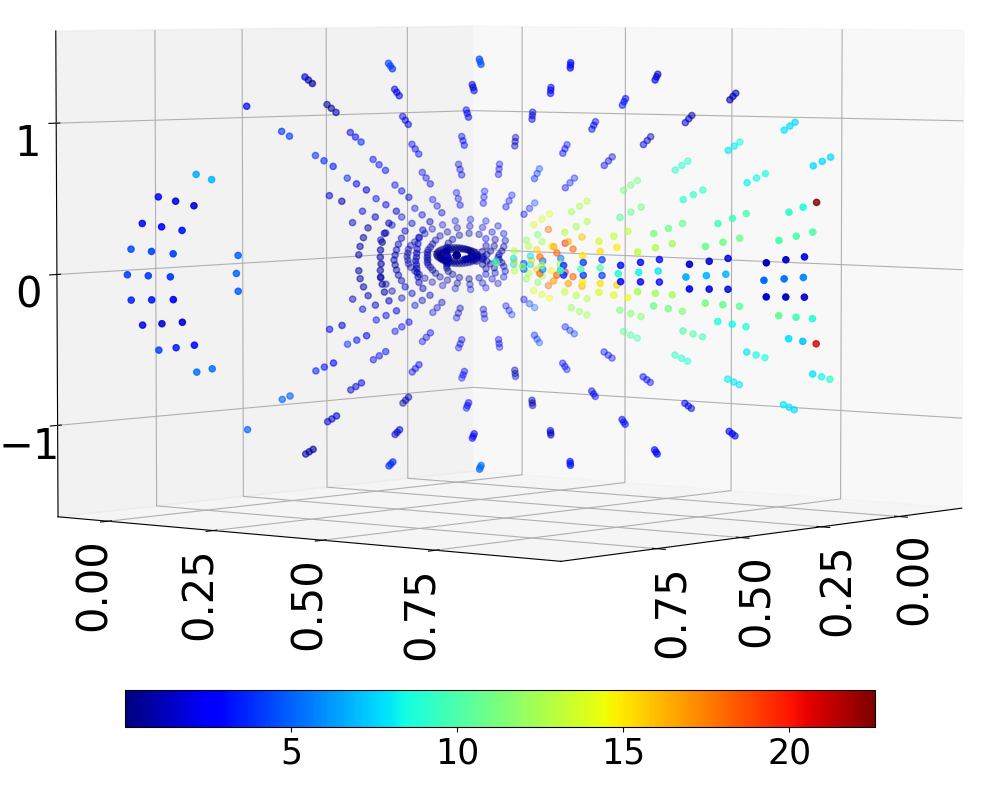}& \includegraphics[width=28mm]{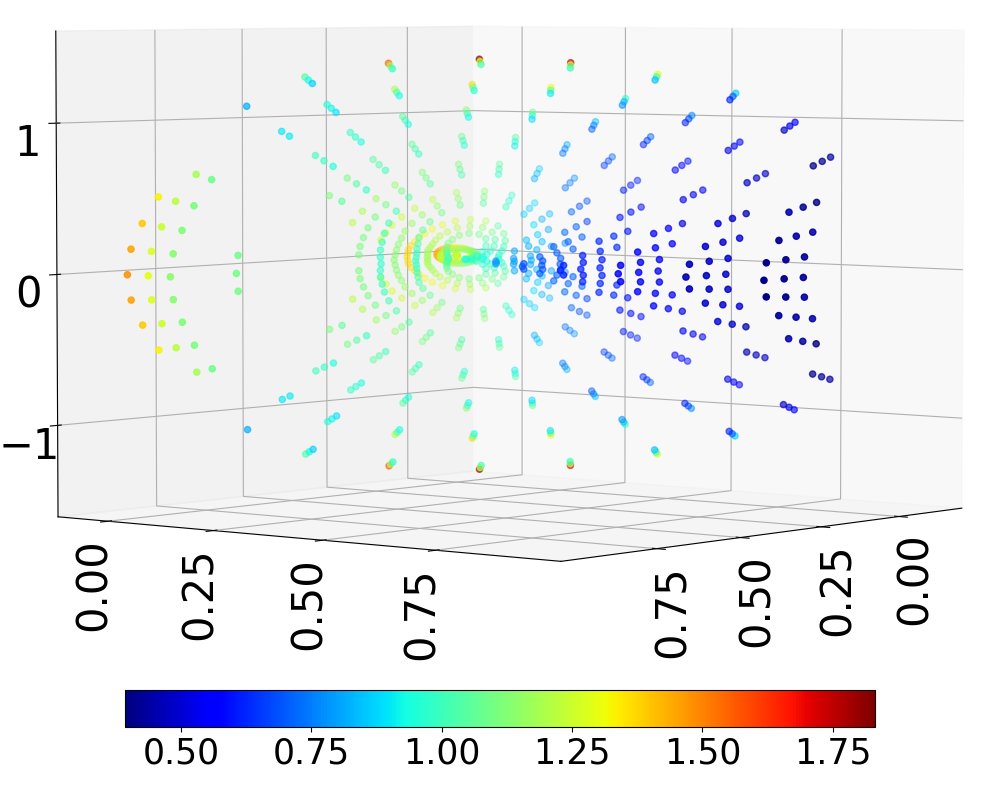}& \includegraphics[width=28mm]{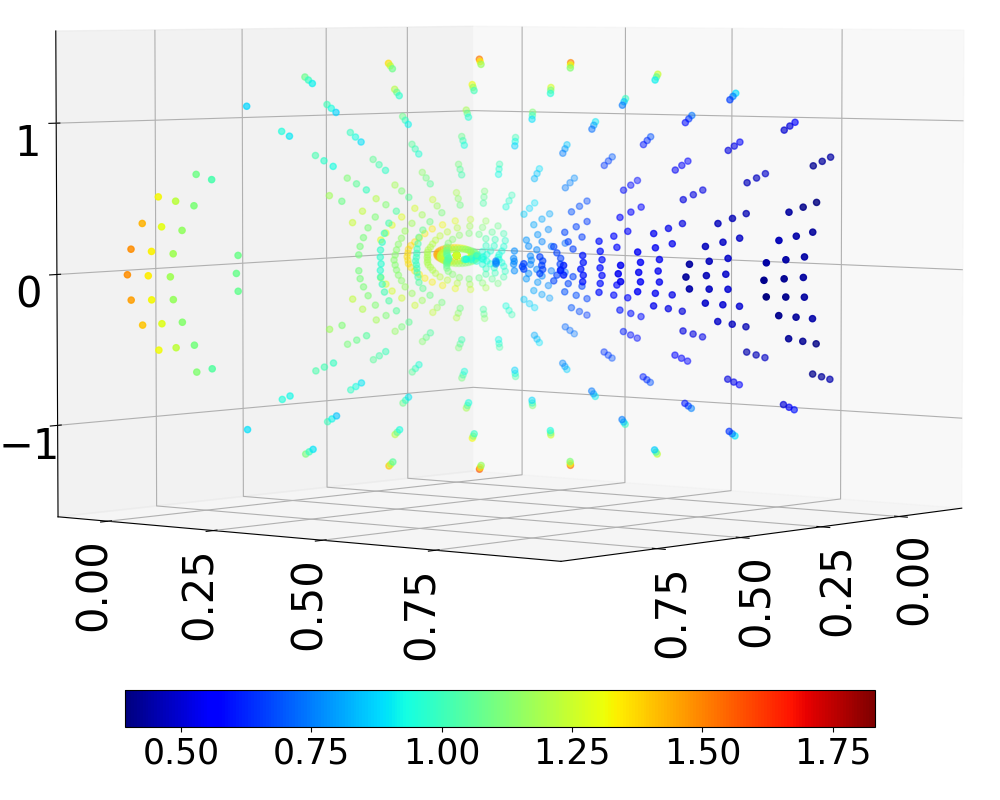}&
    \includegraphics[width=28mm]{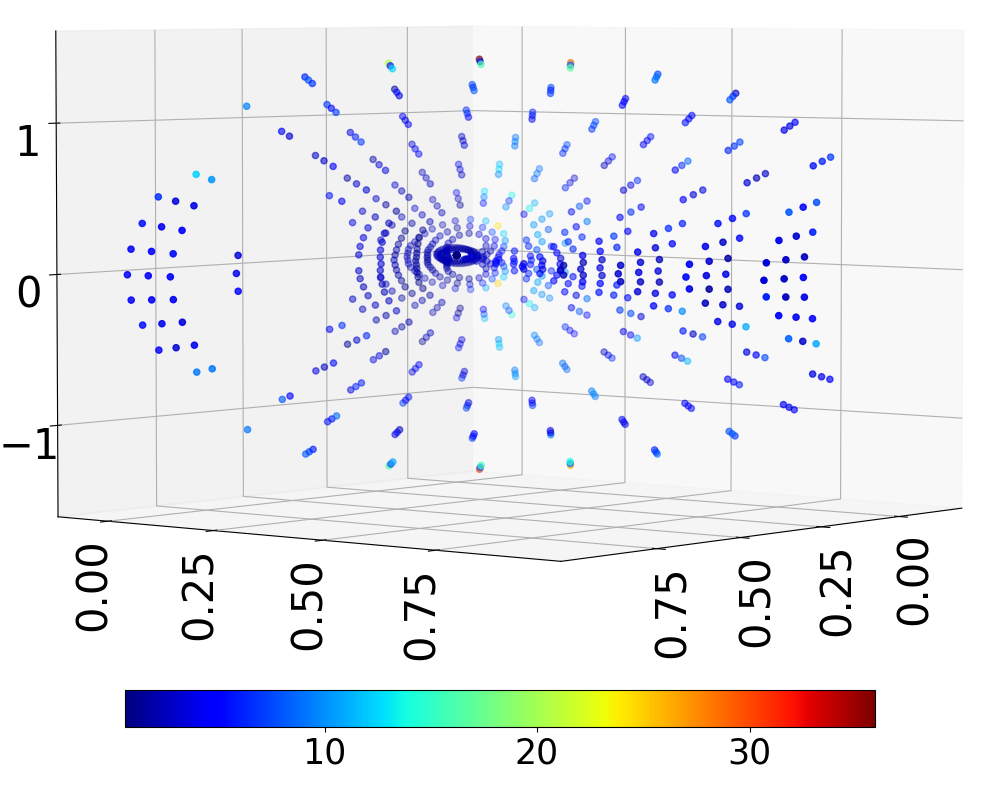}\\
    \hline
    M5 &
    \includegraphics[width=28mm]{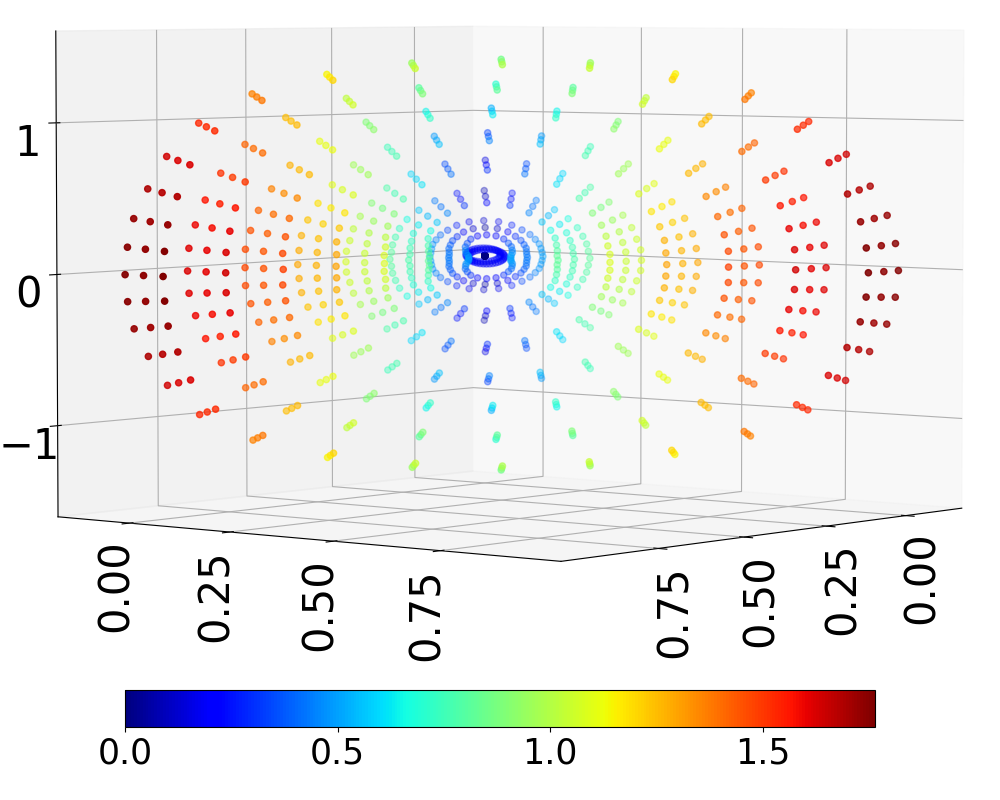}& \includegraphics[width=28mm]{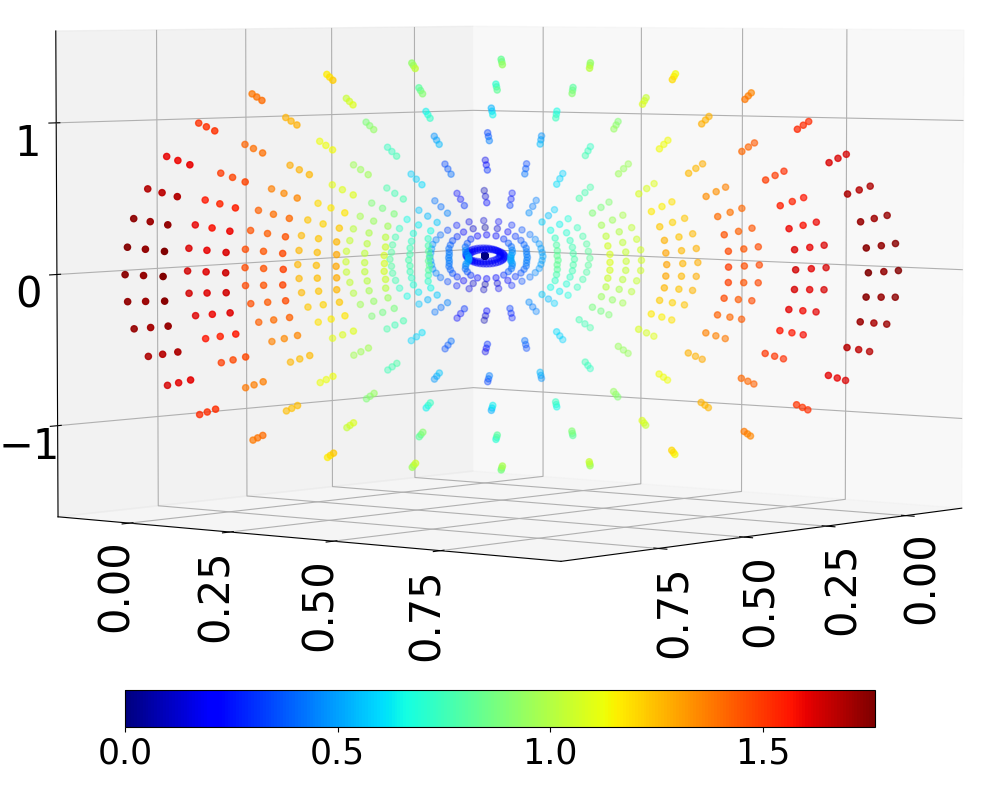}& \includegraphics[width=28mm]{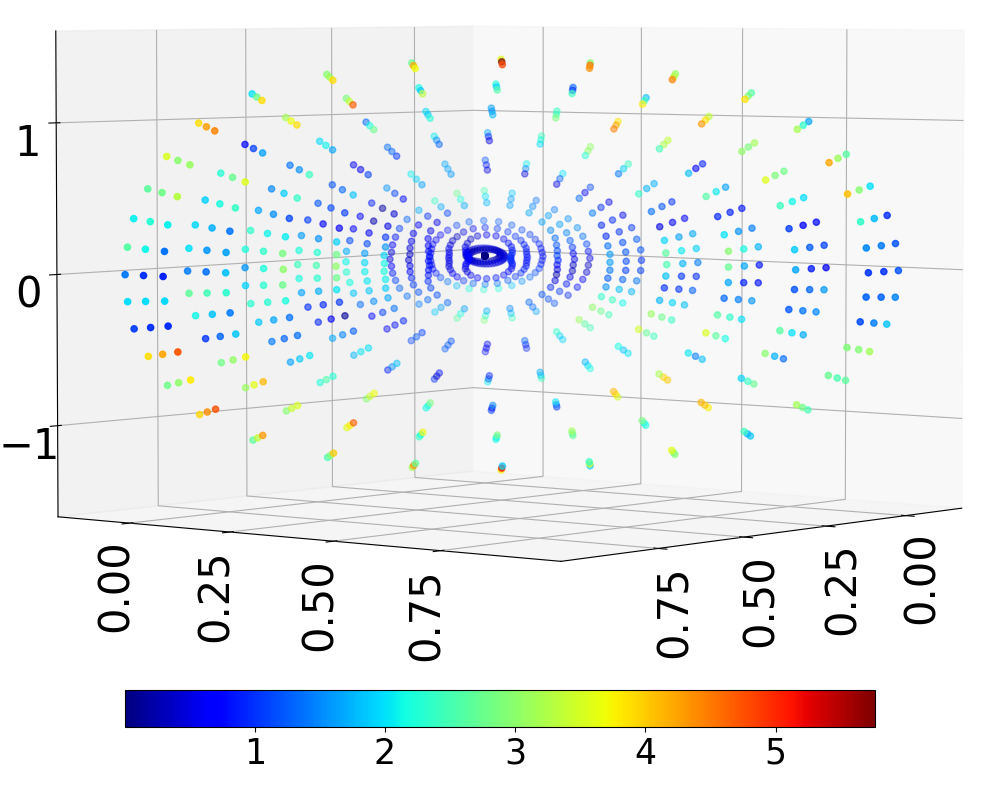}& \includegraphics[width=28mm]{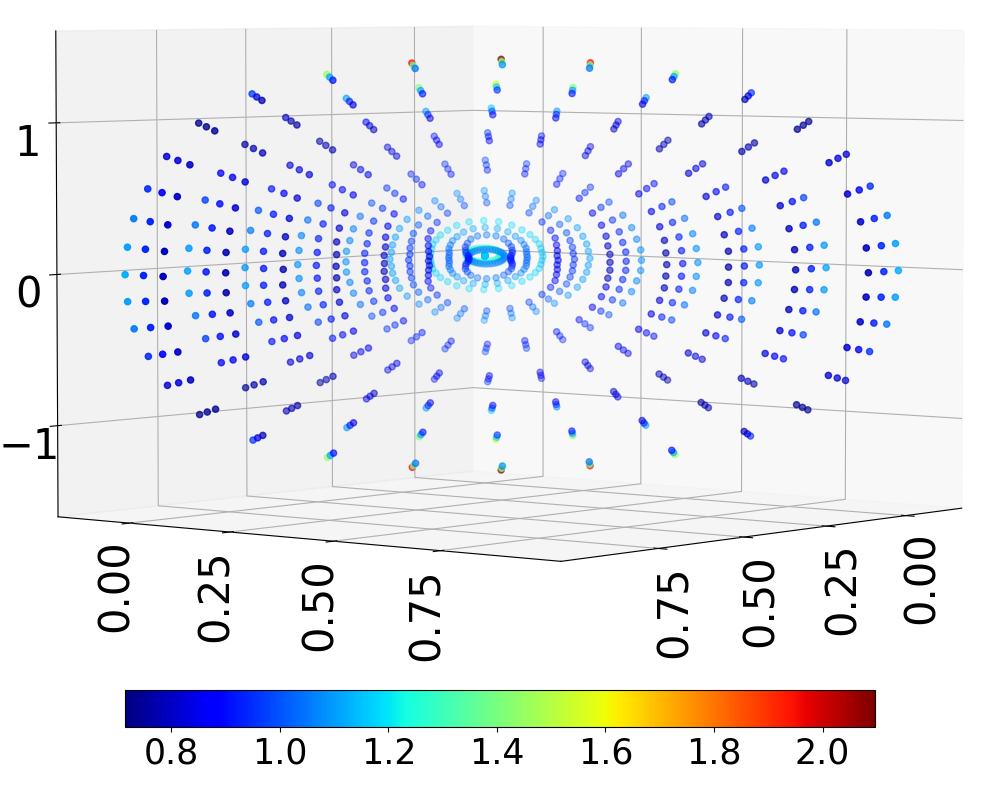}& \includegraphics[width=28mm]{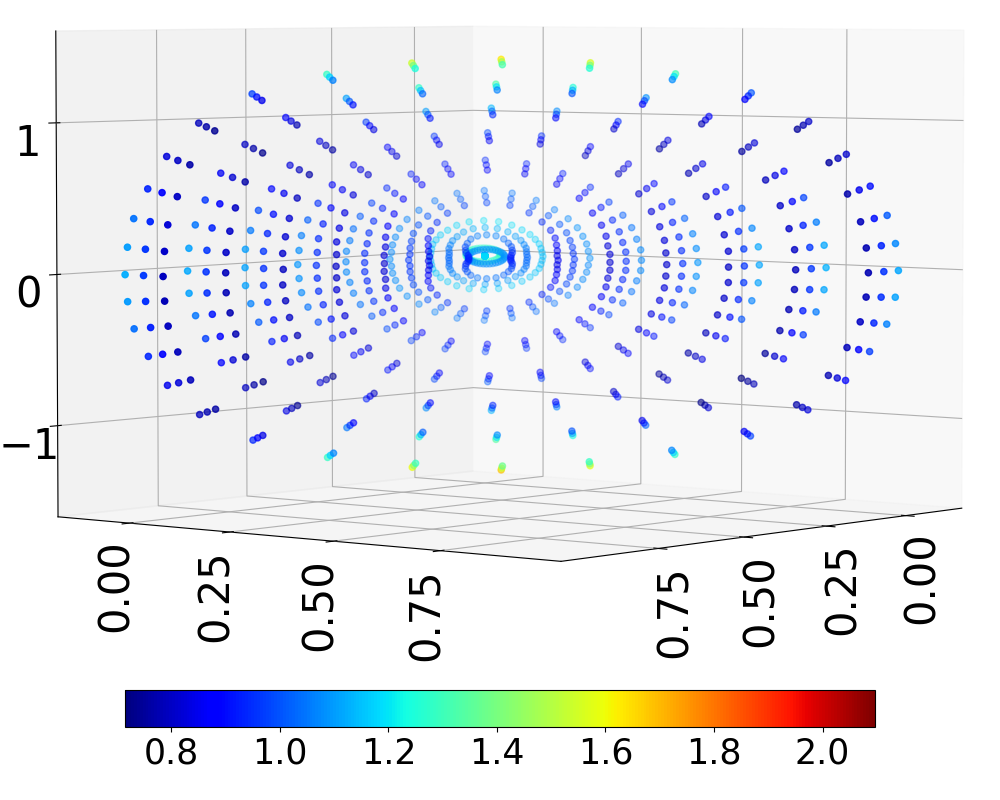}&
    \includegraphics[width=28mm]{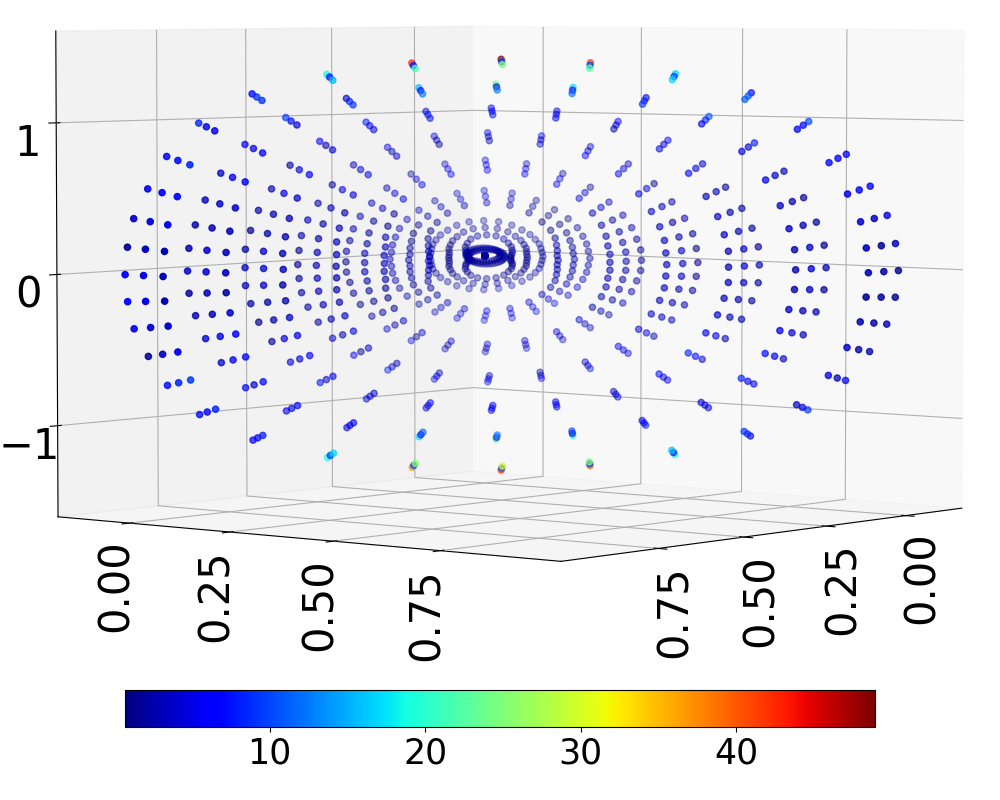}\\
    \hline
    M6 &
    \includegraphics[width=28mm]{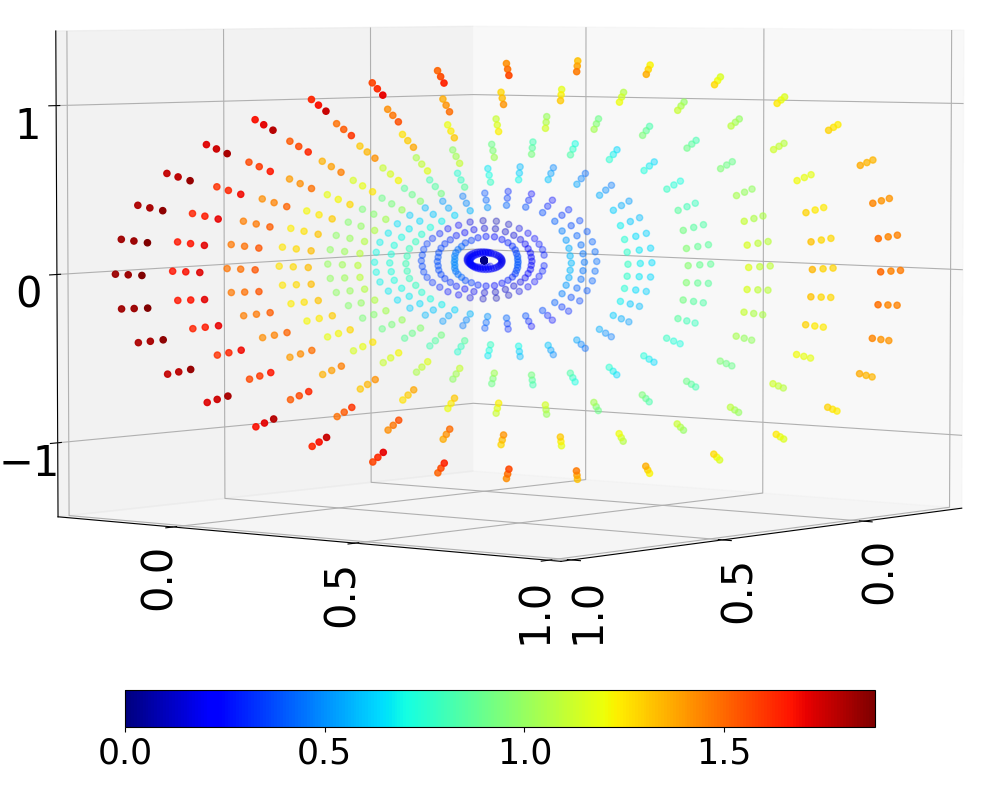}& \includegraphics[width=28mm]{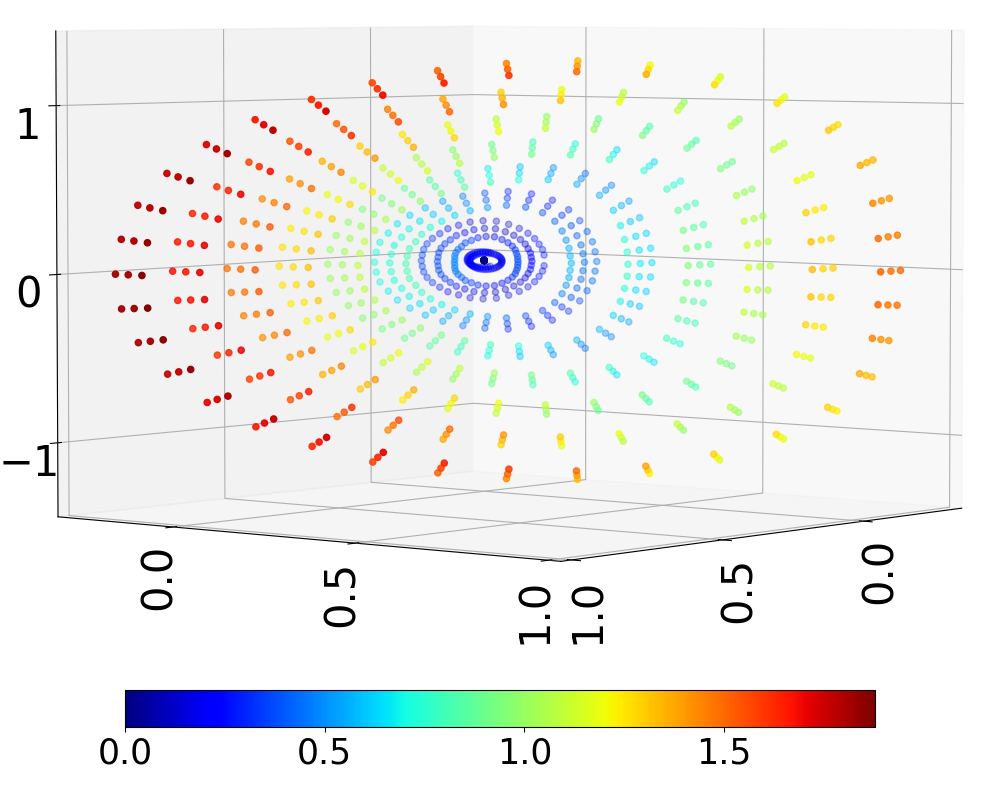}& \includegraphics[width=28mm]{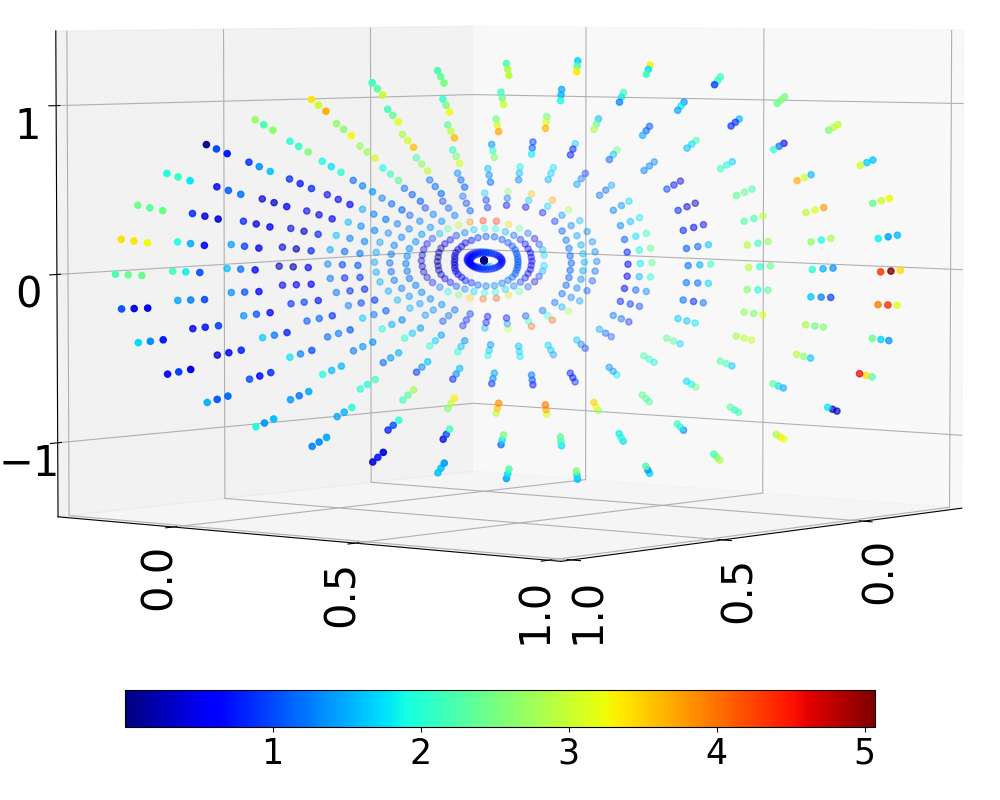}& \includegraphics[width=28mm]{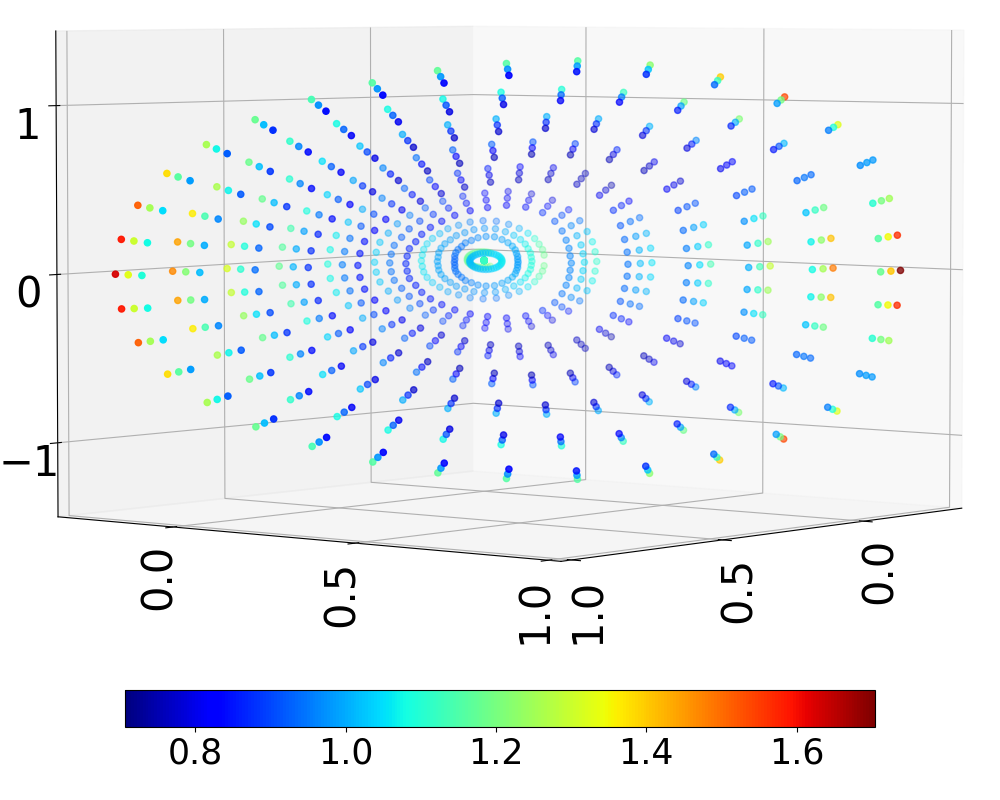}& \includegraphics[width=28mm]{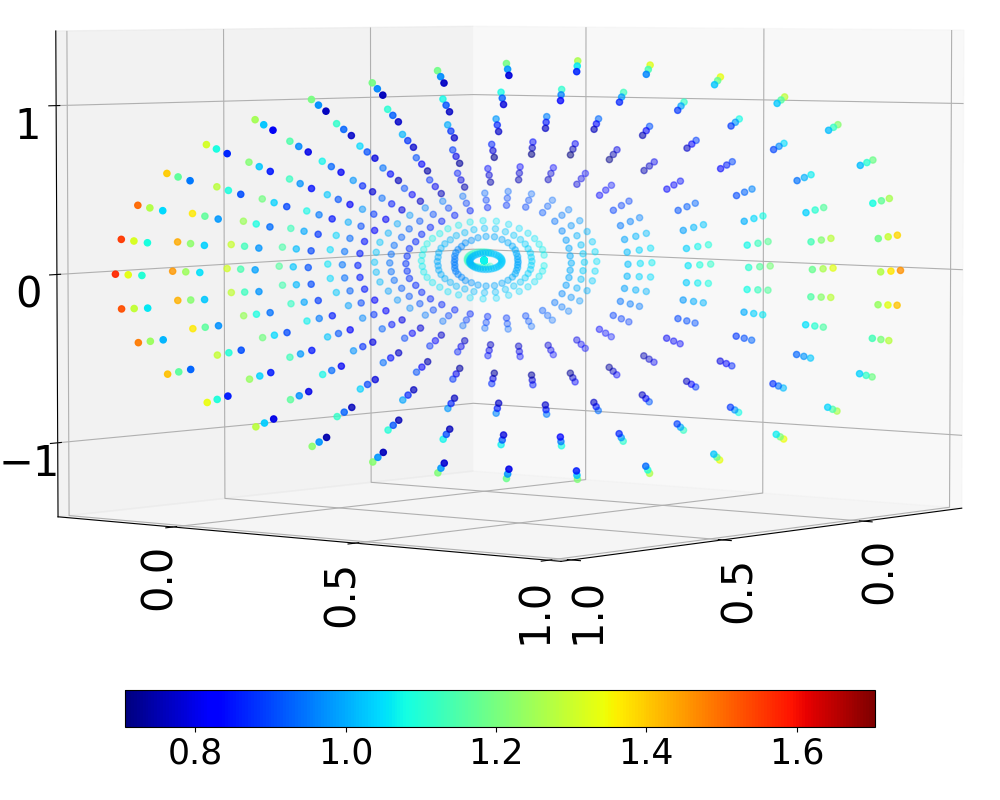}&
    \includegraphics[width=28mm]{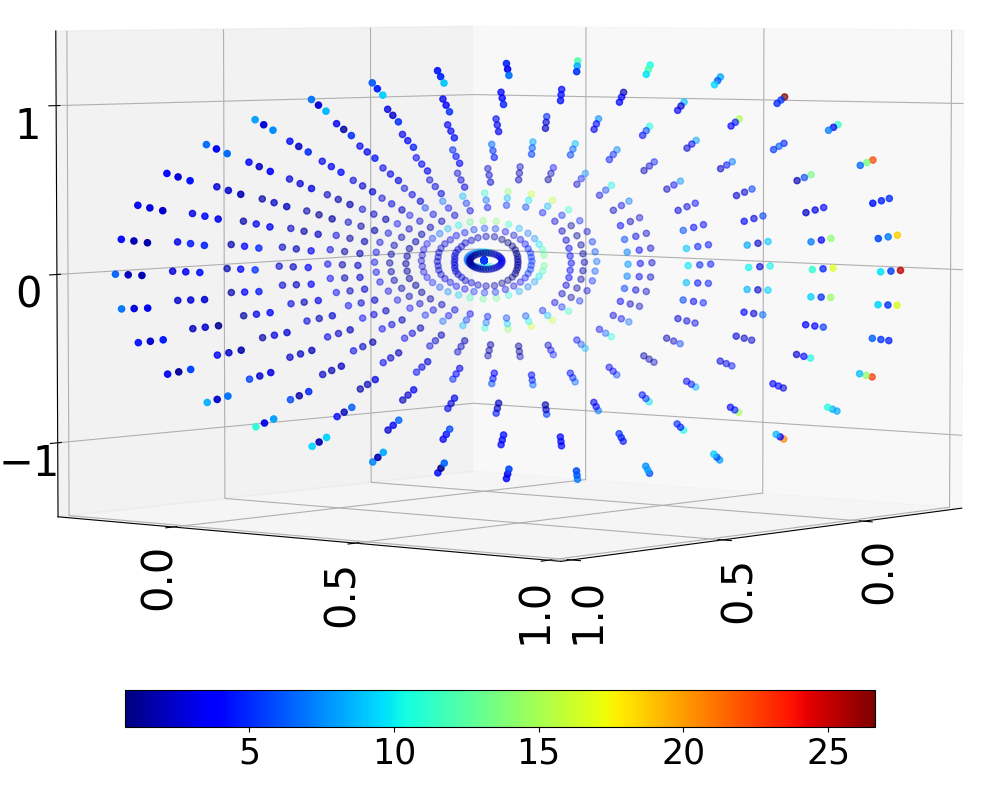}\\
\end{tabular}
\caption{Comparison of target and fitted stress and stiffness for materials 1 to 6. The colors indicate the norm of stress, stiffness, or error. \label{tab:bigtable}}
\end{table*}

\begin{table*}[t!]
    \begin{tabular}{m{0.5cm}  m{2.4cm}  m{2.4cm}  m{2.4cm}  m{2.4cm} m{2.4cm} m{2.4cm} }
    \thead{ID} &
    \thead{Target stress} &
    \thead{Fitted stress} &
    \thead{Stress error (\%)} &
    \thead{Target stiffness} &
    \thead{Fitted stiffness} &
    \thead{Stiff. error (\%)}\\
    \hline
    M7 &
    \includegraphics[width=28mm]{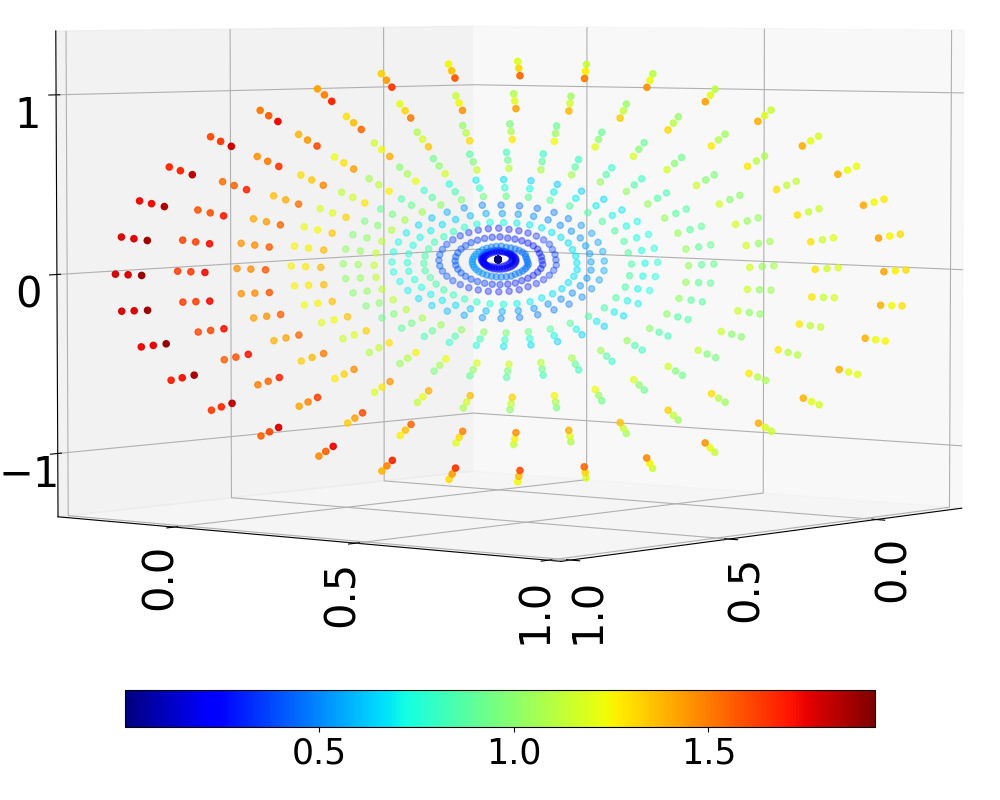}& \includegraphics[width=28mm]{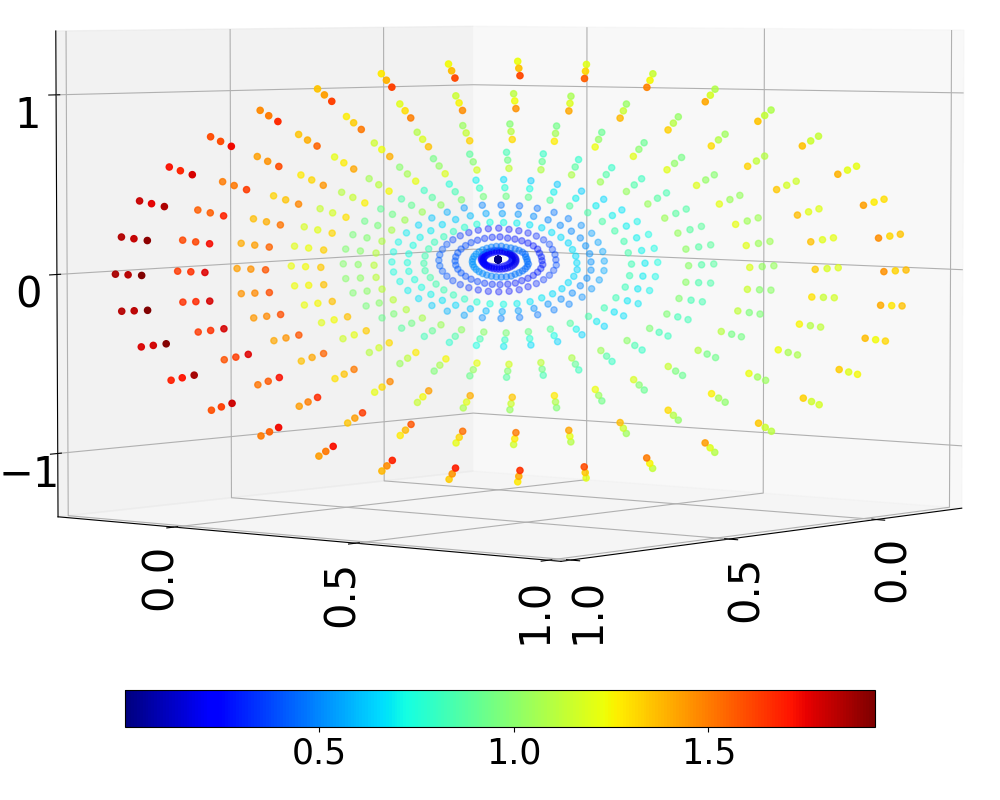}& \includegraphics[width=28mm]{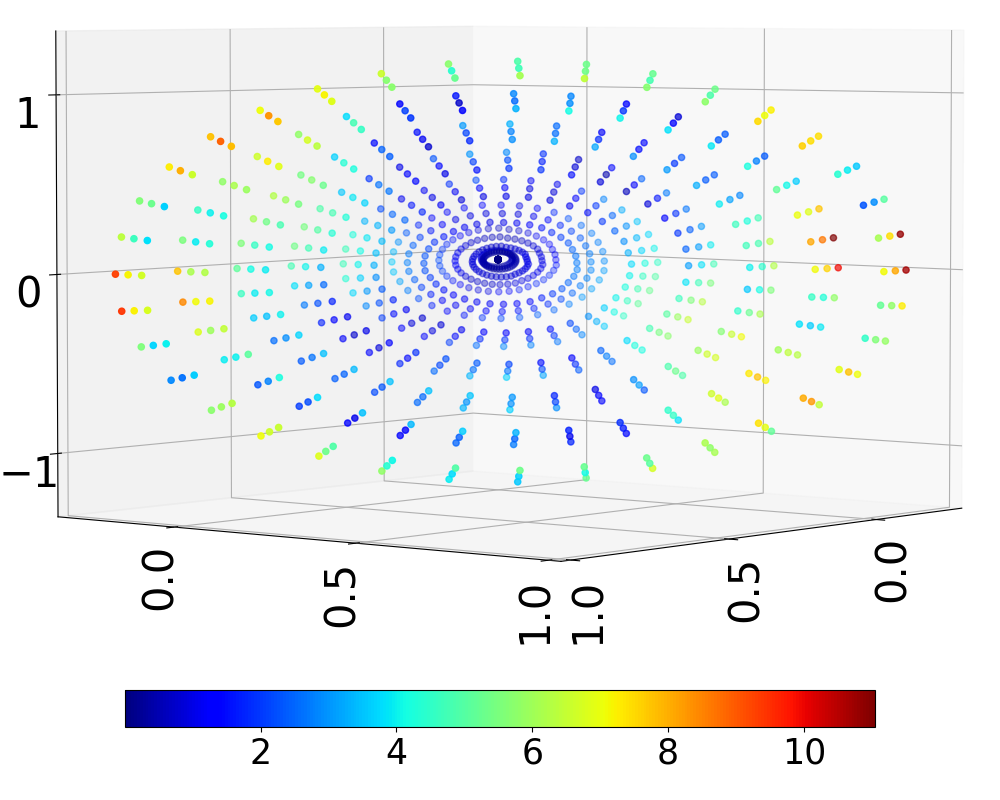}& \includegraphics[width=28mm]{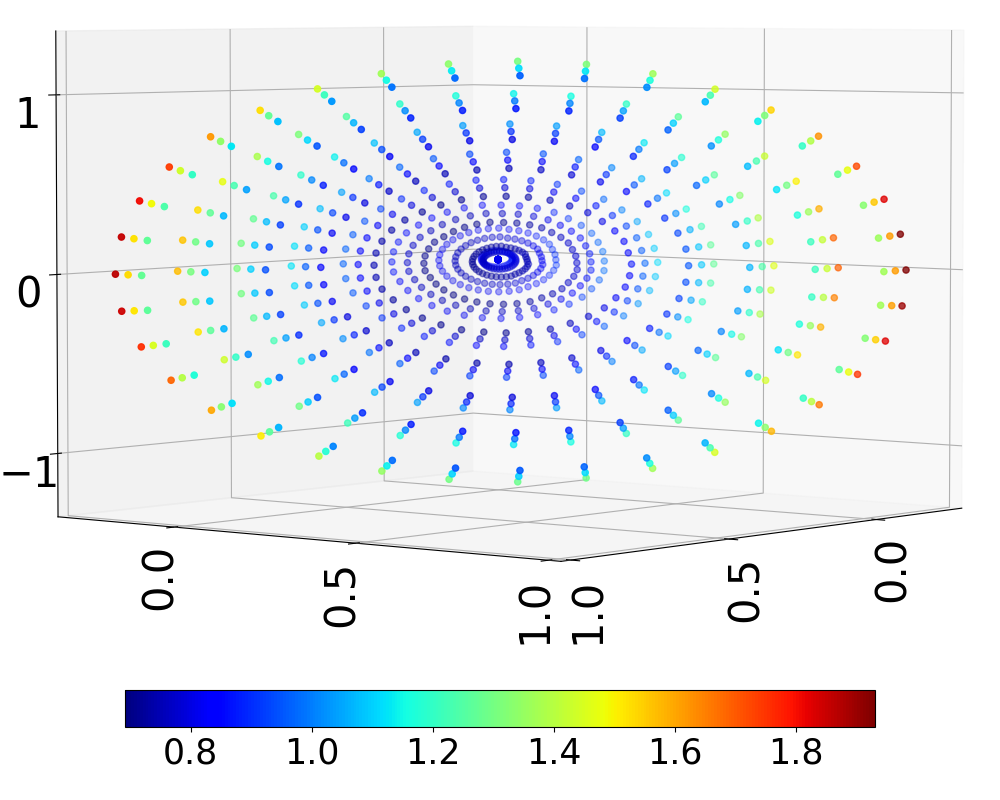}& \includegraphics[width=28mm]{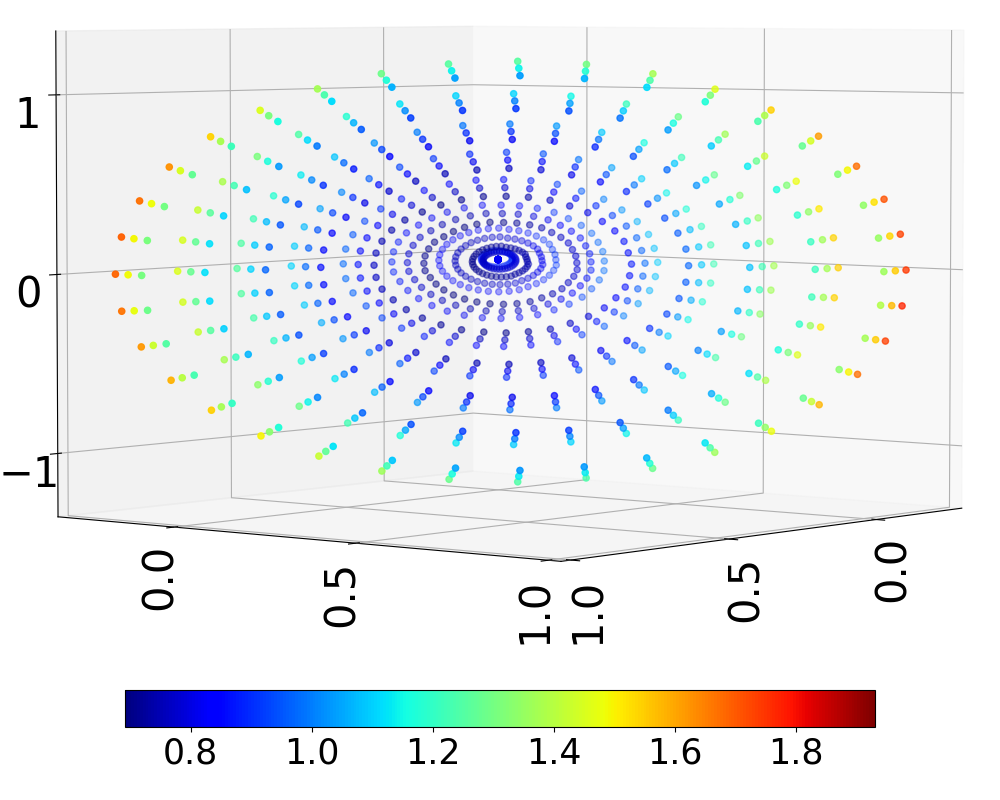}&
    \includegraphics[width=28mm]{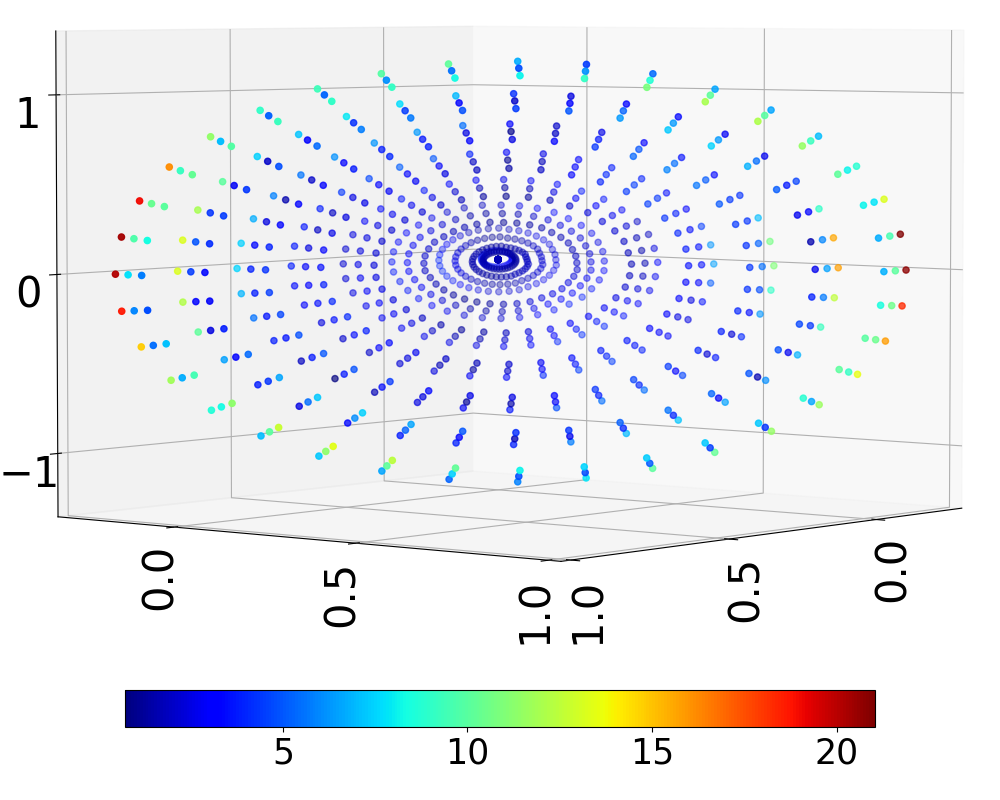}\\
    \hline
    M8 &
    \includegraphics[width=28mm]{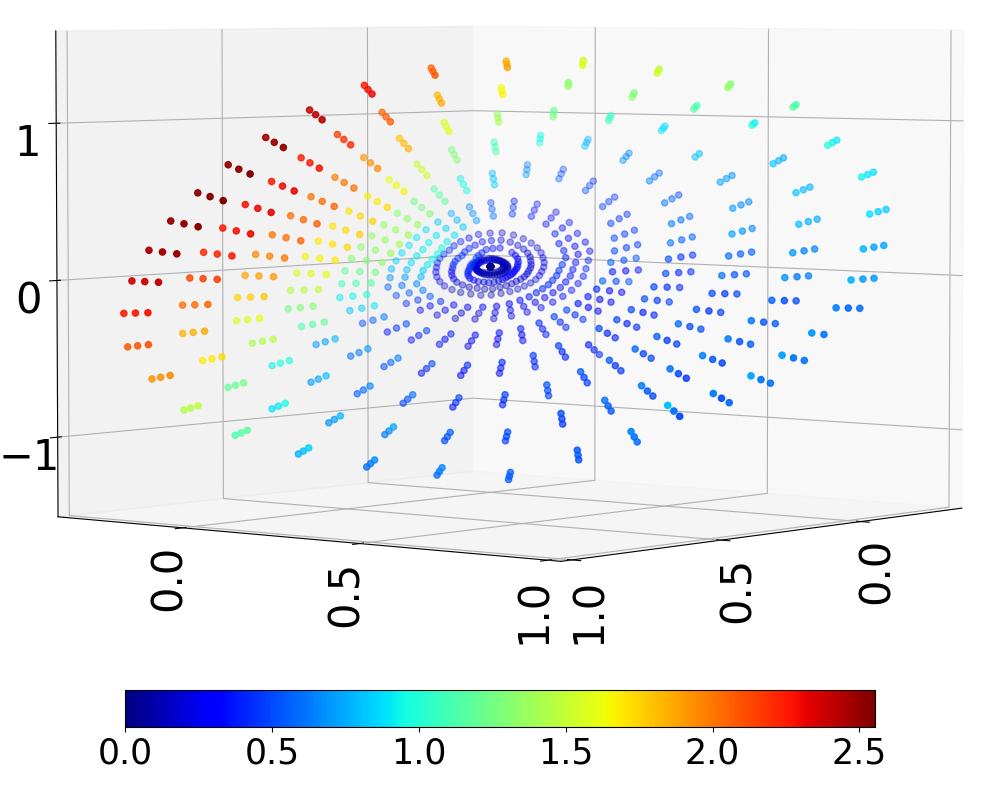}& \includegraphics[width=28mm]{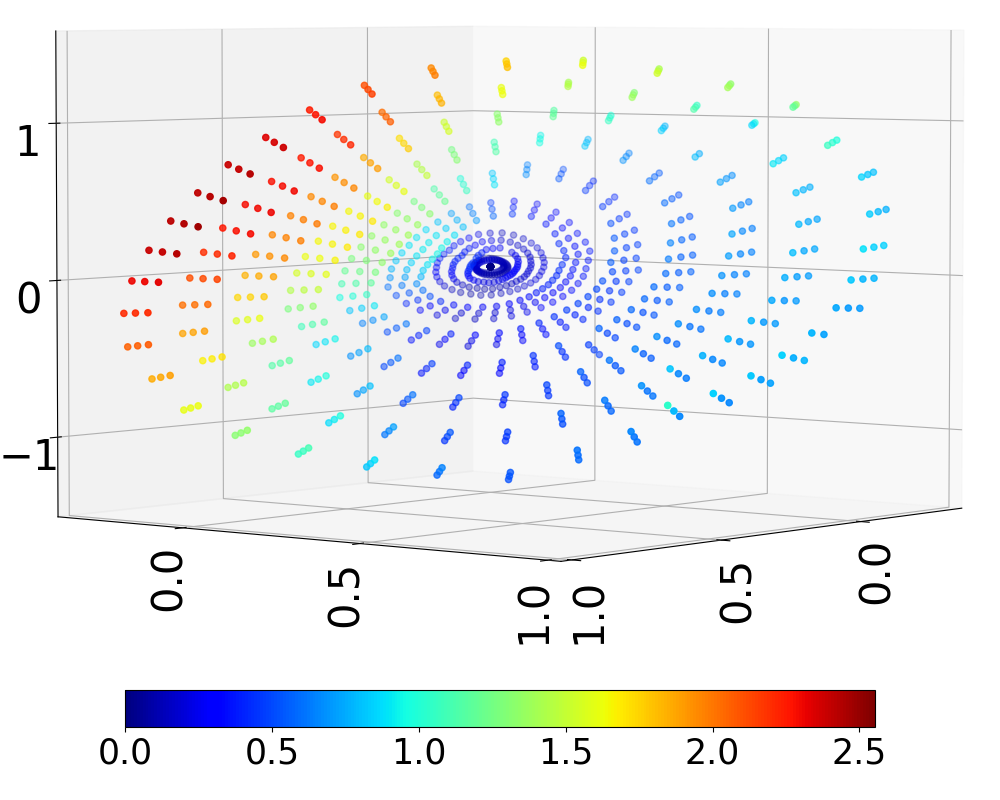}& \includegraphics[width=28mm]{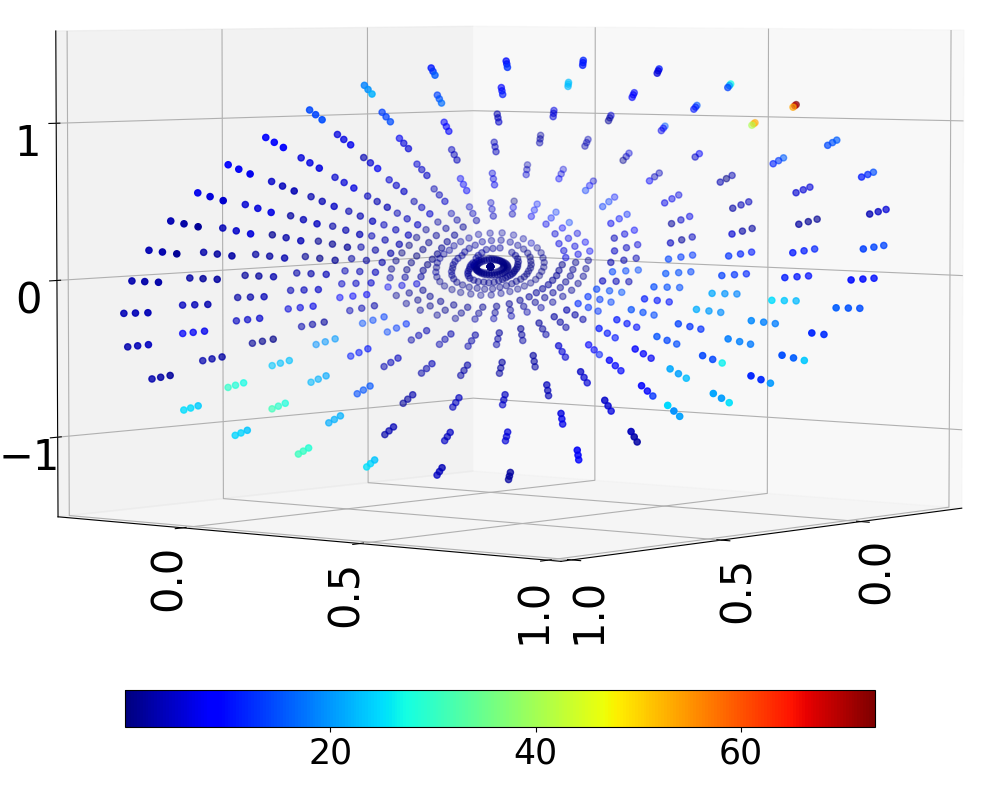}& \includegraphics[width=28mm]{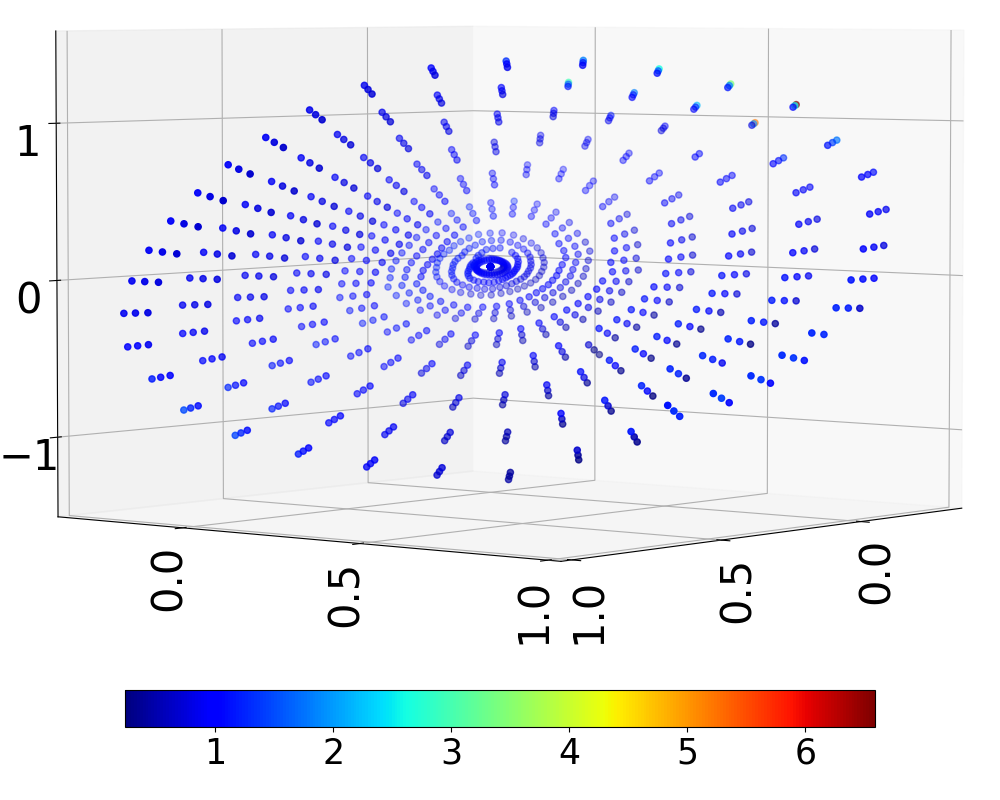}& \includegraphics[width=28mm]{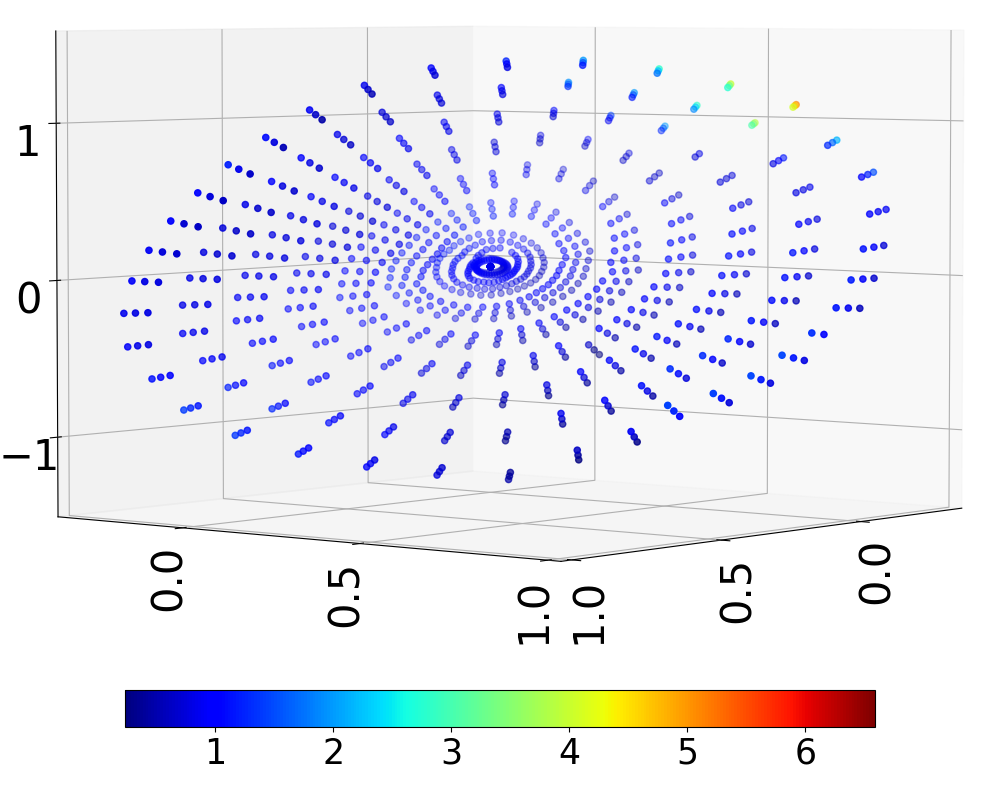}&
    \includegraphics[width=28mm]{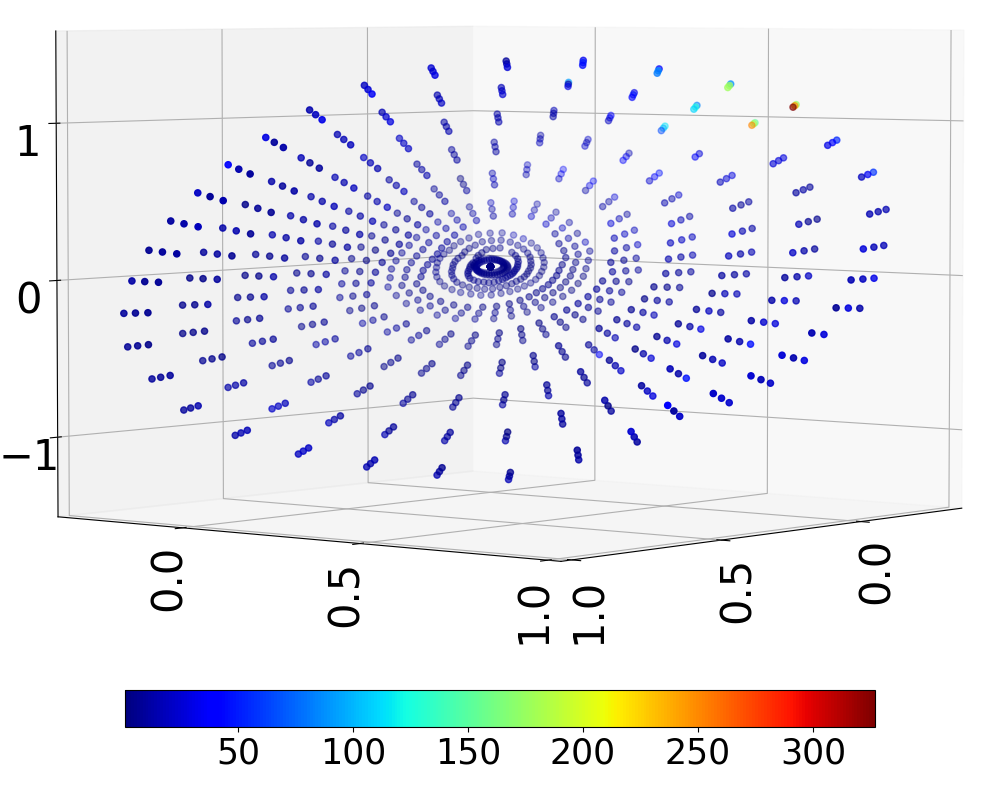}\\
    \hline
    M9 &
    \includegraphics[width=28mm]{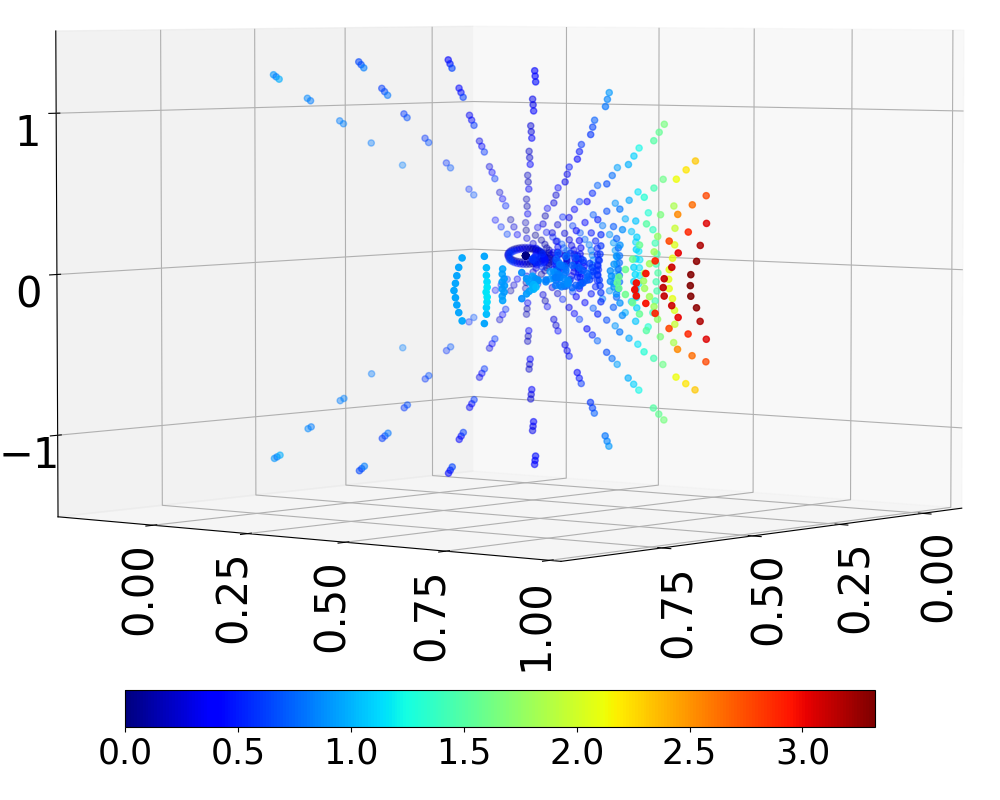}& \includegraphics[width=28mm]{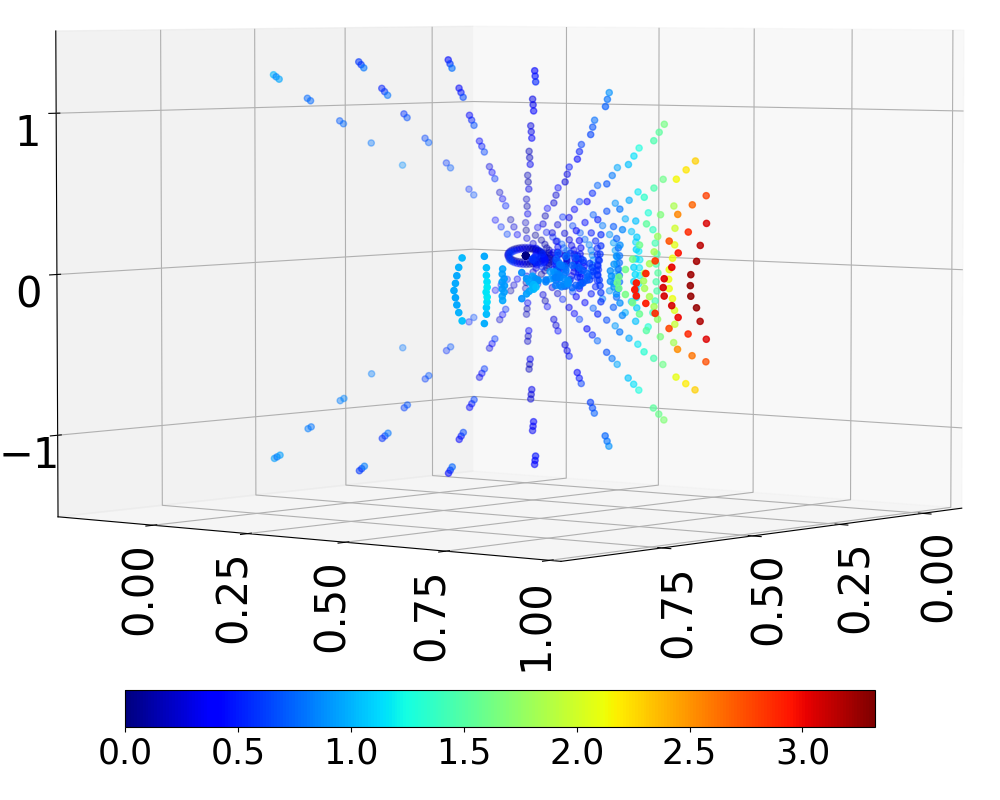}& \includegraphics[width=28mm]{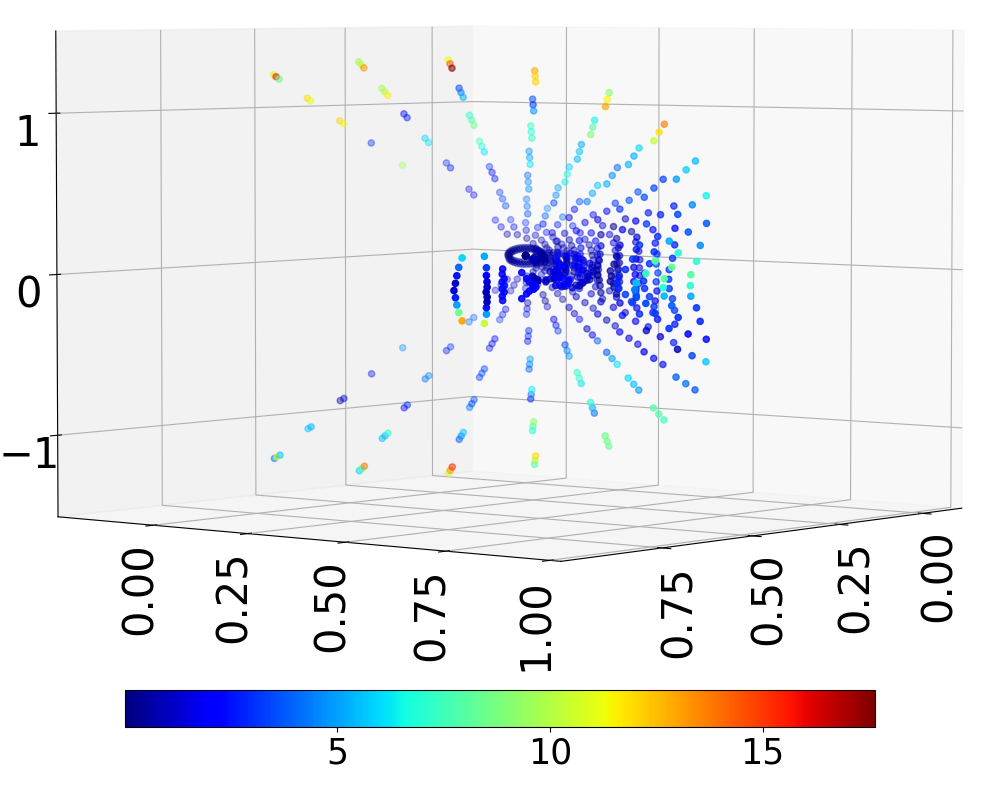}& \includegraphics[width=28mm]{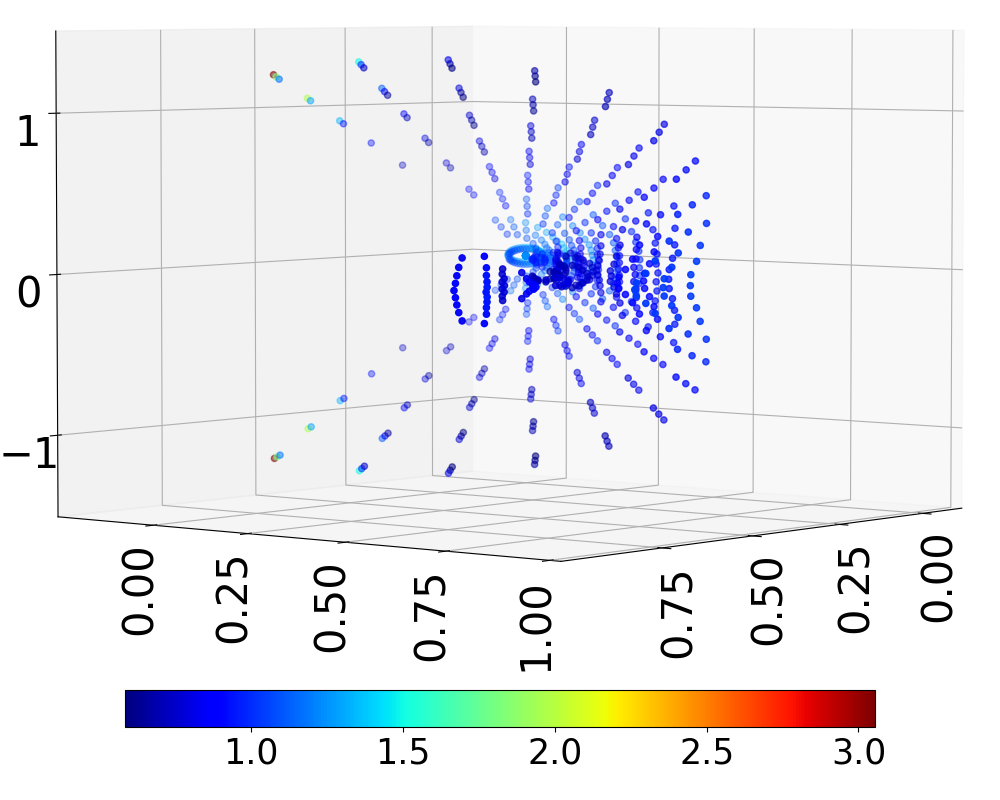}& \includegraphics[width=28mm]{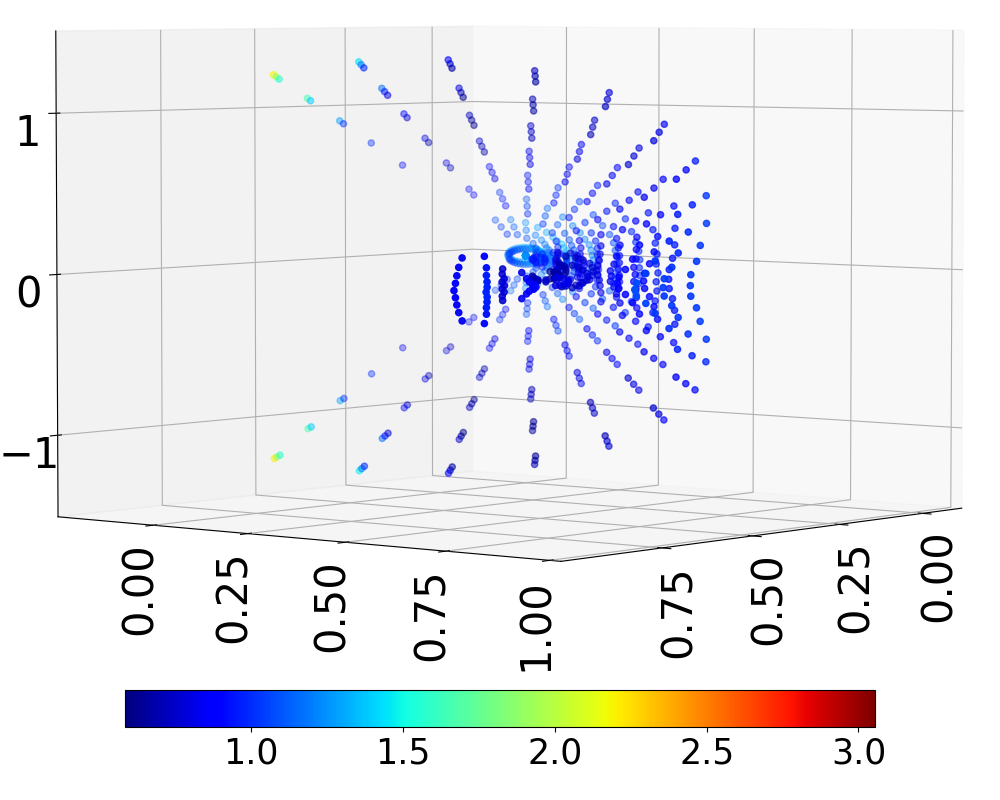}&
    \includegraphics[width=28mm]{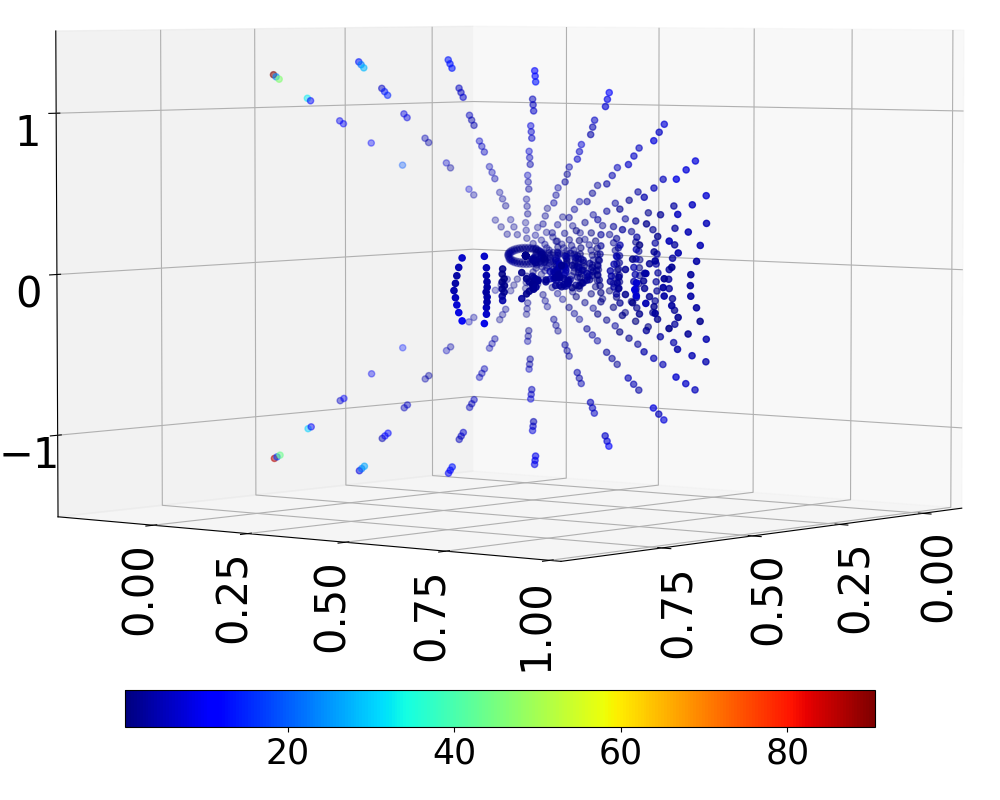}\\
    \hline
    M10 &
    \includegraphics[width=28mm]{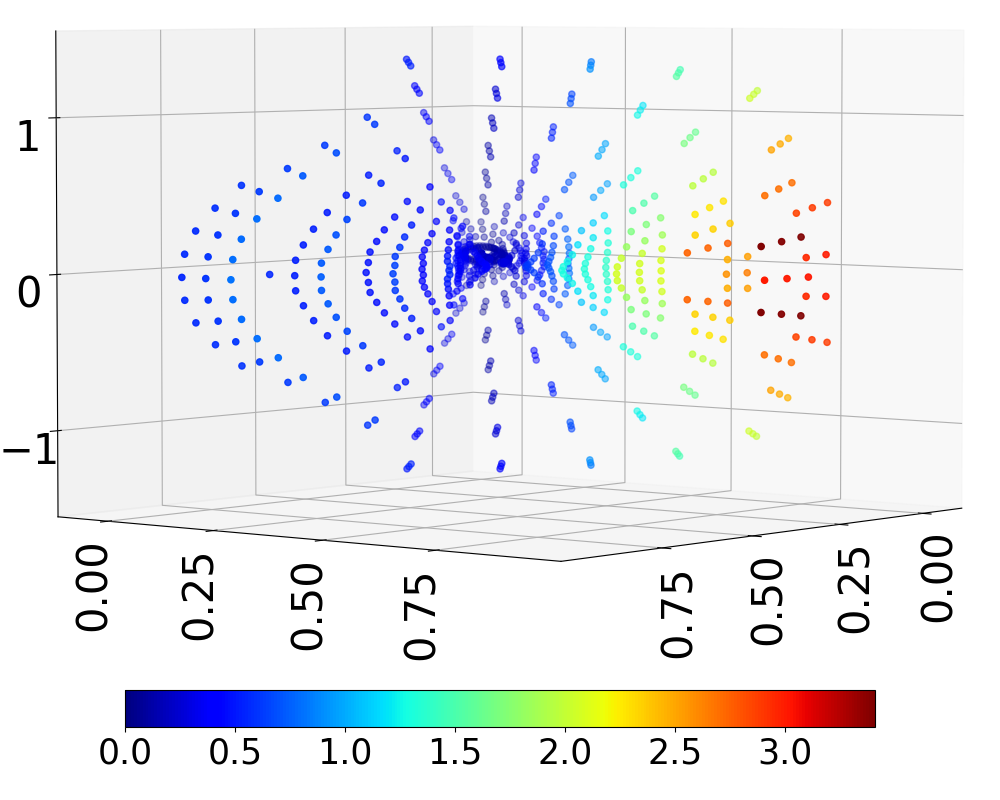}& \includegraphics[width=28mm]{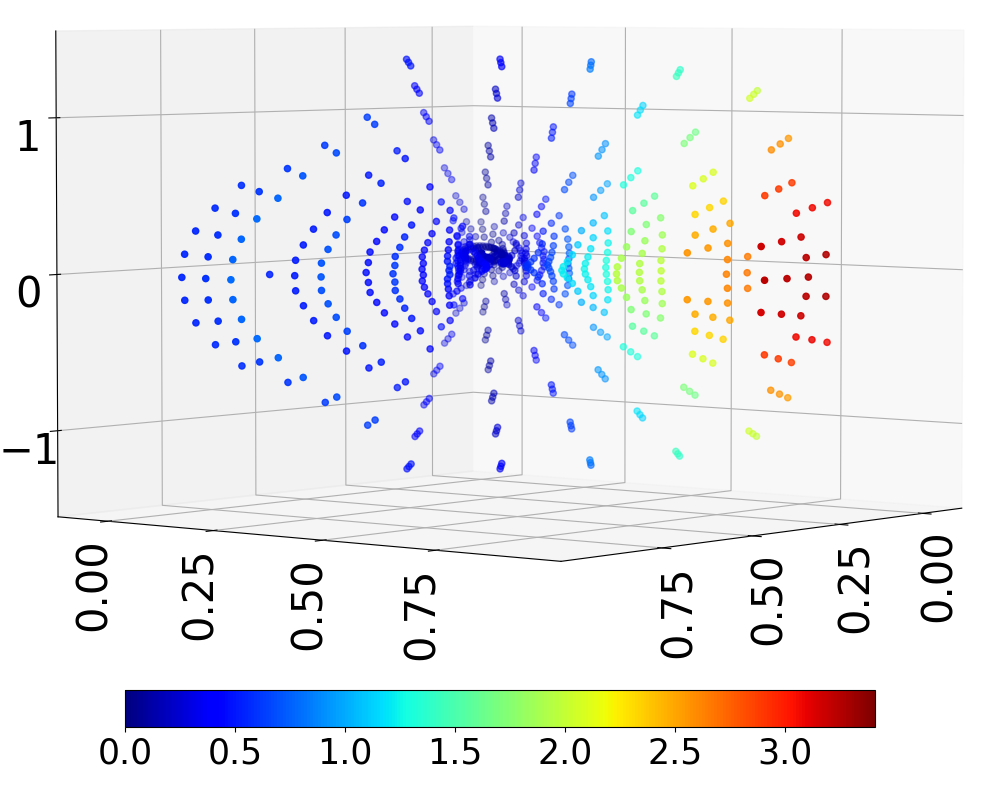}& \includegraphics[width=28mm]{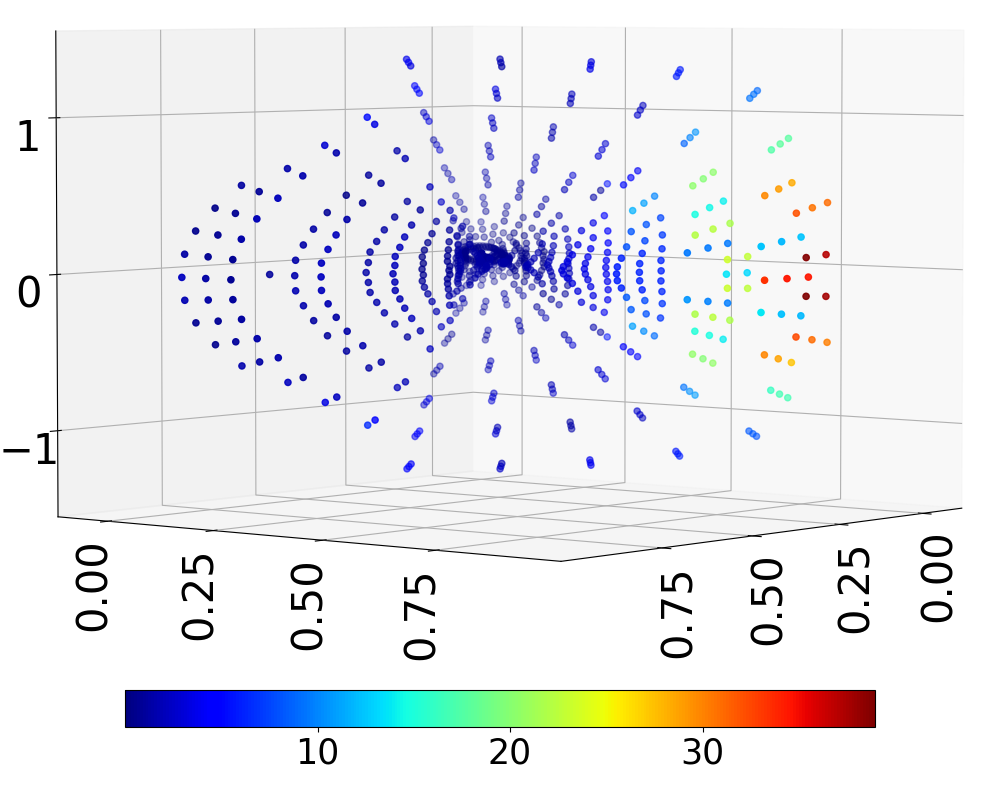}& \includegraphics[width=28mm]{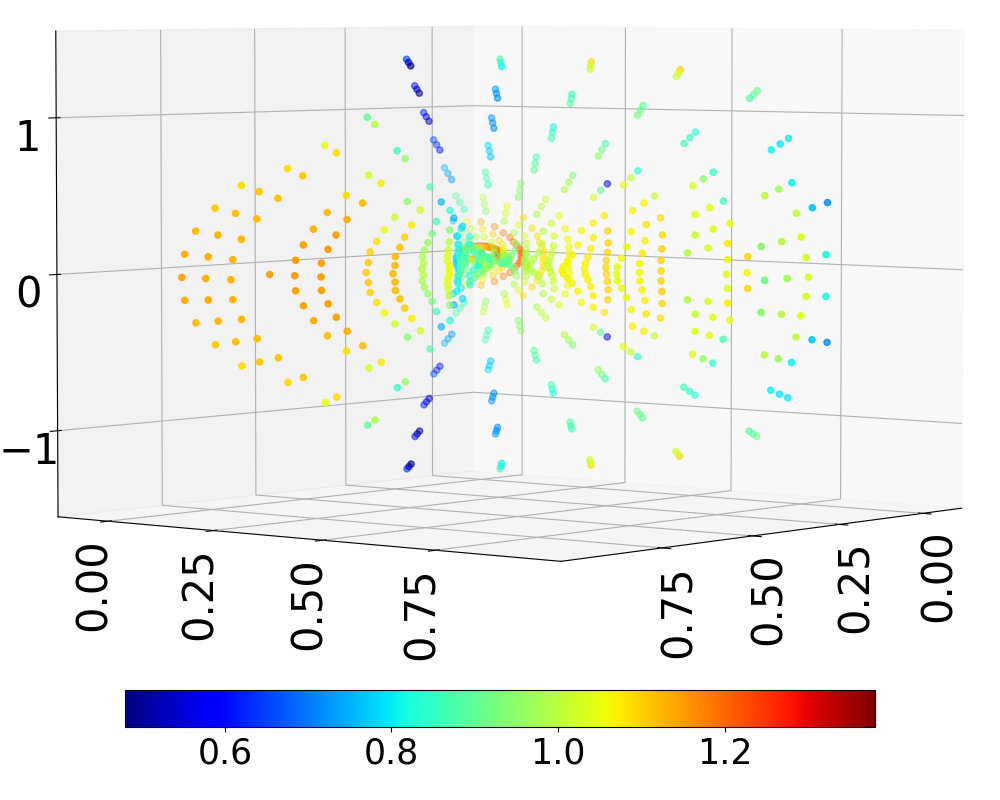}& \includegraphics[width=28mm]{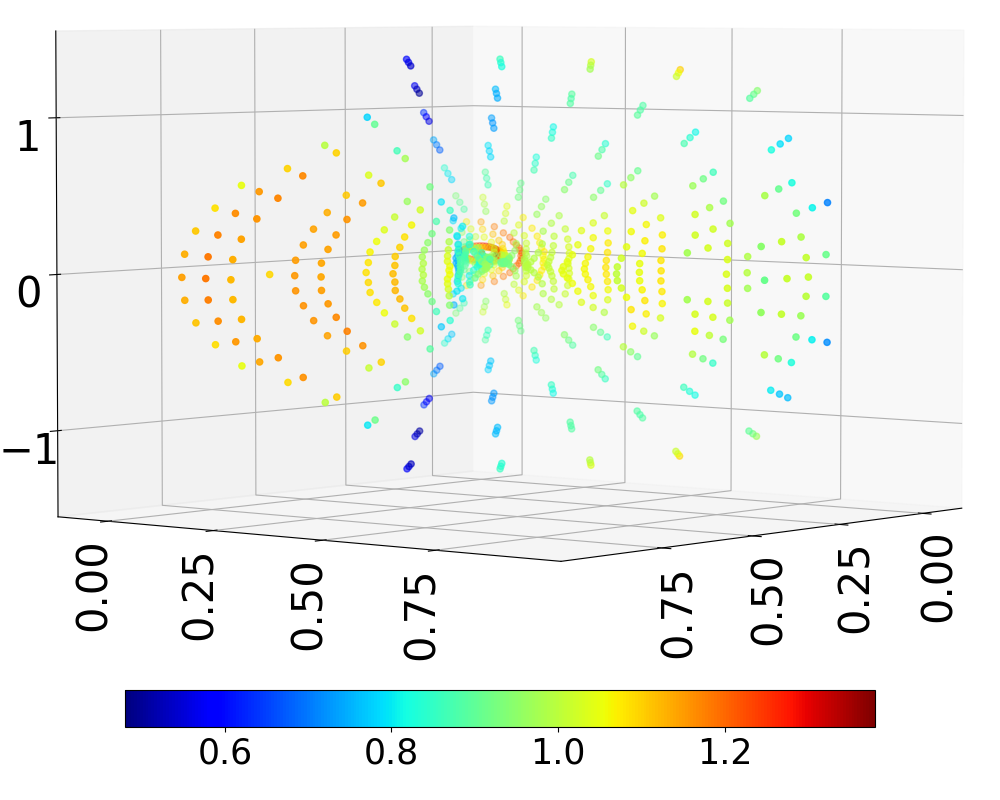}&
    \includegraphics[width=28mm]{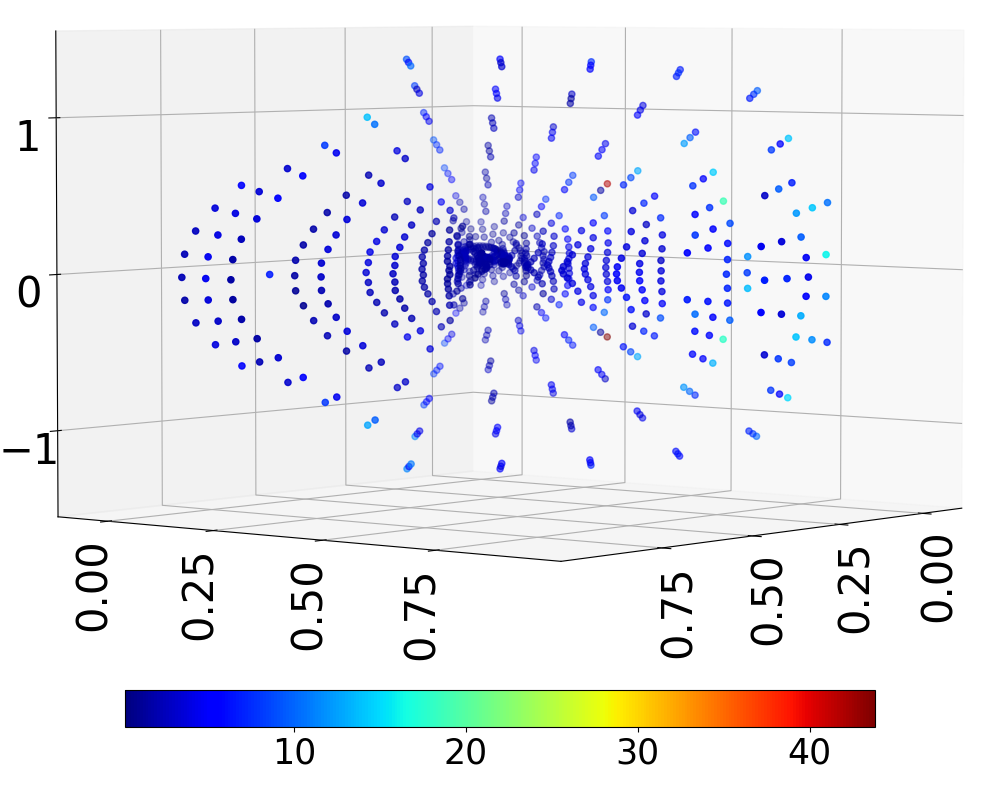}\\
    \hline
    M11 &
    \includegraphics[width=28mm]{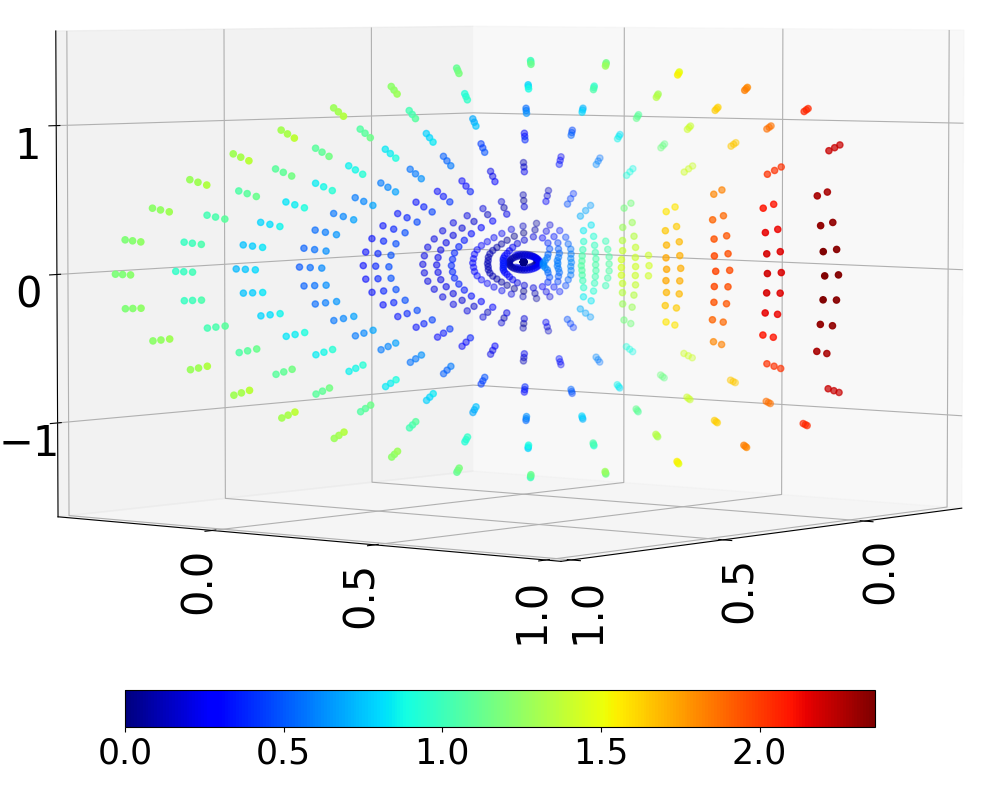}& \includegraphics[width=28mm]{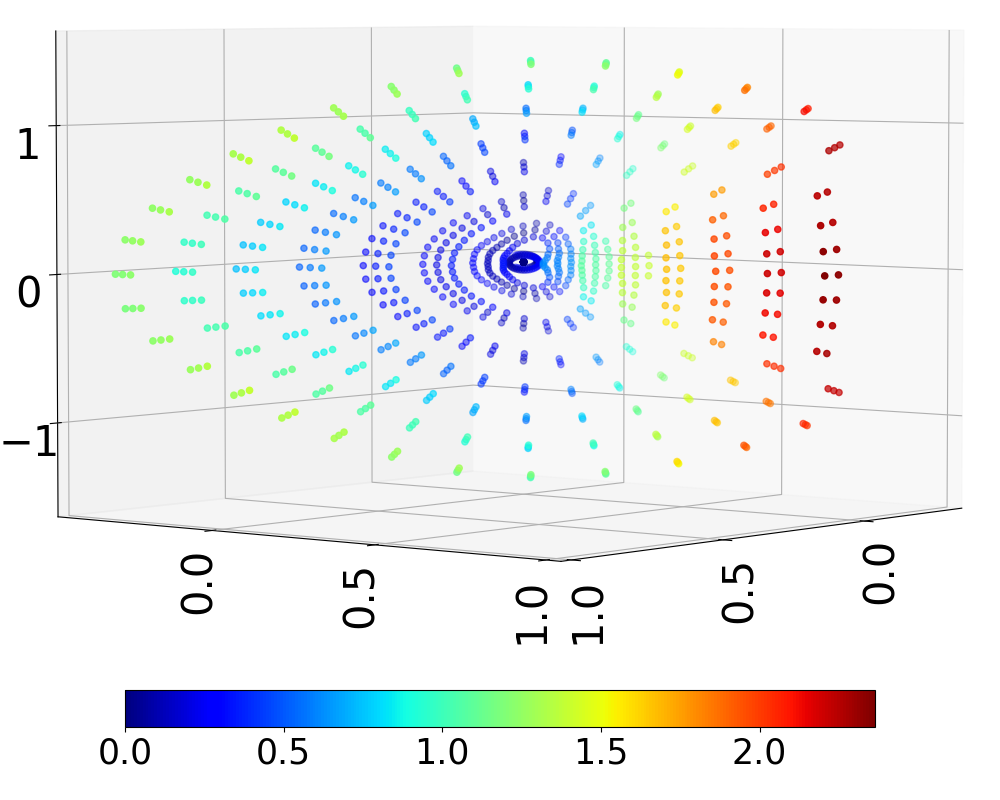}& \includegraphics[width=28mm]{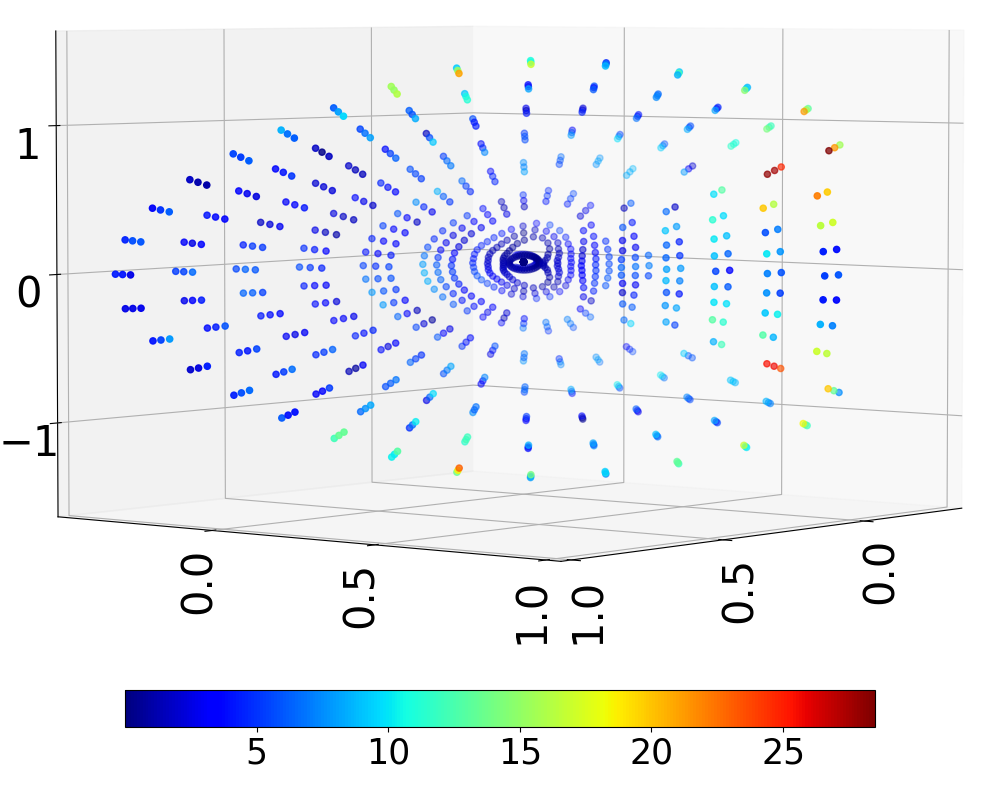}& \includegraphics[width=28mm]{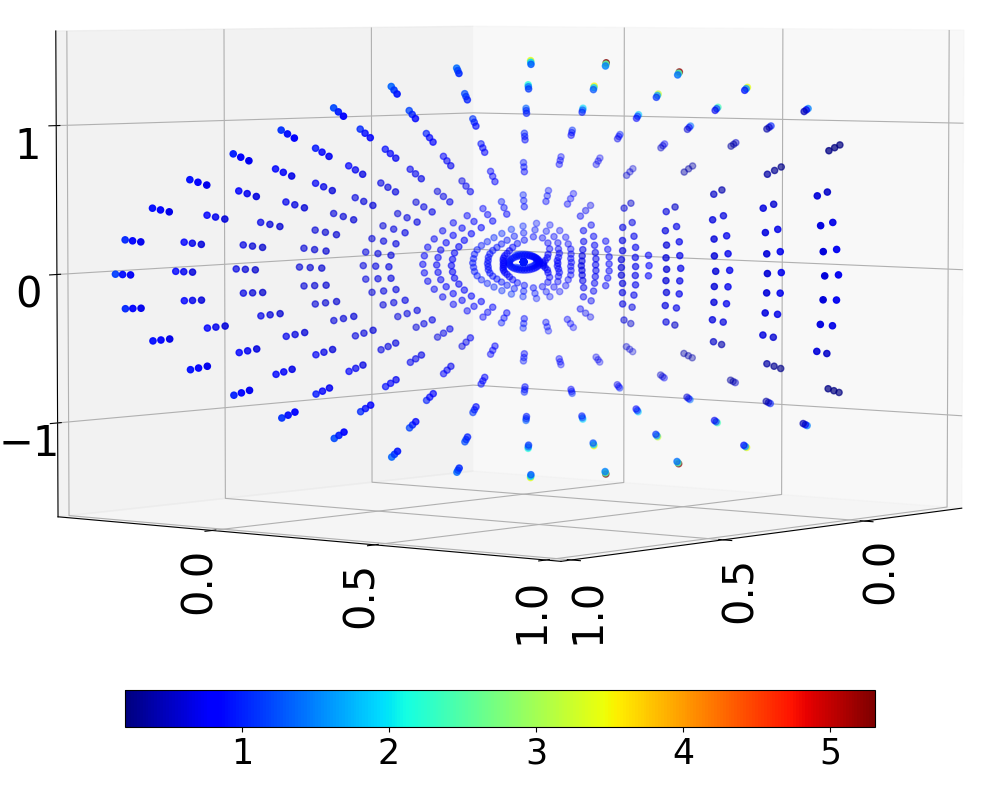}& \includegraphics[width=28mm]{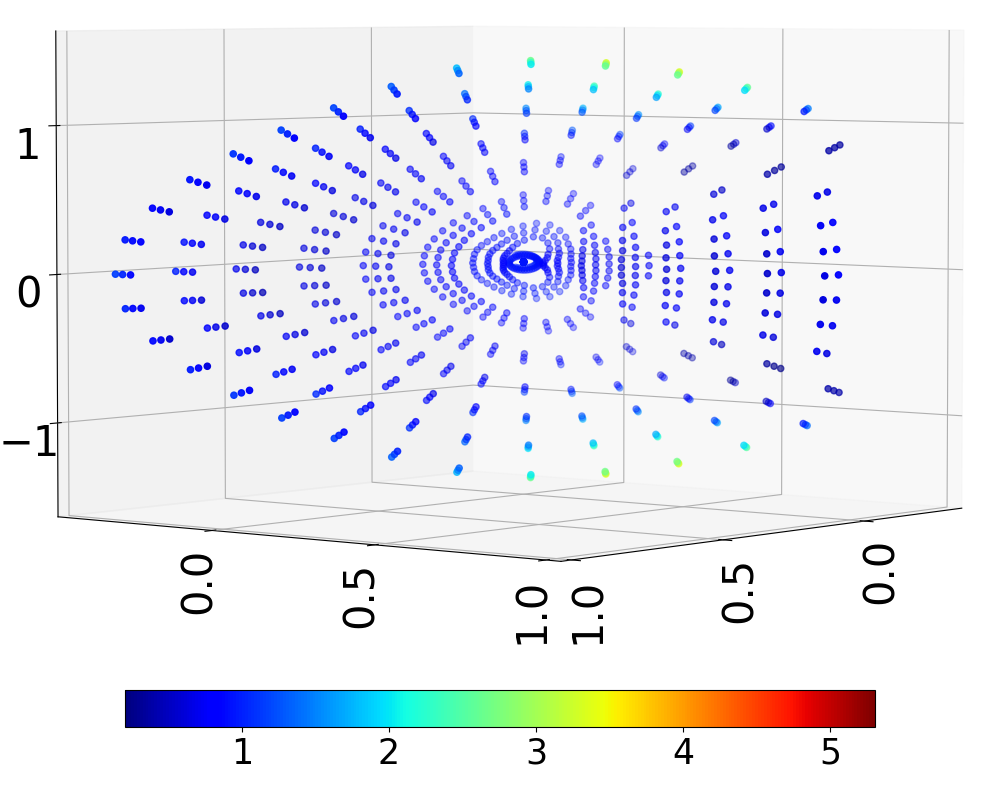}&
    \includegraphics[width=28mm]{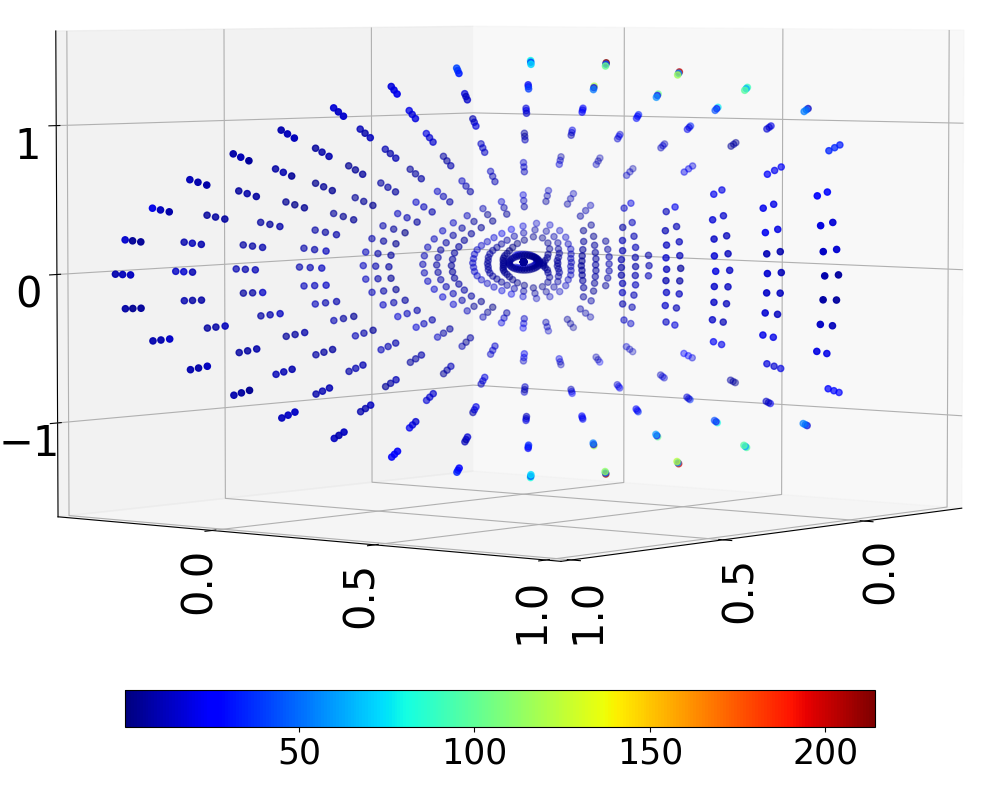}\\
\end{tabular}
\caption{Comparison of target and fitted stress and stiffness for materials 7 to 11. The colors indicate the norm of stress, stiffness, or error. \label{tab:bigtable2}}
\end{table*}

\section{Conclusions and Future Work}
\label{sec:concl}

In this paper, we have presented a novel formulation of elastic energy models based on high-order interpolants.
The interpolants extend scalar RBFs to provide local control over derivatives of the energy function, namely stress and stiffness.
We have shown that, when applied to the homogenization of 2D microstructures, our formulation provides higher accuracy than previous approaches.
The design of optimal high-order RBF interpolants is still an active research topic in numerical analysis~\cite{Drake2021}, and our methodology could see applicability in general high-order interpolation problems, beyond elastic simulation.
\added{To help with reproducibility, a sample implementation is available in the project webpage 
\url{http://mslab.es/projects/HiOInterp}
.}

We have also identified limitations that could motivate future work.
In particular, our current estimation methods appear limited when the stress or stiffness have strong local discontinuities.
This could be addressed by distributing RBF centers with non-uniform density and non-uniform radius.
\added{Similarly, it would be beneficial to sample the deformation range in an adaptive manner, adding training samples where nonlinearity appears higher.}
In general, it would be advantageous to find ways to make the parameterization of the resulting energy more compact.

We have applied our formulation and methodology only to in-plane deformation of 2D microstructures.
The possible extensions include: 3D microstructures, the bending response of thin shells \added{(necessary to apply the method to 3D cloth simulation)}, plasticity, and/or viscosity.
Some of the extensions may be straightforward, such as 3D microstructures or modeling viscosity by interpolating dissipation potentials~\cite{Sanchez2018}; others are unclear.

Finally, it would be interesting to use our methodology in the context of other applications beyond example-based homogenization.
These could include estimating materials from other types of data (e.g., force-deformation examples, or sparse observations of space-time deformations), or using the model in the context of material exploration.
\added{Obtaining homogenized strain from real-world force-deformation examples is straightforward. Stress can be obtained based on boundary forces~\cite{Schumacher2018}. Stiffness is not immediate, but it could be obtained through finite-difference approximation using incremental deformations.}

\ifdefined\final
\paragraph*{Acknowledgments.}
We would like to thank the anonymous reviewers for their feedback. 
We also want to thank Igor Santesteban for help with the rendering pipeline.
This work was funded in part by the European Research Council (ERC-2017-CoG-772738 TouchDesign).
\fi

\bibliographystyle{eg-alpha-doi}  
\bibliography{refs}        


\appendix

\section{RBF Derivatives}

We denote the derivatives of an RBF $\phi_i$ wrt its radius $r_i$ as:
\begin{equation}
\label{eq:RBFderivatives1}
    \phi'_i \equiv \DD{\phi_i}{r_i}, ~~~~ \phi''_i \equiv \DDTwo{\phi_i}{r_i}.
\end{equation}

For additional derivatives, it is convenient to define a radial unit vector $u_i = \frac{1}{r_i} \, \Delta x_i$.
Then, the derivatives of radius $r_i$ wrt the domain $x$ are:
\begin{equation}
\label{eq:radiusderivative}
    \DD{r_i}{x} = u_i^T, ~~~~ \DDTwo{r_i}{x} = \DD{u_i}{x} = \frac{1}{r_i} \, \left( I - u_i \, u_i^T \right).
\end{equation}

And following the chain rule, \refeq{RBFderivatives1} and \refeq{radiusderivative}, the derivatives of an RBF $\phi_i$ wrt the domain $x$ are:
\begin{equation}
\label{eq:RBFderivative}
    \DD{\phi_i}{x} = \phi'_i \, \DD{r_i}{x} = \phi'_i \, u_i^T.
\end{equation}
\begin{align}
\nonumber
    \DDTwo{\phi_i}{x} &= \phi''_i \, \DD{r_i}{x}^T \, \DD{r_i}{x} + \phi'_i \, \DDTwo{r_i}{x} \\
    &= \phi''_i \, u_i \, u_i^T + \frac{\phi'_i}{r_i} \, \left( I - u_i \, u_i^T \right).
\end{align}

\section{Equivalence of RBF Gradient}

Given an RBF $\phi_i$ defined by some choice of function $\phi$ and center $x_i$, there is some other RBF $\psi_i$, with choice of function $\psi$, such that:
\begin{equation}
    \nabla \phi_i = \psi_i \, \Delta x_i.
\end{equation}
As a corollary, any RBF interpolation based on RBF gradients can also be expressed as an RBF interpolation based directly on RBFs multiplied by radial vectors. This equivalent definition largely simplifies the computation of RBF derivatives.

To prove the equivalence, based on \refeq{RBFderivative} we have:
\begin{equation}
    \nabla \phi_i = \DD{\phi_i}{x}^T = \phi'_i \, u_i = \frac{\phi'_i}{r_i} \, \Delta x_i.
\end{equation}
And it follows that
\begin{equation}
    \psi_i = \frac{\phi'_i}{r_i}.
\end{equation}

\end{document}